\documentclass[12pt,a4paper,final]{iopart}

\usepackage{graphicx}
\pdfoutput=1
\expandafter\let\csname equation*\endcsname\relax
\usepackage{comment}

\usepackage[perpage]{footmisc}
\usepackage{hyperref}
\usepackage{doi}
\hypersetup{
    colorlinks=true,
    linkcolor=blue,
    filecolor=magenta,      
    urlcolor=blue,
}

\expandafter\let\csname endequation*\endcsname\relax
\usepackage{amsfonts,amsmath,amsthm,amssymb,amsopn}
\usepackage{dsfont,bbm}
\usepackage{mathrsfs}
\usepackage{leftidx}
\usepackage{cite}
\usepackage{booktabs}
\usepackage[retainorgcmds]{IEEEtrantools}
\usepackage{soul,xcolor}
\usepackage{color}
\usepackage{esint}

\def\no{\nonumber}
\def\be{\begin{equation}}
\def\ee{\end{equation}}

\newcommand{\ave}[1]{{\langle #1\rangle}}
\newcommand{\aveI}[1]{{\langle #1\rangle}}
\newcommand{\aveII}[1]{{\langle\langle #1\rangle\rangle}}

\newcommand{\ii}{ {\rm i} }
\newcommand{\dr}{\mathrm{dr}}
\newcommand{\p}{\partial}
\newcommand{\dd}{ {\rm d} }
\newcommand{\eff}{{\rm eff}}

\newcommand{\RaR}{\mathbb{R}}
\newcommand{\CC}{\mathbb{C}}

\newcommand{\half}{{\textstyle\frac{1}{2}}}

\def\tr{{\,{\rm tr}}}

\newcommand{\bfp}{\boldsymbol{\theta}^+}
\newcommand{\bfh}{\boldsymbol{\theta}^-}

\newcommand{\bfI}{\mathbf{I}}
\newcommand{\bft}{\boldsymbol{\theta}}
\newcommand{\bfk}{\boldsymbol{k}}
\newcommand{\bfx}{\boldsymbol{x}}
\newcommand{\bfy}{\boldsymbol{y}}
\newcommand{\bfz}{\boldsymbol{z}}

\eqnobysec
\definecolor{darkpink}{rgb}{1,0.08,0.5}

\begin{document}

\title{Correlation functions and transport coefficients in generalised hydrodynamics}

\author{Jacopo De Nardis$^*$, Benjamin Doyon$^\clubsuit$, Marko Medenjak$^\diamondsuit$, Mi{\l}osz Panfil$^\sharp$\\[0.3cm]
\it\footnotesize
$^*$ Laboratoire de Physique Théorique et Modélisation, CNRS UMR 8089,CY Cergy Paris Université, 95302 Cergy-Pontoise Cedex, France.\\
$^\clubsuit$ Department of Mathematics, King's College London, Strand, London WC2R 2LS, UK\\
$^\diamondsuit$ Institut de Physique Th\'eorique Philippe Meyer, \'Ecole Normale Sup\'erieure, 
PSL University, Sorbonne Universit\'es, CNRS, 75005 Paris, France\\
$^\sharp$ Faculty of Physics, University of Warsaw, ul. Pasteura 5, 02-093 Warsaw, Poland
}

\date{January 2021}

\begin{abstract}
We review the recent advances on exact results for dynamical correlation functions at large scales and related transport coefficients in interacting integrable models. We discuss Drude weights, conductivity and diffusion constants, as well as linear and nonlinear response on top of equilibrium and non-equilibrium states. We consider the problems from the complementary perspectives of the general hydrodynamic theory of many-body systems, including hydrodynamic projections, and form-factor expansions in integrable models, and show how they provide a comprehensive and consistent set of exact methods to extract large scale behaviours. Finally, we overview various applications in integrable spin chains and field theories.
\end{abstract}

\maketitle
\pagestyle{empty}
\tableofcontents
\pagestyle{headings}

\flushbottom
\clearpage

\section{Introduction}

Discerning the properties of strongly interacting quantum systems out-of-equilibrium is one of the most formidable challenges in theoretical physics. It is also particularly important from the point of view of recent experimental advances, which are now able to probe such regimes \cite{PhysRevLett.122.090601,malvania2020generalized,scheie2021}. The main difficulty is that understanding the time evolution of local quantities in quantum systems requires extracting an exponentially large (in the system size or time) amount of information contained in generic quantum many-body wave functions. Indeed, even for simple initial wave functions, quantum time evolution results in a linearly growing quantum entanglement \cite{Calabrese2005,Pasquale-ed,Eisert2015}, naively associated to an exponential increase of complexity. 
Arguably, the most interesting questions are related to the behaviours emerging from this complexity in many-body systems on large space and time scales. Before full equilibration or relaxation, a universal, hydrodynamic behaviour is expected to emerge, and on these large scales the complexity is typically reduced to much fewer degrees of freedom. Understanding how hydrodynamics emerges from microscopic reversible dynamics is related to some of the most fundamental questions of out-of-equilibrium statistical physics, such as irreversibility and the increase of entropy, or the emergence of the phenomenological laws of diffusion and transport. In this review we will address these aspects. Focusing on one-dimensional systems, we discuss both the general hydrodynamic theory of wide applicability, and the special class of \emph{integrable} systems.

In one spatial dimension a number of paradigmatic many-body interacting systems, such as the Heisenberg spin-1/2 chain \cite{Heisenberg1928}, the Fermi-Hubbard model \cite{Essler2005_Hubbard} for interacting electrons and the Lieb-Liniger model \cite{PhysRev.130.1605} for interacting bosons, are integrable. At the technical level, this allows us to develop advanced theoretical methods to better understand the inner workings of quantum (and also classical) statistical mechanics. One of the defining features of integrable many-body systems is that multi-particle scatterings can be decomposed in terms of pair-wise scattering events. This property can be used to construct eigenstates \cite{Bethe1931,PhysRev.130.1605,hep-th/9605187}, study the thermodynamics of such systems \cite{Yang1969,Takahashi1999} and calculate the expectation values of single point observables in stationary states \cite{Delfino2001,Kitanine2002,LeClair1999,Kormos2010,Pozsgay:2010xd}. Despite these notable results, computing multi-point dynamical correlations, which are of extreme experimental relevance \cite{Schlappa2012,PhysRevLett.115.085301,Schweigler2017,Langen2013,PhysRevA.91.043617,2102.08376,scheie2021}, is far from being resolved. Several approaches, directly based on microscopic formulations, have been employed in the past decades, including numerical summations of the spectral K\"all\'en–Lehmann sum \cite{2009_Caux_JMP_50,Schlappa2012,PhysRevA.89.033605}, analytical partial summations of matrix elements \cite{POZSGAY2008209,Pozsgay:2010cr,Pozsgay2018,Essler:2009zz,1742-5468-2012-09-P09001,doi:10.1063/1.5094332,10.21468/SciPostPhys.9.6.082,granet2020lowdensity,10.21468/SciPostPhys.9.3.033} and quantum transfer matrix  \cite{Dugave_2013,G_hmann_2017,2011.12752}. While this progress gives rise to formulae for correlation functions in certain integrable systems either in their ground state or at finite temperatures, a full characterisation, and the extraction of large-scale behaviours, in generic stationary states (thermal or not) is nowadays still lacking. For this purpose, methods based on hydrodynamics and on the emergent thermodynamic degrees of freedom seem to be more powerful.

Generalised hydrodynamics (GHD) was invented in 2016 \cite{PhysRevX.6.041065,PhysRevLett.117.207201} in order to describe the non-equilibrium behaviour of inhomogeneous states in integrable models, and provides a full and intuitive understanding of the dynamics of quantum and classical integrable models on large space and time scales. In particular, it has provided a handy set of tools to access the asymptotics of dynamical correlation functions and transport coefficients. The latter were part of an intense theoretical research in the past decade, which led to the development of effective approaches \cite{SD97,PhysRevB.57.8307} and some exact results, especially in spin chains \cite{PhysRevB.55.11029,PhysRevLett.82.1764,PhysRevB.84.155125,Ilievski2012,1742-5468-2014-9-P09037}. The purpose of this review article is to show how GHD, together with recent advances on the general theory of hydrodynamic projections, and on matrix elements in the thermodynamic limit (the so-called thermodynamic form factors \cite{Doyon_2007,DeNardisP2018,Bootstrap_JHEP}), provides a unifying description of the asymptotic dynamical correlation functions and transport coefficients in generic stationary states, putting together many previous results in one framework. Notably the combination of hydrodynamic theory and thermodynamic form factors, allowed the computation of diffusion and conductivity coefficients, representing the first instance where these coefficients can be obtained from the microscopic dynamics at arbitrary temperatures, in both quantum and classical systems. A number of general results going beyond the standard hydrodynamic theory of many-body systems are briefly reviewed, including nonlinear response, higher-point correlation functions and full counting statistics, linear response on top of non-stationary states, and projection mechanisms for diffusion.

The main ideas that come out from combining the GHD and form factor perspectives, and which we endeavour to explain in this review, are as follows.
\\

\noindent {\em Euler scale:} In general, the leading algebraic correlations at ballistic scales of space and time are controlled by small waves propagating on top of the thermodynamic state. Mathematically, such ballistic waves are identified with extensive conserved quantities, and span the tangent space to the state manifold. They form the Hilbert space of emergent degrees of freedom onto which local observables are projected. In integrable models, the space of such waves is infinite-dimensional. It can be seen as particle-hole pairs co-propagating at a continuum of possible velocities. The result of the projection is given by appropriate thermodynamic form factors.\\

\noindent {\em Diffusive scale:} Diffusive corrections may have many sources, but one of them is the scattering of ballistic waves, which is accessed by nonlinear response theory. In general, this provides a lower bound on diffusion constants. This is the only source of diffusive corrections in integrable models, so the bound is saturated. In this case, diffusion is related to the scattering of two particle-hole pairs, and is again given by appropriate thermodynamic form factors.

\subsection{Content}

The review is organised as follows. The first two sections of the review deal with one-dimensional many-body systems in general. In sec. \ref{sec:hydro} we review general concepts appearing in hydrodynamic theory, clarify the assumptions and briefly discuss how the hydrodynamic description emerges in integrable models. In sec. \ref{sec:hydroCorr} we discuss the phenomena of transport, diffusion and entropy production, establish the relations between these concepts, and show how explicit lower bounds can be obtained by projections onto appropriate conserved quantities. We also discuss how the higher point functions, coding for fluctuations, can be computed. From the fourth section onward we focus on integrable systems. In sec. \ref{sec:corrFF} we show how the dynamic correlation functions can be accessed via expansions over intermediate quasiparticle excitations with the thermodynamic form factors. We proceed by reviewing how these results can be used on different hydrodynamic scales, in particular  on the Euler scale, in sec. \ref{sec:euler}, and on the diffusive scale in sec. \ref{sec:diffusion}. Finally in sec. \ref{sec:applications} we provide a review of some notable results on correlation functions and transport coefficients that have been obtained with the present formalism in a number of systems, ranging from $T\bar{T}$-deformed conformal field theories, quantum Heisenberg spin chains, Lieb-Liniger model to the classical Sinh-Gordon field theory.

\section{Basics of Hydrodynamics}\label{sec:hydro}

Hydrodynamics is an extremely general framework for the large-scale dynamical properties of statistical mechanics away from equilibrium. It is applicable to a wealth of seemingly very different physical setups: lattices of classical or quantum degrees of freedom in interaction, gases of particles \cite{Spohn1982,Spohn1991} or other emergent objects such as solitons \cite{PhysRevLett.120.144101}, field theories, or even cellular automata \cite{Medenjak17,medenjak2019two,klobas2018exactly,Klobas2019,10.21468/SciPostPhysCore.2.2.010,PhysRevLett.123.170603}, under deterministic or stochastic time evolution. In the context of the present review, we focus on one-dimensional quantum systems with Hamiltonian evolution, although most results within this section hold, with minimal adjustments, in other setups in one dimension.

The starting point of hydrodynamics is the set of {\em local conservation laws} admitted by the many-body model under consideration, and a closely related concept of {\em homogeneous, stationary, ergodic states}.

On one hand, the conservation laws are the continuity equations for local densities and currents $q_i(x,t)$, $j_i(x,t)$ at space-time coordinates $x,t$ (under Heisenberg time evolution):
\begin{equation}\label{continuity}
    \partial_t q_i(x,t)+\partial_x j_i(x,t)=0.
\end{equation}
with the densities that comprise the local conserved quantites, $[H, Q_i]=0$, as 
\begin{equation}\label{eq:densitiesQ}
    Q_i = \int \dd x\, q_i(x),
\end{equation}
where the integral over $x$ can be either a sum over a discrete infinite lattice or a continuous integration over the real line, depending on the definition of the microscopic model [this does not affect any of the concepts or results discussed in this review]. 
If space and/or time is discrete, the partial derivatives are to be replaced by appropriate finite-difference expressions in \eqref{continuity}. A local density or current at $x$ is an observable mainly supported on a neighbourhood of $x$. It is beyond the scope of this review to fully specify the concept of locality, but we mention that we include the quasi-local observables discussed for instance in \cite{ilievski2016quasilocal,Essler_2015_GGE}.

The definition of charge densities \eqref{eq:densitiesQ} requires the choice of a gauge, as these are always defined up to a total derivative with respect to space 
\begin{equation}
    q_i(x) \to q_i(x) + \partial_x g(x)
\end{equation}
with $g(x)$ a generic local operator. Partial gauge fixing is achieved by assuming, as we do throughout this review, $PT$ symmetry, meaning that both charge densities and currents are invariant under simultaneous inversions of the signs of space and time\footnote{$PT$ symmetry is an anti-linear involution that reverses the sign of space and preserves the algebra of observables.}. This assumption can be fulfilled in a wide family of models with time-reversal invariant Hamiltonian dynamics. We shall see later how this also implies the standard Einstein's relation between conductivity and diffusion constant. To our knowledge, the relevance of $PT$ symmetry to hydrodynamics was first pointed out in \cite{De_Nardis_2019}.

On the other hand, the stationary, homogeneous, ergodic states are the states $\langle \cdots\rangle =  \tr(\rho \cdots)$ (with $\tr \rho = 1$), supported on the line $\RaR$,  which are invariant under space-time translations, for instance \begin{equation}\label{homostat}
\ave{o_1(x,t)o_2(y,s)} = \ave{o_1(x+z,t)o_2(y+z,s)} = \ave{o_1(x,t+z)o_2(y,s+z)},
\end{equation}
and clustering at large distances \footnote{Here and below, if the time coordinate is not written, time is set to 0.},
\begin{equation}\label{clustering}
\ave{o_1(x)o_2(y)} \xrightarrow[|x-y|\to \infty]{} \ave{o_1(x)}\ave{o_2(y)}.
\end{equation}
These states are referred to as ``ergodic" following the nomenclature in the context of quantum statistical mechanics \cite{convexity}, where a weak condition of clustering is referred to as ergodicity. Clustering is loosely connected to the more usual notion of ergodicity, the equivalence of the time average with the ensemble average. Indeed, a theorem in quantum spin chains shown in \cite{hydroprojectionsEuler} says that, if uniform clustering holds outside of a finite-velocity light-cone (for instance as a consequence of the Lieb-Robinson bound \cite{Lieb1972}), for any local observable $o(x,t)$, the time-averaged observable $ t^{-1} \int_0^t \dd s\,o(x,s)$ has vanishing variance at large $t$ in the state $\ave{\cdots}$. Mathematically, ergodic states are important because they are extremal states: they cannot be written as convex linear combinations of other states, and can always be used to decompose other states. Physically, they appear to be the states which arise under relaxation.

Crucially, according to what is sometimes referred to as the ``ergodic principle", there is a deep connection between the local conservation laws and the homogeneous, stationary, ergodic states admitted by the system: the latter are all states that maximise entropy under the conditions of fixed averages of all local conserved densities. Thus, formally, they are those with density matrices of the Gibbs form,
\begin{equation}\label{eq:gge}
    \rho = \frac1Z e^{-\sum_i \beta^i Q_i}.
\end{equation}
These are conventionally referred to as Gibbs ensembles in non-integrable systems, where a finite number of conservation laws exist, and generalised Gibbs ensembles (GGEs) in integrable systems \cite{Vidmar2016}, admitting an infinite number of conservation laws. We will often refer to them as maximal entropy states.

\begin{figure}
    \centering
    \includegraphics[width=6cm]{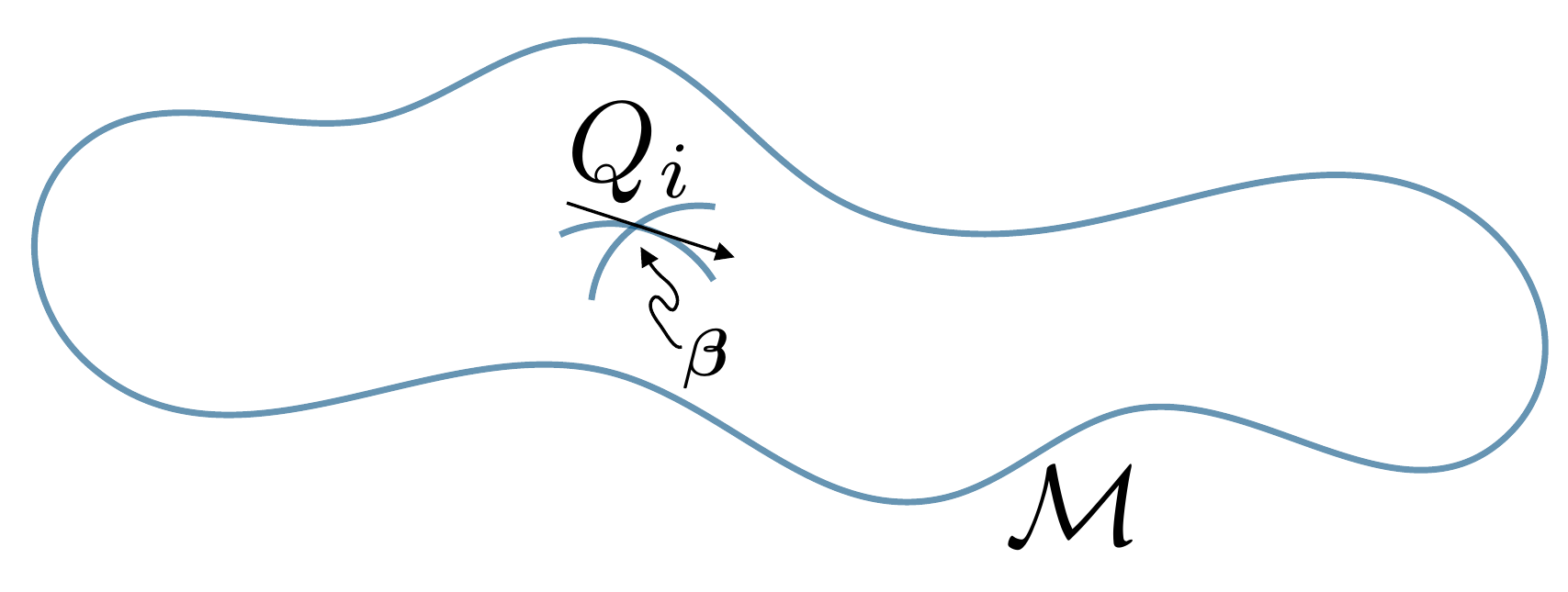}
    \caption{The maximal entropy states form a manifold $\mathcal M$. The extensive conserved quantities $Q_i$ describe its tangent spaces $T\mathcal M_{\boldsymbol\beta}$, and form a Hilbert space. The specific description of the Hilbert space $T\mathcal M_{\boldsymbol\beta}$ in terms of local observables may depend on the point $\boldsymbol\beta$.}
    \label{fig:manifold}
\end{figure}

Although \eqref{eq:gge} is a convenient formal representation of the maximal entropy states, for a universal theory of hydrodynamics, it does not matter how many conservation laws there are, finite or infinite, and it does not matter if the density matrix may or may not be written in the form \eqref{eq:gge}, in terms of any ``natural" set of conserved quantities with a convergent sum. In all cases, the maximal entropy states are simply the homogeneous, stationary, ergodic states, and their physical meaning is that they are the states that are expected to be reached at very long times after local relaxation has occurred (see, e.g.~\cite{Eisert2015}). Their relation to the conserved quantities is rather a geometric re-interpretation of \eqref{eq:gge}: the conserved quantities $Q_i$ span the tangent space $T\mathcal M_{\boldsymbol\beta}$ to the manifold $\mathcal M$ of maximal entropy states at the point whose coordinates are the conjugate thermodynamic potentials $\boldsymbol\beta$ \cite{pseudolocality}, see fig.~\ref{fig:manifold}. They form a Hilbert space, whose specific description may strongly dependent on the exact way local densities cluster in the state $\boldsymbol\beta$ (this Hilbert space is reviewed in sec. \ref{subsec:projectionsEuler}).

The above are the building blocks of hydrodynamics. In order to describe how hydrodynamics emerges, we consider a state $\langle \cdots\rangle$ of the microscopic model, say supported on the line $\RaR$,  and its evolution in time $\langle e^{\ii Ht} \cdots e^{-\ii Ht}\rangle$. The state is not necessarily stationary nor homogeneous, but should be ergodic \eqref{clustering}. We are looking for a theory describing the profiles of local observables and their correlations in space-time.

\begin{figure}
    \centering
    \includegraphics[width=10cm]{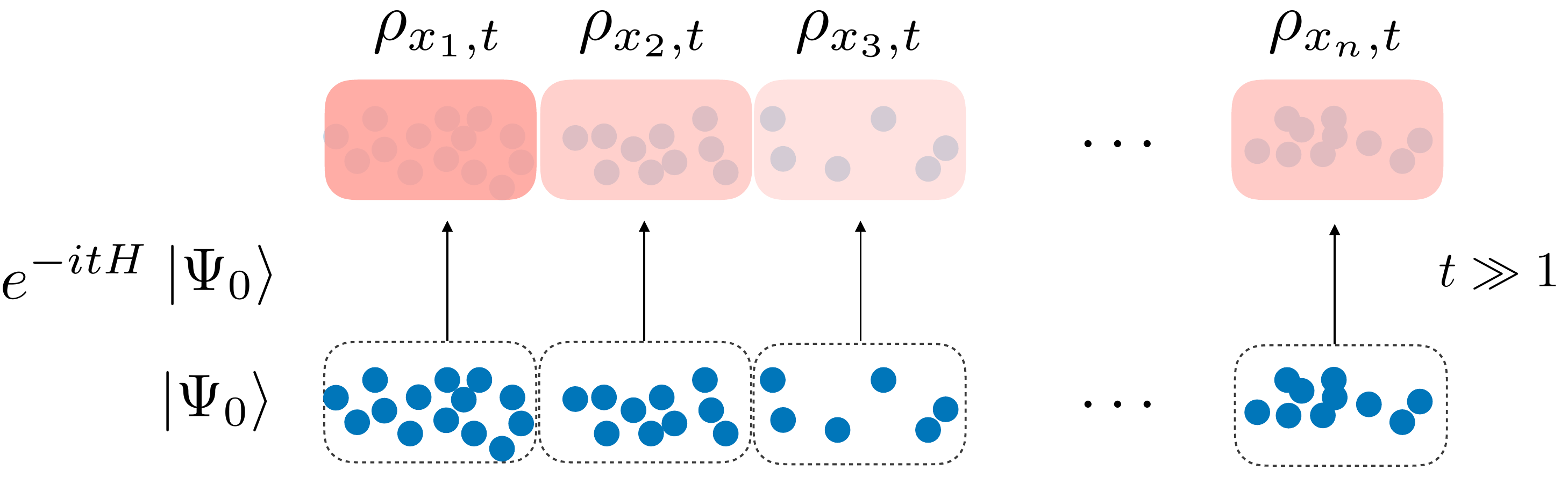}
    \caption{Cartoon representation of the hydrodynamic assumption, Eq. \eqref{replacement}. An initial non-equilibrium state is let to time evolve under Hamiltonian $H$. At large times, the quantum state in each mesoscopic fluid cell (whose size is much larger than microscopic scales and much smaller than the system size) is replaced by a large set of local thermodynamic states. Each thermodynamic state at $x_k$ is determined solely by the expectation values of the local densities of charges ${\tt q}_i(x,t)|_{x=x_k}$ and their spacial derivatives $\partial_x^n {\tt q}_i(x,t)|_{x=x_k}$.}
    \label{fig:hydrof}
\end{figure}

The hydrodynamic theory purports that for expectations of local observables on large space-time scales, after local equilibration has occured, one may replace the initial state and all of its time evolutes by a set of {\em space-time dependent} thermodynamic states, see fig. \ref{fig:hydrof},
\begin{equation}\label{replacement}
    \{\langle e^{\ii Ht}\cdots e^{-\ii Ht}\rangle:t\in\RaR\} \to \{\langle\cdots\rangle_{x,t}:(x,t)\in\RaR^2\}.
\end{equation}
The thermodynamic state $\ave{\cdots}_{x,t}$ represents what the time-evolved initial state ``looks like" at sufficiently large space-time point $x,t$. Hydrodynamics further purports that we may use the manifold of homogeneous, stationary, ergodic states to describe the map $x,t\mapsto \ave{\cdots}_{x,t}$: we  associate to it the density matrices
\begin{equation}\label{localGGE}
    \rho_{x,t} = \frac1Z \exp\left[-
    \sum_i \beta^i(x,t) Q_i\right]
\end{equation}
or equivalently the Lagrange parameters $\beta^i(x,t)$'s. More precisely, the state $\ave{\cdots}_{x,t}$ is determined by the point $\beta^i(x,t) \in \mathcal M$ and its local variation $\in T\mathcal M_{\boldsymbol\beta(x,t)}$ as function of $x$.

Physically, $\ave{\cdots}_{x,t}$ is the state in the mesoscopic fluid cell surrounding $x,t$. It is a state supported on the line, which is the extension of the fluid cell around $x,t$. The idea is that relaxation has nearly occurred in the cell, so the local thermodynamic state is nearly homogeneous and stationary, and entropy has nearly maximised, while keeping averages of extensive conserved quantities $Q_i$ fixed as they do not change under mesoscopic-time evolution in the cell.

The actual mechanism for (near) local entropy maximisation depends on the type of system at hand. In a classical system with non-fluctuating initial state, this is usually seen as a consequence of ergodicity in time: one needs to perform a mesoscopic time average, which then reproduces the ensemble average of the local thermodynamic state. In a quantum system with pure initial state, the phenomenon of decoherence is involved instead, without the need for a mesoscopic time average. An ensemble is generated by virtue of the entanglement between the local cell and the rest of the system. The entanglement entropy \cite{Calabrese2005} of the mesoscopic cell measures the entropy of this local ensemble. Quantum fluctuations are sufficient for entropy to maximise, without the need for classical state fluctuations. If the initial state is itself an ensemble, then no time average is needed in the classical case, and the concept of decoherence does not have to be employed in the quantum case.

In the replacement \eqref{replacement}, although the state $\langle \cdots \rangle_{x,t}$ is near to a maximal entropy state \eqref{localGGE}, it should be emphasised that it is {\em not equal to} a maximal entropy state $\langle \cdots \rangle_{x,t} \neq \tr( \rho_{x,t}\cdots)$. Instead, the expectation value of every local observable is subject to the {\em hydrodynamic expansion}, which is an expansion in the spatial derivatives of $\beta^i(x,t)$'s. That is, the expectation value of an observable at the point $x$ and time $t$ does not depend only on the state at that particular point, but also on the nearby points. In place of $\beta^i(x,t)$'s, one may take the average densities ${\tt q}_i(x,t) = \langle q_i\rangle_{x,t}$'s as a parametrisation, since they are physically more meaningful. Conveniently, one  {\em defines} the $\beta^i(x,t)$'s by inverting the defining relation for density averages
\begin{equation}\label{qiaverage1}
{\tt q}_i(x,t) =\tr(q_i \rho_{x,t}).    
\end{equation}
By thermodynamics, the ${\tt q}_i(x,t)$ are fixed by the local fluid-cell specific free energy $\mathsf f(x,t)$. Then, one postulates for the average currents ${\tt j}_i(x,t) = \langle j_i\rangle_{x,t}$:
\begin{equation}\label{hydroexpansion}
    {\tt j}_i(x,t) =
    \mathsf J_i[{\tt q}_\bullet(x,t)]
    - \frac12 \sum_j \mathfrak D_i^{~j}[{\tt q}_\bullet(x,t)] \partial_x {\tt q}_j(x,t) + \ldots,
\end{equation}
which reflects that the expectation values of the currents at point $x$ can be expressed in terms of expectation values of the complete set of conserved charge densities ${\tt q}_\bullet(x,t)$ in the vicinity.

The leading order, the {\em Euler scale}, is obtained from the local entropy maximisation principle, because \eqref{hydroexpansion} should hold, as a particular case, in homogeneous, stationary states. That is,
\begin{equation}
\mathsf J_i[{\tt q}_\bullet(x,t)] = \tr(j_i \rho_{x,t}).
\end{equation}
Here the right-hand side is a function of the ${\tt q}_i(x,t)$'s via the change of variables from $\beta^i(x,t)$'s to ${\tt q}_i(x,t)$'s given by \eqref{qiaverage1}. If the state $\rho_{x,t}$ involves few conserved quantities, say the number of particles, the momentum and the energy, then the associated currents $\tr(j_i \rho_{x,t})$ are also fixed by the specific free energy $\mathsf f(x,t)$ \cite{freeenergyfluxes} thanks to the Kubo-Martin-Schwinger relation. More generally, however, it is a nontrivial problem to evaluate the currents.

\begin{figure}
    \centering
    \includegraphics[width=7cm]{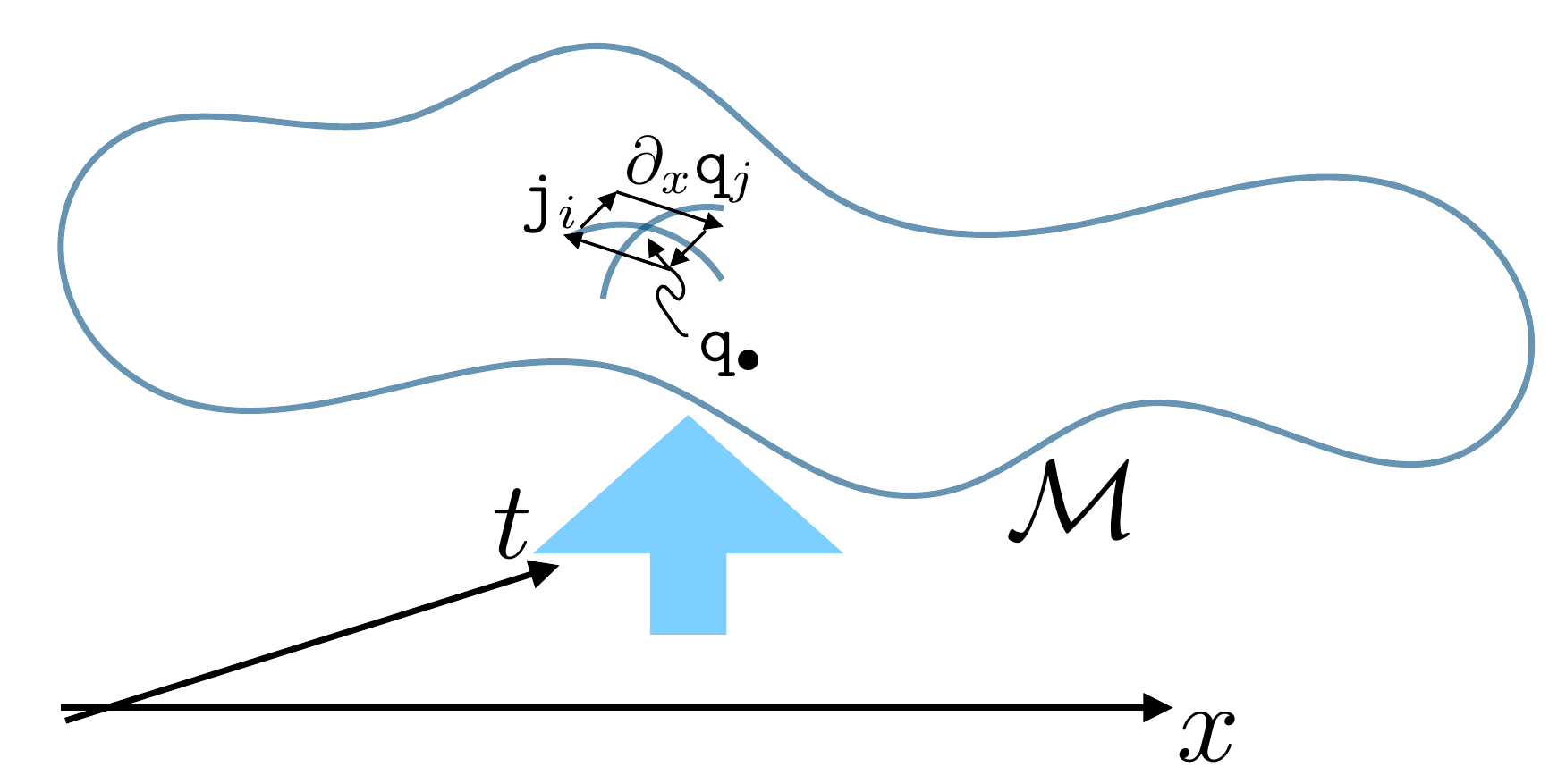}
    \caption{In the hydrodynamic expansion, the current is determined not only by the state in the fluid cell, but also by its local variation. This, geometrically, is related to the curvature of the state manifold.}
    \label{fig:manifold2}
\end{figure}

The next order in \eqref{hydroexpansion} is even more tricky, and will be one of the main topics of our discussion. The matrix $\mathfrak D_i^{~j}$ is the {\em diffusion matrix}, and is one of the transport coefficients. As it relates the current, which describes the time variation of local coordinates, to the spatial derivative of the state's coordinates, it is a coupling between tangent vectors and thus is related, geometrically, to the curvature of the state manifold (this is discussed in \cite{Ben}); see fig.~\ref{fig:manifold2}. Similar expansions hold for any local observables, and  characterise the dynamics of the nearly-entropy-maximised states $\ave{\cdots}_{x,t}$.

Once the hydrodynamic approximation \eqref{hydroexpansion} is made, the emergent large-scale dynamics on $\ave{\cdots}_{x,t}$ can be deduced by using Eq.~\eqref{continuity}. By averaging over the fluid cells, one can show that these imply that continuity holds for the averages of densities and currents taken within the space-time dependent thermodynamic states:
\begin{equation}\label{hydroequations}
    \partial_t {\tt  q}_i(x,t)+\partial_x  {\tt j}_i(x,t)=0.
\end{equation}
That is, the dynamics is transferred from that of the initial state $\ave{\cdots}$ on the line $\RaR$, with a very large number of degrees of freedom, to a dynamics on maximal entropy states, i.e.~on $\beta^i(x,t)$'s. Thus, the space of states at any given point $x,t$ is much smaller than the space of local observables there. The space of states is ``as large as" the space of conserved densities, and Eq.~\eqref{continuity} gives, in principle, enough equations for the emergent dynamics to be a well-posed initial-value problem. The reduction of the number of degrees of freedom implied by the hydrodynamic approximation is the basis for its predictive power.

Finally, restricting to the case of quantum and classical integrable models, an important idea is that, thanks to factorised and elastic scattering, it is possible to pass from a description in terms of many conservation laws to the  {\em quasiparticles} or modes of quantum integrability (or the solitons and radiative modes of classical integrability). Quasiparticles are identified with the asymptotic objects forming the scattering states of the many-body models. For each conserved quantity $Q_i$ there exists a function $h_i(\theta)$, where $\theta$ is the quasiparticle rapidity labelling its momentum, which is the value of $Q_i$ carried by the asymptotic state $\theta$ (in models with bound states, multiple species or auxiliary nested indices, $\theta$ is replaced by $(\theta, s_1, s_2, \ldots)$, where $s_i$ are discrete or continuous extra quasiparticle labels). A maximal entropy state is then described by a single function $\rho_{\rm p }(\theta; x ,t )$ characterising the distribution of rapidities in the state. This determines the local charges (or equivalently, by thermodynamic relations, the $\beta_i(x,t)$, see sec. \ref{subsec:TBA_hydro}), 
\begin{equation}\label{qiaverage}
{\tt q}_i(x,t) = \int d\theta \   \rho_{\rm p }(\theta; x ,t ) \ h_i(\theta).
\end{equation}
In integrable models the above relation can always be inverted as the set $h_i(\theta)$ is a ``complete" set of functions (in an appropriate sense).

This identification clearly also allows, in principle, to express the expectations of current operators or other observables as functional of the root density $\rho_{\rm p }(\theta; x ,t )$, which we will clarify in the coming sections. These functionals depend only on few details of the model, as follows. In integrable models, quasiparticles give rise to stable excitations on top of the maximal entropy state. These are characterized by their energy $\varepsilon(\theta)$ and momentum $k(\theta)$. Further, the interaction in the model is characterised by the scattering shift $T(\theta,\alpha)$ incurred in vacuum two-body scattering: this is the log-derivative of the two-body scattering phase $S(\theta,\alpha)$, as 
\begin{equation}
   T(\theta,\alpha) = \frac{\dd}{\dd \theta} \frac{\log S(\theta,\alpha)}{2 \pi \ii } .
\end{equation}
These quantities define the particular integrable model under consideration. Finally, one of the central quantities, characterizing the hydrodynamics of integrable models, is the effective velocity
\begin{equation}
    v^{\rm eff}(\theta) = \frac{\partial \varepsilon (\theta)}{\partial k (\theta)} = \frac{\partial_\theta  \varepsilon(\theta)}{\partial_\theta k(\theta)}.
\end{equation}
Quasiparticles with a given rapidity $\theta$ can be seen as quasi-local wave packets, moving with their group velocity  $v^{\rm eff}(\theta)$. Importantly, since there is a direct connection between quasiparticles and local or quasilocal charges, the hydrodynamics of integrable models can be seen in terms of kinetic flows of quasiparticles. For instance, the Euler-scale currents are written in terms of the effective velocity, and the Euler-scale hydrodynamic equation takes the kinetic form
\begin{equation}
    \partial_t \rho_{\rm p}(\theta;x,t)
    +
    \partial_x (v^{\rm eff}(\theta;x,t) \rho_{\rm p}(\theta;x,t)) = 0.
\end{equation}
It turns out that such a ``hydro-kinetic" duality is also present at the diffusive level and plays an important role in the study of correlation functions, as we will explain.

\section{The hydrodynamic perspective on correlation functions}\label{sec:hydroCorr}

If the steady state or the dynamics of the system is only slightly perturbed, its out-of-equilibrium behaviour and hydrodynamics can be reduced to the multi-point time dependent correlation functions at equilibrium.
In this section we review the fundamental concepts underlying the linear (and nonlinear) response regime and related correlation functions in interacting many-body systems.
This section is concerned with the general hydrodynamic viewpoint, and does not depend on integrability. It overviews structures and concepts based solely on general properties of many-body systems and, at times, on the local entropy maximisation assumption at the basis of hydrodynamics. We define basic transport coefficients, explain how they are related to conductivity and the hydrodynamic equations, and recall their main properties. We then explain what these imply for correlation functions and related quantities at large scales of space and time. This is expected to be valid in a wide variety of many-body systems admitting many conservation laws, integrable or not.

The hydrodynamic equations \eqref{hydroequations} with \eqref{hydroexpansion} form the basis for the study of many phenomena far from equilibrium. For the purpose of studying correlation functions and transport, an extremely fruitful method is to apply linear response theory to this hydrodynamic description.
Linear response theory assumes, of course, that the dominant change of physical quantities is linear in the perturbation strength. In combination with ideas of hydrodynamics, linear response is a very powerful method. It gives quite precise predictions, for instance, for the large-scale profile of correlations in space-time, based solely on the knowledge of local thermodynamic averages and few space-time integrated correlation functions -- the transport coefficients. This is the reduction of the number of degrees of freedom, at the level of correlation functions. The transport coefficients are to be evaluated by microscopic calculations (analytically if possible, or numerically), and are the model-dependent quantities. The large-scale predictions then take universal forms in terms of transport coefficients, fixed by hydrodynamics.

In order to implement this approach, one considers two separate ways of applying linear response theory. Firstly, by doing a linear response analysis on the microscopic dynamics of the model, using, in particular, the microscopic conservation laws Eq.~\eqref{continuity}. This gives expressions for transport coefficients, which in turn determine quantities important for non-equilibrium transport, for instance the conductivity and the hydrodynamic equations. The second is by doing a linear response analysis of the emergent hydrodynamic equations themselves, Eq.~\eqref{hydroequations}. This is the theory of linearised hydrodynamics. It is this theory that gives universal results for correlation functions at large scales, based on the linear response transport coefficients.

In sec. \ref{subsec:conductivity}, linear response theory is used in order to derive the main transport coefficients: the Drude weights and other Euler-scale hydrodynamic matrices, and the Onsager matrix characterising the diffusive scale. In sec. \ref{subsec:einstein}, we discuss Einstein's relations for dynamical correlation functions, which give further intuition about the transport coefficients and, by applying the linear response to the hydrodynamic equations, predictions for the behaviours of correlation functions at large scales of space-time. In sec. \ref{subsec:projectionsEuler} and \ref{subsec:projectionsdiffusive}, transport coefficients and correlation functions are analysed using the notion of hydrodynamic projection, which give rise to properties of positivity, to lower bounds, and to the Boltzmann-Gibbs principle for two-point correlation functions. sec. \ref{subsec:higher} explains how to take into account some nonlinear effects, giving rise to general results for higher-point functions.

Under certain conditions, the linearised hydrodynamic description fails beyond the Euler scale. This results in anomalous transport dynamics, which is considered in a separate review in this special issue \cite{2103.01976}.

\subsection{Conductivity from linear response: Drude weights and Onsager matrix} \label{subsec:conductivity}

Two fundamental transport coefficients, which play a pivotal role in our analysis, are the {\em Drude weight} and the {\em Onsager matrix}. As we will see, they are directly related to the two orders of the hydrodynamic expansion of the average current in \eqref{hydroexpansion}.

But perhaps the most physically transparent way of introducing them is by studying the {\em conductivity matrix} $\sigma_{ab}(\omega)$. It describes the linear response of the current $j_b$ to the infinitesimal perturbation of the dynamics by a time-dependent ``generalised" force, coupled to a charge density $q_a$. Consider an extended one-dimensional many-body system with Hamiltonian $H$. Assume that it is in a  homogeneous, stationary (for $H$), ergodic state. The evolution from this state is now performed with the time dependent Hamiltonian $H^{\delta h}(t) = H-\delta h(t) \int \dd x \,x q_a(x)$. The term $\delta h(t) \int \dd x \,x q_a(x)$  represents the external force applied to the system, a linear potential with time-dependent strength $\delta h(t)$. Because of this force, one expects currents to develop in time -- that is, additional contributions to currents potentially already present in the original state. Time evolution of observables is given by the Heisenberg time evolution, $o^{\delta h}(x,t) = {\cal T}\exp\big( \int_0^t \dd s\, \ii [H^{\delta h}(s),\cdot]\big) o(x)$, where ${\cal T}$ denotes time ordering.

We will assume, for simplicity, that the initial state is thermal, $\rho\propto \exp(-\beta H)$, and consider a general current $j_b^{\delta h}(0,t)$. Note that the current may have a nonzero value in the thermal state -- for instance, the current of momentum, which is the pressure, is generically nonzero. Let us consider the frequency dependence of the conductivity, describing the response to the harmonic modulation of the external force $\delta h(t)= \mathrm{Re}(\delta h \exp(\ii\omega t))$ for some $\delta h\in\CC$. The response of the current can be obtained by  perturbation theory, giving
\begin{equation}
\ave{j_b^{\delta h}(0,t)}- \ave{j_b(0,0)}\sim  \mathrm{Re}\big[\delta h \exp(\ii\omega t)\sigma_{ab}(\omega)\big]+\mathcal{O}(\delta h^2)\quad (t\to\infty)
\end{equation}
where
\begin{equation}
    \sigma_{ab}(\omega)=\frac\beta2\int_{-\infty}^\infty\dd s \exp(-\ii\omega |s|)
    \int_{-\infty}^\infty \dd x \int_0^1 
    \dd\lambda\, \ave{j_a(x,s-\ii\beta \lambda)j_b(0,0)}^{\rm c}.
\end{equation}
In order to extend the time integral to the whole real line, we have used $PT$ symmetry \cite{De_Nardis_2019}. Further, we used the Kubo-Martin-Schwinger (KMS) relation \cite{robinson1997operator}: this allows the commutator, coming from the linear expansion of ${\cal P}\exp\big( \int_0^t \dd s\, \ii [H^{\delta h}(s),\cdot]\big)$, to be recast into an imaginary time evolution. The two-point function with such an imaginary time evolution is sometimes referred to as the  Kubo-Mori-Bogoliubov (KMB) inner product, $(o_1(x,t),o_2(0,0)) = \int_0^1
\dd\lambda\,\ave{o_1(x,t-\ii\beta\lambda)o_2(0,0)}^{\rm c}$.

The real part of the conductivity comprises two contributions, the Drude delta-function peak at zero frequency, and the regular part:
\begin{equation}
    \mathrm{Re}(\sigma_{ab}(\omega))=\pi\beta  \mathsf D_{ab}\delta(\omega)+\sigma^{\rm reg}_{ab}(\omega).
\end{equation}
The Drude weight $\mathsf D_{ab}$ corresponds to the asymptotic, or time-averaged, value of the current-current auto-correlation function,
\begin{equation}
\label{drude}
    \mathsf D_{ab}=\lim_{t\to\infty}\frac{1}{2t}\int_{-t}^t \dd s \int_{-\infty}^\infty\dd x\, \ave{j_a(x,s)j_b(0,0)}^{\rm c}.
\end{equation}
On the right-hand side, the KMB inner product has been reduced to the connected corraltion function. This is argued as follows: the two-point function $\ave{j_a(x,s-\ii\beta \lambda)j_b(0,0)}^{\rm c}$ is uniformly bounded, and assuming that it is analytic in $s$ in the appropriate domain, we may deform the real-time contour by introducing vanishingly small terms at ${\rm Re}(s) = \pm t$,  and therefore cancel the $\lambda$ dependence.
The regular part at $\omega=0$, the so-called DC conductivity, is obtained in the $\omega\to 0$ limit after the Drude weight is subtracted. One defines the Onsager matrix $\mathfrak{L}_{ab} =(2/\beta)\lim_{\omega\to 0}\sigma^{\rm reg}_{ab}(\omega)
$, giving
\begin{equation}
\label{onsager}
    \mathfrak{L}_{ab}=\lim_{t\to\infty}\int_{-t}^t 
    \dd s \left(\int_{-\infty}^\infty\dd x\, \ave{j_a(x,s)j_b(0,0)}^{\rm c}-\mathsf D_{ab}\right).
\end{equation}
Again on the right-hand side, the KMB inner product has reduced to the connected correlation function. The argument is similar: assuming that the integrand of the $s$ integral has a finite, time independent large-$|s|$ asymptotic value (given by $\mathsf D_{ab}$), then we can deform the contour to obtain the usual connected correlation function.

Conductivity in thermal states is the most physically relevant. However, it is possible to extend this calculation to GGEs instead of thermal states. Two major differences appear: in the use of the KMS relation, currents associated not just to time evolution by the Hamiltonian, but also by all conserved charges -- generating ``generalised times" -- appear; and the KMB inner product involves imaginary evolution with respect to all generalised times whose generators are involved in the GGE (More precisely, the KMB inner product involves imaginary evolution with respect to the generalised time that generates the GGE). Such generalised currents $j_{ia}(x,t)$, satisfying $\ii [Q_i,q_a(x,t)] + \partial_x j_{ia}(x,t) = 0$, exist if the charges are in involution, $[Q_i,Q_a]=0$, and are discussed in \cite{SciPostPhys.2.2.014}. Thus we get
\begin{equation}
    \sigma_{ab}(\omega)=\sum_i\frac{\beta^i}2\int_{-\infty}^\infty\dd s \exp(-\ii\omega |s|)
    \int_{-\infty}^\infty \dd x
    (j_{ia}(x,s),j_b(0,0))
\end{equation}
where the (generalised) KMB inner product, for a GGE $\exp\big[-\sum_i\beta^iQ_i\big]$ is
\begin{equation}\label{genKMB}
    (o_1(x,t),o_2(0,0)) =
    \int_0^1 
    \dd\lambda\,
    \ave{o_1(x,\boldsymbol e t-\ii\boldsymbol{\beta} \lambda)o_2(0,0)}^{\rm c}.
\end{equation}
Here the vector notation is used for the set of generalised times (conserved quantities), and $\boldsymbol e$ is the direction of the ordinary time (generated by $H$). From this, generalised three-indices Drude weights $\mathsf D_{iab}$ and Onsager matrix $\mathfrak L_{iab}$ give the delta-peak and regular parts, $\mathrm{Re}(\sigma_{ab}(\omega))=\sum_i \beta^i \big(\pi  \mathsf D_{iab}\delta(\omega)+\frac12 \mathfrak L_{iab}\big) + o(\omega)$, with
\begin{eqnarray}\label{drudegeneralised}
    \mathsf D_{iab}&=&\lim_{t\to\infty}\frac{1}{2t}\int_{-t}^t \dd s \int_{-\infty}^\infty\dd x\, (j_{ia}(x,s),j_b(0,0)) \\
    \mathfrak{L}_{iab}&=&\lim_{t\to\infty}\int_{-t}^t 
    \dd s \left(\int_{-\infty}^\infty\dd x\, (j_{ia}(x,s),j_b(0,0))-\mathsf D_{iab}\right).
    \label{onsagergeneralised}
\end{eqnarray}
In these expressions, it is not possible to reduce the KMB inner product by time-integral contour deformations, because generalised times are involved. Omitting the first index $i$, one recovers the ordinary Drude weights, Onsager matrix and currents.

Five comments are here in order.

First, requesting $PT$ symmetric charges is necessary in order to obtain the above expression for the Onsager matrix. If the conserved densities and currents, and the state, are not $PT$-symmetric, a different expression for the regular part of the conductivity arises. 
While for generic Hamiltonian systems it is expected to be always possible to choose such a gauge, this may not be the case in Non-Hermitian and open quantum systems, where the relation between Onsager and diffusion matrix is indeed exptected to be more involved. 

Second, the Onsager matrix is always zero in free systems, that is whenever the corresponding quasiparticles scatters trivially, with zero differential phase shift. A positive non-zero Onsager matrix is therefore a clear sign of the presence of non-trivial interactions within the system. It has been suggested that it can be used to distinguish between interacting and non-interacting integrable systems \cite{Spohn_JMP}. We will see below how this relates to other ways of characterising the presence of interactions in many-body systems.

Third, although the Drude weights are expected to be finite or zero in a large family of systems, there is no guarantee that the Onsager matrix is finite. In fact, in non-integrable systems with a conserved momentum, it is argued to be infinite by nonlinear fluctuating hydrodynamics \cite{Spohn2014}, or by projection and the structure of three-point functions  \cite{Ben}. In integrable spin systems with non-abelian symmetry, it is also found to be infinite \cite{Ilievski2018,1812.02701,PhysRevLett.123.186601,PhysRevB.101.041411,PhysRevLett.124.210605}, and the ensuing superdiffusion is the suject of another review \cite{2103.01976}. For most, or all currents in most integrable systems, however, it is finite, and exactly calculable.

Fourth, as discussed in other sections in this review, one can bound and/or evaluate both for the Drude weight and the Onsager matrix by projecting the currents onto conserved quantities. Using projection, provided that considered charges are in involution, the reduction of the KMB inner product to the connected correlation function follows even in GGEs. Such projections evaluate both the (generalised) Drude weights and Onsager matrix in integrable systems, but only bound the latter away from integrability.

For the hydrodynamic theory and for the study of correlation functions, which are the focus of this review, the relevant objects are the ordinary (non-generalised) Drude weights and Onsager matrix elements, evaluated within GGEs:
\begin{eqnarray}\label{drudeKMB}
    \mathsf D_{ab}&=&\lim_{t\to\infty}\frac{1}{2t}\int_{-t}^t \dd s \int_{-\infty}^\infty\dd x\, (j_{a}(x,s),j_b(0,0)) \\
    \mathfrak{L}_{ab}&=&\lim_{t\to\infty}\int_{-t}^t 
    \dd s \left(\int_{-\infty}^\infty\dd x\, (j_{a}(x,s),j_b(0,0))-\mathsf D_{ab}\right).
    \label{onsagerKMB}
\end{eqnarray}

Finally, let us mention that expression for Drude weights can be obtained in two alternative ways, which do not require the time dependent driving force. Firstly, we can consider the rate at which the current increase in the presence of a small, time independent external force $\delta h(t)=\delta h$ \cite{Ilievski2012}, which yields
\begin{equation}
   \beta \mathsf{D}_{ab}=\lim_{t\to\infty}\frac{1}{t} \left [\partial_{\delta h} \ave{j_b^{\delta h}(0,t)}\right]_{\delta h=0}.
\end{equation}
Secondly, the Drude weight correspond to the total asymptotic current $j_b$, in the bi-partition protocol in which the left and the right side of the system are prepared in states with a slightly different values of generalised temperatures $\delta \beta^a$, associated with charge $q_a$ \cite{SciPostPhys.3.6.039}
\begin{equation}
    \mathsf{D}_{ab}=\frac{\partial}{\partial \delta \beta^a}\left(\lim_{t\to\infty}\frac{1}{t}\int\dd x\left<j_b(x,t)\right>\right).
\end{equation}

\subsection{Hydrodynamics from linear  response: Euler currents and diffusion} \label{subsec:hydro}

Coupling the dynamics to a gradient potential implements a force, leading to the conductivity matrix. But it is also possible to generate currents by evolving with the unperturbed dynamics $H$ from a deformed state. This is more natural in the context of connecting transport coefficients to the hydrodynamic equations. Indeed, the homogeneous and stationary state $\ave{\cdots}$ evolves trivially (as it is stationary!), and the most basic context for hydrodynamics is the emergent dynamics for $H$, from a state that is not stationary and homogeneous, but that varies only on large scales. One way of deriving the hydrodynamic equations is therefore to deform the stationary and homogeneous state $\ave{\cdots}$ to a slowly varying state, with a gradient in the Lagrange parameters. 

Therefore, consider
\begin{equation}\label{initrhohydro}
    \rho^{\delta \beta}
    = \exp\Big[-\sum_i \int \dd x\,(\beta^i + \delta \beta^i x)q_i(x)\Big].
\end{equation}
Physically, for finite $\delta \beta^i$'s, at large times currents become large if the system admits ballistic transport for the corresponding charges, because of the large (infinite) amount of discrepancy between the charges present in the left and right semi-infinite regions of the initial state \eqref{initrhohydro}. Subleading long-time contributions are controlled by diffusion. In order to see this at leading order in $\delta \beta^i$ we perform a simple perturbative calculation, which shows that
\begin{eqnarray}\label{hydrolinejq}
    \ave{j_i(0,t)}^{\delta \beta}&=&
    \ave{j_i} + \frac{\delta \beta^l}2
    \int_{-t}^t \dd s\,\int \dd x\,
    (j_l(x,s),j_i(0,0)) + O\big((\delta \beta)^2\big)\\
    \ave{q_i(0,t)}^{\delta \beta}&=&
    \ave{q_i} + t\, \delta \beta^l \mathsf B_{li} + O\big((\delta \beta)^2\big).
\end{eqnarray}
Here and at various instances below, for lightness of notation we use implied summation over repeated indices (unless otherwise specified). These expressions involve the generalised KMB inner product \eqref{genKMB}, as a consequence of perturbation theory applied to the exponential \eqref{initrhohydro}. We again used $PT$ symmetry in order to symmetrically extend the time integral.  We have introduced the hydrodynamic $\mathsf B$ matrix
\begin{equation}\label{eq:Bmatrix}
    \mathsf B_{li} = \int \dd x\,
    (j_l(x,s),q_i(0,0)) 
    = -\frac{\partial}{\partial \beta^i}
    \ave{j_l}
\end{equation}
The independence on $s$  follows by using homogeneity, stationarity, the conservation law \eqref{continuity} and clustering of correlation functions. By comparing \eqref{hydrolinejq} with the expression for the Drude weight \eqref{drudeKMB}, we see that the current indeed grows linearly in time, at this order in $\delta\beta^l$'s, if  $\delta\beta^l\mathsf D_{li}$ is nonzero. The charge density also grows linearly in time if it overlaps, in the sense of having nonzero susceptibility, with $\delta\beta^l j_l$. We note that, as a consequence of general properties of clustering states, the $\mathsf B$ matrix is symmetric \cite{toth2003onsager,grisi2011current,PhysRevX.6.041065,De_Nardis_2019,10.21468/SciPostPhys.6.6.068}
\begin{equation}\label{eq:Bsymm}
    \mathsf B_{ij} = \mathsf B_{ji}.
\end{equation}

In order to obtain the hydrodynamic equations, we need to recover the expansion of the average current \eqref{hydroexpansion}. For this purpose, we imagine that at long time in the above linear-response problem, we have reached the hydrodynamic limit, and we simple need to express $\ave{j_i(0,t)}^{\delta \beta}$ in terms of $\ave{q_i(0,t)}^{\delta \beta}$. More precisely, the hydrodynamic expansion of the current is obtained in the limit where $\delta\beta^i$'s are sent to zero, before $t\to\infty$, much like in the conductivity analysis. This is to be done order by order: we {\em keep only the zeroth and first order in $\delta \beta^i$'s}, on which we take the infinite-time limit. By construction, as the unperturbed state is a maximal entropy state, we have $\ave{j_i} = \mathsf J_i[\ave{q_\bullet}] = \mathsf J_i[\ave{q_\bullet(0,t)}^{\delta\beta} - t\delta\beta^l\mathsf B_{l\bullet}]$ (see \eqref{qiaverage}), which we expand in $\delta \beta^l$. The result ${\tt j}_i = \lim_{t\to\infty} \ave{j_i(0,t)}^{\delta \beta}\big|_{\mbox{\small zeroth, first orders}}$, in terms of ${\tt q}_i = \lim_{t\to\infty} \ave{q_i(0,t)}^{\delta \beta}\big|_{\mbox{\small zeroth, first orders}}$, is then
\begin{equation}\label{expansion1}
    {\tt j}_i =
    \mathsf J_i[{\tt q}_\bullet] + \frac{\delta \beta^l}2
    \lim_{t\to\infty} \int_{-t}^t \dd s\,\Big(\int \dd x\,
    (j_l(x,s),j_i(0,0)) 
    - \mathsf A_i^{~j} \mathsf B_{lj}\Big).
\end{equation}
Here we have introduced yet another hydrodynamic matrix, the flux Jacobian, the variation of the average currents in a maximal entropy state with respect to the average conserved densities:
\begin{equation}\label{eq:Amatrix}
    \mathsf A_i^{~j} = \frac{\partial \mathsf J_i[{\tt q}_\bullet]}{\partial {\tt q}_j}.
\end{equation}

Comparing the right-hand side of \eqref{expansion1} with the expression for the Onsager matrix \eqref{onsagerKMB}, we see that the integral may exist only if we have the identity
\begin{equation}\label{drudeAB}
    \mathsf D_{il} = \mathsf A_i^{~j}\mathsf B_{lj}.
\end{equation}
We will explain below how this identity follows from projection principles. Then, with the expression for the Onsager matrix \eqref{onsagerKMB}, we obtain
\begin{equation}\label{expansion2}
    {\tt j}_i =
    \mathsf J_i[{\tt q}_\bullet] +
    \frac12 \mathfrak L_{il} \delta \beta^l.
\end{equation}

In order to recover the more standard form of the hydrodynamic current \eqref{hydroexpansion}, in terms of the charge density gradients, we perform perturbation theory for $\ave{q_i(x,t)}^{\delta \beta}$ and set ${\tt q}_i(x) = \lim_{t\to\infty} \ave{q_i(x,t)}^{\delta \beta}\big|_{\mbox{\small zeroth, first orders}}$. The result is
\begin{equation}
    \partial_x {\tt q}_i(x) =
    -\delta \beta^l \mathsf C_{li}
\end{equation}
where we have introduced our final hydrodynamic matrix, the static covariance matrix
\begin{equation}\label{eq:Cmatrix}
    \mathsf C_{li}
    = \int \dd x\,(q_l(x,0),q_i(0,0)) = -\frac{\partial}{\partial \beta^i}\ave{q_l}
\end{equation}
which is symmetric and positive. The hydrodynamic current is obtained under the definition
\begin{equation}\label{onsagerLC}
    \mathfrak D_i^{~j}
    = \mathfrak L_{il}\mathsf C^{lj}
\end{equation}
for the diffusion matrix, where we use Einstein's notation for inverse tensors,
\begin{equation}
    \mathsf C_{il}\mathsf C^{lj} = \mathsf C^{jl}\mathsf C_{li} = \delta_i^j.
\end{equation}
We note that using the static covariance matrix and the chain rule of differentiation, we have, in matrix notation, $\mathsf B = \mathsf A \mathsf C$ and therefore
\begin{equation}\label{drudeAsquareC}
    \mathsf D = \mathsf B\mathsf C^{-1}\mathsf B = \mathsf{A^2C}.
\end{equation}

Using the hydrodynamic expansion for the current \eqref{hydroexpansion}, along with the flux Jacobian and diffusion matrix defined above, one then obtains the ``quasi-bilinear" form of the diffusive-order hydrodynamic equations,
\begin{equation}\label{hydroquasi}
    \partial_t {\tt q}_i
    + \mathsf A_i^{~k}
    \partial_x {\tt q}_k
    -\frac12 \mathfrak D_i^{~k}\partial_x^2 {\tt q}_k
    -\frac12 \mathfrak D_i^{~kl}
    \partial_x{\tt q}_k
    \partial_x{\tt q}_l=0
\end{equation}
where
\begin{equation}
    \mathfrak D_i^{~kl}
    = \frac{\partial \mathfrak D_i^{~k}}{\partial {\tt q}_l}.
\end{equation}

The hydrodynamic equations have been discussed in many textbooks. Their ``universal" form Eqs.~\eqref{hydroexpansion} and \eqref{hydroequations}, or equivalently Eq.~\eqref{hydroquasi}, has been discussed in the context of particle systems with many conservation laws, see for instance the textbook \cite{Spohn1991} and the important studies \cite{toth2003onsager,grisi2011current}. The GHD lecture notes \cite{10.21468/SciPostPhysLectNotes.18} give a presentation nearer to the one shown here. The properties satisfied by the hydrodynamic matrices, in particular by the flux Jacobian $\mathsf A$, are crucial for the consistency of the hydrodynamic equations, as discussed in \cite{toth2003onsager,grisi2011current} (see also  \cite{10.21468/SciPostPhysLectNotes.18}), and, with additional terms representing external forces, in \cite{freeenergyfluxes}. Hydrodynamic matrices and their physical applications are also discussed in \cite{Spohn2014,SciPostPhys.3.6.039} (some of which is reviewed  below). It has been suggested that one may characterise the microscopic model as admitting nontrivial interactions by properties of the hydrodynamic matrices. As mentioned in sec. \ref{subsec:conductivity}, in \cite{Spohn_JMP} it is suggested that the absence of diffusion indicates that there is no nontrivial interaction. Other ways of characterising the presence / absence of interactions have been proposed. In \cite{Fagotti_2016} it is proposed that in the absence of interactions, the current observables are themselves, in an appropriate gauge, conserved densities, see sec \ref{subsec:projectionsEuler} for a more precise statement. In \cite{doyonFluctuations2020}, it is suggested that in non-interacting models, by contrast to interacting models, the flux Jacobian $\mathsf A$ is independent of the state, and thus the hydrodynamic equations are linear. Both notions are clearly related, and both are related to the absence of diffusion, as we will see in sec. \ref{subsec:projectionsdiffusive}.

Finally, we mention that by the symmetry of the static covariance $\mathsf C$, and of the current susceptibility $\mathsf B$ (Eq.~\eqref{eq:Bsymm}), two functions of the Lagrange parameters are guaranteed to exist: the specific free energy $\mathsf f$, which generates averages of conserved densities in maximal entropy states, and the free energy flux $\mathsf g$, which generates average currents:
\begin{equation}\label{eq:fg}
    \frac{\partial \mathsf f}{\partial \beta^i} = \ave{q_i},\quad
    \frac{\partial \mathsf g}{\partial \beta^i} = \ave{j_i}\quad \mbox{($\ave{\cdots}$ a maximal entropy state).}
\end{equation}
In particular, if the state involves only, say, the number of particles $Q_0$, the momentum $Q_1$ and the energy $Q_2$, then $\mathsf g = -\beta^1 \mathsf f / \beta^2$ \cite{freeenergyfluxes}. For integrable systems, see sec. \ref{subsec:TBA_hydro}.

\subsection{Einstein relation and the dynamical structure factor}\label{subsec:einstein}

The connection between the response of the system to external forces and diffusion of Brownian particles was first established by Einstein \cite{Einstein1905}. In this section we shall show that the transport coefficients are indeed related to the dynamical correlation functions of the system, which express the linear response to external perturbations.

In the previous section we have seen that similar response functions are obtained to either an external force, or to a charge gradient in the initial state. We now consider the response of the system to a {\em local} change in the initial state. In particular, we explain how the late-time dynamics, under such a local perturbation, is governed by the propagation velocities, and by the nature of charge spreading around these velocities, in analogy with Brownian particles. We show that these are encoded by the flux Jacobian $\mathsf A$ and the Onsager matrix $\mathfrak L$.

We consider the dynamical structure factor
\begin{equation}\label{eq:defS}
S_{ij}(x,t) =  \langle  q_i(x,t) q_j(0,0) \rangle^c,\quad \mbox{or}\quad
S_{ij}(x,t) =  (  q_i(x,t), q_j(0,0)).
\end{equation}
The calculations below hold for both definitions, in terms of the connected correlation function and of the KMB inner product \eqref{genKMB}. The latter definition is more appropriate in the generic quantum case -- in particular, it is a real quantity -- but, as per the discussion in sec. \ref{subsec:conductivity}, in integrable systems, with all charges being in involution, all conclusions below hold for the former as well. Note that by definition of the static correlation matrix and time-independence of the conserved quantities, the dynamical structure factor is normalised as

\begin{equation}\label{Snorm}
    \int \dd x\,S_{ij}(x,t) = \mathsf C_{ij}.
\end{equation}

The dynamical structure factor characterises the distribution of charge at time $t$, given that a small, localised perturbation was applied to the initial state at space-time point $(0,0)$. Here what is important is not the explicit implementation of the physical process of modifying the initial state, as done in the previous sections, but rather the representation of this change by the insertion of a local density field at $(0,0)$. In sec. \ref{subsec:projectionsEuler} we will explain how the dynamical structure factor is a crucial building block for the general question of correlations between local perturbations and local probes.

Much like for the Brownian particle, let us consider the second moment of the distribution  \eqref{eq:defS}.
The dependence of the second moment on time characterizes the dynamical universality class -- the transport coefficients. Through the continuity equations, and using $PT$ symmetry, one can establish the relation between the current-current correlation function and the second moment of the dynamical structure factor (here written for the KMB inner product) \cite{Spohn1991,De_Nardis_2019}:
\begin{equation} \label{eq:sumrule}
\frac{1}{2}\int \dd x \,  x^2\,  \big(S_{ij}(x,t) + S_{ij}(x,-t) - 2S_{ij}(x,0)\big)  = \int_0^t \dd s \int_0^t   \dd s' \int  \dd x\, ( j_i(x,s), j_j(0,s')).
\end{equation}
Using the definitions for the Drude weights \eqref{drudeKMB}
and Onsager matrix \eqref{onsagerKMB}, the asymptotic at long times of the right-hand side is readily obtained:
\begin{equation} \label{eq:S-largetime}
\frac{1}{2}\int \dd x \,  x^2\,  \big(S_{ij}(x,t) + S_{ij}(x,-t)\big)   = \mathsf D_{ij} t^2 + \mathfrak{L}_{ij} t + o(t).
\end{equation}
This expression can be understood as a generalization of the Einstein relation. The term that is proportional to $t^2$ indicates a linear growth of the distribution support in space, representing the ballistic  propagation (along straight lines of fixed velocities) of charges from the initial disturbance. This is controlled by the Drude weights. The term proportional to $t$ represents the diffusive expansion around this ballistic propagation. 

One can build more intuition about the hydrodynamic manifestation of transport coefficients by considering the space-time profile of the function $S_{ij}(x,t)$. Consider first the simple case of diffusive propagation of a single mode. Standard arguments suggest that the distribution arising from the Brownian motion in a linear flow is a Gaussian centered around the velocity $v$ of linear propagation, with a width that grows as a square-root in time:
\begin{equation}\label{Sbrownian}
S(x,t)=\frac{\chi}{\sqrt{2\pi\mathcal{D}|t|}}\exp\Big(-\frac{(x-v t)^2}{2\mathcal{D} |t|}\Big).
\end{equation}
Using the normalisation \eqref{Snorm}, the coefficient $\chi$ equates the (1 by 1) static covariance matrix $\chi = \mathsf C$. For this distribution, the left-hand side of \eqref{eq:S-largetime} then evaluates to $\mathsf Cv^2 t^2 + \mathsf C \mathcal D |t|$. Using the form of the Onsager matrix \eqref{onsagerLC}, we identify the diffusion constant with the (1 by 1) diffusion matrix $\mathcal D = \mathfrak D$, and using the form \eqref{drudeAsquareC} of the Drude weight, we identify the propagation velocity with the (1 by 1) flux Jacobian $v = \mathsf A$. 

\begin{figure}
    \centering
    \includegraphics[width=10cm]{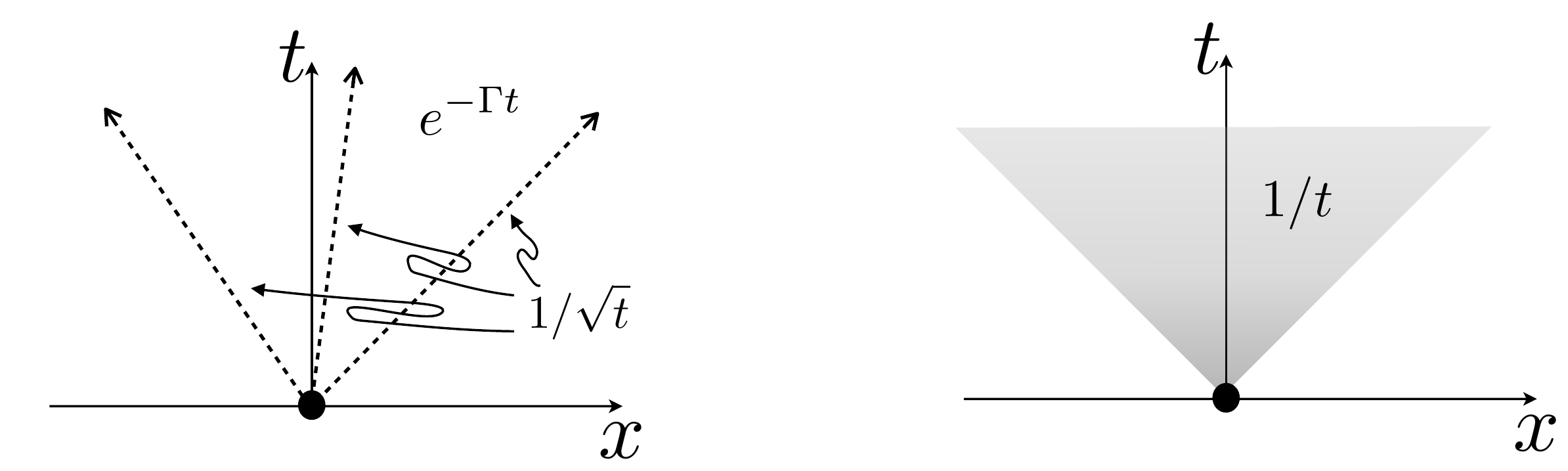}
    \caption{Decay of correlation functions in the cases of few diffusive hydrodynamic modes (left) and a continuum of hydrodynamic modes (right).}
    \label{fig:decay}
\end{figure}

In the case of multi-components fluids,  correlation functions \eqref{Sbrownian} are instead given by a superposition of all terms with different velocities (that can take arbitrarily small values) and different diffusion constants. Thus, the decay is in $1/\sqrt{t}$ along these velocities, and exponential otherwise. In particular, if a continuum of modes exist, as in integrable systems, then by contrast from the single or few-mode cases, correlation functions decay as $1/t$ throughout the space-time region where these modes propagate, which can be deduced by taking into account the conservation law $\int \dd x\, S(x,t)=\text{const.}$. We shall see this in more details in the following. See fig.~\ref{fig:decay}.

The simple form \eqref{Sbrownian} can be extended to multiple-component hydrodynamics, $S_{ij}(x,t)$, by {\em applying linear response theory to the hydrodynamic equations themselves}. That is, consider the form \eqref{hydroquasi} of the diffusive-order hydrodynamic equations, and the response of the average charge density ${\tt q}_i(x,t)$ to a small, local perturbation at $(0,0)$, on top of a homogeneous initial state. In general the small perturbation of strength $\epsilon$ has the form ${\tt q}_i^\epsilon(x,t) = {\tt q}_i + \epsilon (q_i(x,t),o(0,0))$ for some local observable $o(0,0)$. Suppose $o=q_j$. Then to leading order in $\epsilon$ one only keeps the terms of Eq.~\eqref{hydroquasi} that are linear in densities, giving
\begin{equation}\label{eq:twopointfunction}
\partial_t S_{ij}(x,t) +  \left(\mathsf A_{i}^{~k} \partial_x - {\rm sgn}(t) \frac{\mathfrak{D}_{i}^{~k}}2 \partial^2_x \right)  S_{kj}(x,t)  = 0\quad\mbox{(at large scales)}.
\end{equation}
Note how the irreversibility of the hydrodynamic equation had to be taken into account: the response of ${\tt q}_i(x,t)$ can be evaluated for a perturbation ``in the past" only. In the single-component case, the solution is \eqref{Sbrownian}. Eq.~\eqref{eq:twopointfunction} is the emergent dynamical equation for the propagation of correlations at large scales of space-time. Recall how the transport coefficients in \eqref{eq:S-largetime} and those in \eqref{eq:twopointfunction} are related: $\mathsf D = \mathsf A^2 \mathsf C$, and $\mathfrak L = \mathfrak D \mathsf C$. We also emphasise that, in general, $\mathsf A$ and $\mathfrak D$ depend on the background state on which the dynamical structure factor is evaluated.

Eq.~\eqref{eq:twopointfunction} is to be interpreted in an appropriate fashion. It was {\em not} derived from arguments based on a microscopic analysis; instead, it was derived by linear response of the hydrodynamic equation, hence it is based on the hydrodynamic principles. Thus, it is to be interpreted as being valid {\em at large scales of space and time only}. It is not so simple to make a precise statement about what this large-scale regime is. In general, the large-scale limit should include an appropriate fluid-cell averaging, so as to take away possible oscillations which are not described by Euler- and diffusive-scale hydrodynamics. One way of doing this is to consider the Fourier transform
\begin{equation}\label{eq:Sfourier}
    S_{ij}(k,t) = \int \dd x\,
    e^{\ii kx} S_{ij}(x,t),
\end{equation}
in an appropriate expansion as $k\to0$ and $t\to\infty$.

In particular, the Euler scale is with $\kappa=kt$ fixed. More precisely, the Euler scaling limit \eqref{eq:Sooeulertime} below, which is the time average on all $t$ of \eqref{eq:Sfourier} at $\kappa=kt$ fixed, is shown rigorously in \cite{hydroprojectionsEuler} to satisfy the Fourier transform of the Euler-scale part of Eq.~\eqref{eq:twopointfunction}. In \cite{hydroprojectionsEuler} it is in fact shown that the existence of the limit in this formulation is not necessary in order for results to apply; one may consider the notion of generalised (or Banach) limit. There are other ways of formulating the Euler scaling limit, which may be simpler or more explicit. For instance, one formulation is an explicit fluid-cell averaging on the space-time cell $[-a\ell,a\ell]^2$ around the point $(\ell x,\ell t)$, in the limit of large scales $\ell\to\infty$, and then of small fluid cell with respect to this large scale $a\to0^+$ (that is, of a mesoscopic fluid cell):
\begin{equation} \label{eq:Seulerscale}
    \lim_{\rm Eul} S_{ij}(x,t) = \lim_{a\to0^+}\lim_{\ell \to \infty} \ell \int_{-a\ell}^{a\ell} \frac{\dd y}{2a\ell} \int_{-a\ell}^{a\ell} \frac{\dd s}{2a\ell}\,
    S_{ij}(\ell x+y,\ell t+s).
\end{equation}
Note that the large-$\ell$ space-time average is expected to decay as $1/\ell$, hence the factor $\ell$ is introduced to make the Euler scaling limit finite. This is expected to have the form $t^{-1}$ times a function of $\xi = x/t$. Usually, the double limit formulation can be weakened to a formulation where the fluid cell on which the average is taken is the cell $[-\ell^\nu,\ell^\nu]^2$ around $(\ell x,\ell t)$, for appropriate $0<\nu<1$; in this case a single limit on $\ell$ is required. How large $\nu$ is chosen depends on the corrections to the Euler scale; for instance, for diffusive correction, we must have $\nu>1/2$. In some cases, such for particle density correlations in the hard-rod gas, it appears not to be necessary to perform the time averaging. In some situations, in integrable models, it may be that no fluid-cell averaging is actually required at all, in which case we would have
\begin{equation}
    S_{ij}(x,t) \sim \lim_{\rm Eul} S_{ij}(x,t)\quad (x,t\to\infty,\ \xi=x/t\ \mbox{fixed}),
\end{equation}
indicating a decay as $1/t$ of the dynamical structure factor.

It is a simple matter to give a formal solution to \eqref{eq:twopointfunction}, by exponentiation of the matrices involved. This is most clearly expressed for the Fourier transform \eqref{eq:Sfourier}. Taking into account the normalisation \eqref{Snorm},
\begin{equation}\label{eq:Ssolution}
    S_{ij}(k,t) = \Big(\exp\Big[
    \ii \mathsf A kt - \frac{\mathfrak D}2
    k^2 |t|
    \Big]\mathsf C\Big)_{ij}\quad (k\to0,\ t\to\infty).
\end{equation}
As emphasised, this is to be interpreted as an appropriate expansion as $k\to0$, $t\to\infty$. The solution to \eqref{eq:twopointfunction} implies that we may have $k$-dependence, $\mathsf C$ being replaced by $\mathsf C(k)$, with $\mathsf C(0) = \mathsf C$ by the normalisation \eqref{Snorm}. But by $PT$ symmetry, $\mathsf C(k) = \mathsf C + O(k^2)$, thus the expansion \eqref{eq:Ssolution} holds up to the diffusive order. In the Euler scaling limit expressed in \eqref{eq:Seulerscale}, the result is simply
\begin{equation}\label{eq:Ssolutionxt}
    \lim_{\rm Eul} S_{ij}(x,t) = \big(\delta(x-\mathsf At)\mathsf C\big)_{ij}.
\end{equation}

By a change of basis in the space of conserved densities, it is possible to bring the term involving a single derivative in diagonal form,
\begin{equation}
    \mathsf R_i^{~a}\mathsf A_a^{~b}(\mathsf R^{-1})_b^{~j}  = \delta_i^j v_i^{\rm eff}\quad\mbox{(no summation on $i$)}.
\end{equation}
The eigenvalues (or more generally, elements of the spectrum) $v^{\rm eff}_i$ are interpreted as the state-dependent propagation velocities of the fluid's normal modes $n_i$,  \begin{equation}\label{nmb}
   n_i= \mathsf R_{i}^{~b}q_b.
\end{equation} 
Assuming that the velocities are nondegerate it can be shown that normal modes are orthogonal $\int \dd x\,\ave{n_a(x) n_b(0)}^{\rm c}=\delta_{ab}$, implying that
\begin{equation}\label{orto}
    \mathsf R  \mathsf  C  \mathsf  R^\mathrm{T} =  \mathsf 1.
\end{equation}
Using the symmetry \eqref{eq:Bsymm} of the $\mathsf B$ matrix and positivity of $\mathsf C$, it is possible to show that the spectrum of $\mathsf A$ must be real \cite{toth2003onsager,grisi2011current,10.21468/SciPostPhysLectNotes.18}. 

\begin{figure}
    \centering
    \includegraphics[width=8cm]{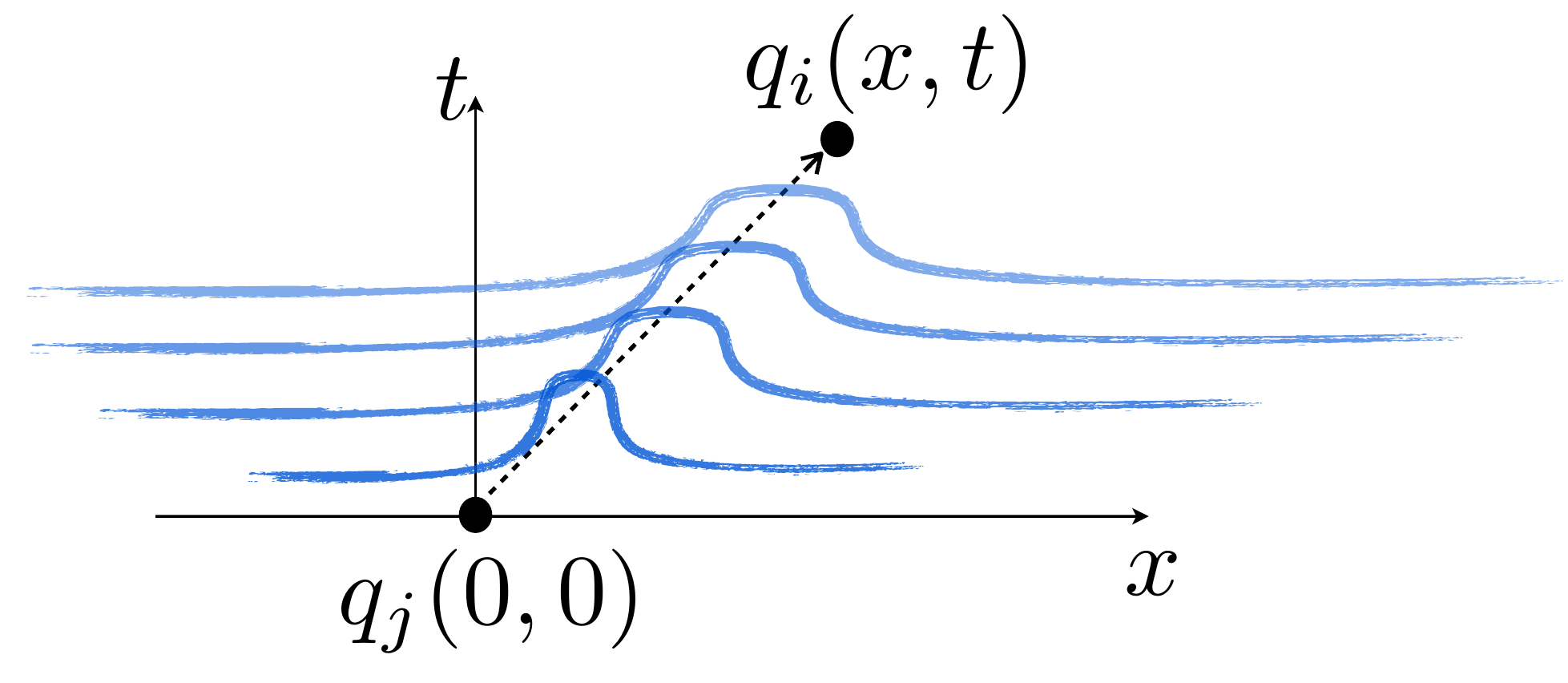}
    \caption{Cartoon representation of Eq.~\eqref{eq:twopointfunction}: the leading correlation between conserved densities comes from waves emanating from the local disturbance, propagating at a velocities in the spectrum of $\mathsf A$ and spreading diffusively as controlled by $\mathfrak D$.}
    \label{fig:wavepropagation}
\end{figure}

Eq.~\eqref{eq:twopointfunction} indicates that correlations are strong along the velocities corresponding to fluid normal modes, and spread diffusively around this. These normal modes are here manifested as Euler-scale ``linear waves" emanating from the disturbance at $(0,0)$ -- waves formed by linear-order perturbations on top of a background. They give rise to the leading correlations at large scales, see fig.~\ref{fig:wavepropagation}. For instance, in conventional (one-dimensional) fluids, with, say, conserved particle number, momentum and energy, there are three normal modes: two sound modes typically with equal and opposite velocities at equilibrium, and the heat mode typically with vanishing velocity. In this case, Eq.~\eqref{eq:Ssolutionxt} shows that the Euler scaling limit is a distribution, indicating that correlation functions vanish faster than $1/t$ away from these velocities (supposedly exponentially), and slower than $1/t$ at these velocities (with the diffusive form of the hydrodynamic equations \eqref{eq:twopointfunction}, it is $t^{-1/2}$). We note that in general, the diffusion matrix is not diagonal in the basis of the Euler-scale normal modes, and thus (super-)diffusive spreading is intricate. See \cite{De_Nardis_2019} for discussions and sec. \ref{subsec:diffusionscrambling} below. As mentioned above, if continuum of modes with distinct velocities is present, the decay of correlation functions inside of the maximal light-cone is governed by a $1/t$ power-law.

Finally, it is worth noting that one can apply linear response theory on top of {\em fluids with large-scale motion}. If the background densities ${\tt q}_i(x,t)$ are space-time dependent, then the linear response at space-time point $x,t$ to a local disturbance at $(0,0)$ can be written in the form ${\tt q}_i^\epsilon(x,t) = {\tt q}_i(x,t) + \epsilon (q_i(x,y)o(0,0)) + O(\epsilon^2)$ for appropriate local observable $o(0,0)$, where the KMB inner product is with respect to the initial state (with Euler scaling of the initial state, it is with respect to the local state at $(0,0)$). Keeping only the Euler scale (first-order derivative) for simplicity, and using the symmetry property $\partial_{{\tt q}_k}\mathsf A_i^{~j} =\partial_{{\tt q}_j}\mathsf A_i^{~k}$, the result from expanding \eqref{hydroquasi} to linear order in $\epsilon$ is
\begin{equation}\label{eq:twopointfunctionnonstat}
\partial_t S_{ij}(x,t) +  \partial_x \left(\mathsf A_{i}^{~k}(x,t)  S_{kj}(x,t)\right) = 0\quad\mbox{(Euler scale, non-stationary)}.
\end{equation}
This is expected to hold for the non-stationary Euler scaling limit
\begin{equation}\label{eq:Snonstateulerscale}
    \lim_{\rm Eul} S_{ij}(x,t) =
    \lim_{a\to0^+}\lim_{\ell \to \infty} \ell \int_{-a\ell}^{a\ell} \frac{\dd y}{2a\ell} \int_{-a\ell}^{a\ell} \frac{\dd s}{2a\ell}\,
    \ave{ q_i(\ell x+y,\ell t+s)q_j(0,0)}^{\rm c}_\ell
\end{equation}
where $\ave{\cdots}_\ell = \tr (\rho_\ell\,\cdots)$, and we may take
\begin{equation}\label{eq:nonstatdensity}
    \rho_\ell \propto\exp\Big[
    -\int \dd x\,\beta^i(x)q_i(\ell x)
    \Big].
\end{equation}
Again, if no fluid-cell averaging is required then we simply have
\begin{equation}
    \ave{q_i(\ell x,\ell t)q_j(0,0)}_\ell^{\rm c}\sim
    \ell^{-1} \lim_{\rm Eul} S_{ij}(x,t)\quad (\ell\to\infty).
\end{equation}
In general, \eqref{eq:twopointfunctionnonstat} is much harder to solve, however we will see in sec. \ref{subsec:nonstatintegrable} how in integrable models there is an integral-equation solution to this initial-value problem.

\subsection{Projections, bounds and the Boltzmann-Gibbs principle at the Euler scale} \label{subsec:projectionsEuler}

Now we will discus how to use the microscopic features of dynamical systems, such as the properties of conservation laws, to obtain explicit expressions for transport coefficients and large scale dynamics of correlation functions at the Euler scale.
One of the most powerful concepts in the non-equilibrium statistical mechanics of many-body systems is that of projections. The basic idea is to describe the emergent dynamics by projecting onto the large-wavelength, low-frequency subspace of observables, and extract the dynamical equations and large-scale physics that pertains to this subspace. This idea can be argued to be the precursor to the renormalisation group in quantum field theory and emergent phenomena. It allows a treatment of linearised hydrodynamics -- the results of linear response theory as applied to hydrodynamics -- that does not actually rely on the physical principle of linear response, and that can be brought to a  mathematically rigorous form. Hydrodynamic projections also give a number of results that we mentioned without proof in the previous sections: the reduction of the KMB inner product in the Drude weights and Onsager matrix to connected correlation functions in sec. \ref{subsec:conductivity}, and the imporant relation \eqref{drudeAB} between hydrodynamic matrices. In this section we describe some of the general results from hydrodynamic projections at the Euler scale.

The starting point is the construction of a new space of observables, that of extensive observables, and their associated inner product. Here we concentrate on connected correlation functions instead of the generalised KMB inner product \eqref{genKMB}, but results are expected to hold in the KMB formulation as well.

Before making this construction, we recall a standard construction in statistical mechanics. On the space of local observables,  connected correlation functions $\ave{o_1^\dagger(x)o_2(0)}^{\rm c}$ form a pre-inner product; it is positive semi-definite by positivity of the state. It becomes an inner product when zero-norm observables are moded out (usually only operators proportional to the identity, which obviously have zero connected correlation functions), and the resulting space of equivalence classes is Cauchy completed, with respect to the norm induced by the pre-inner product, to some Hilbert space $\mathcal H$. This is essentially the Gelfand-Naimark-Segal construction (see e.g.~\cite{convexity}).

The same principles can be used at infinite wavelenght. Consider the ``1$^{\rm st}$ order" hydrodynamic pre-inner product on the space of local observables \cite{Ben}, defined by spatial integration of connected correlation functions
\begin{equation}\label{eq:innereuler}
    \aveI{o_1,o_2} = 
    \int \dd x\,
    \ave{o_1^\dagger(x)o_2(0)}^{\rm c}.
\end{equation}
One can show that this pre-inner product is positive semi-definite \cite{pseudolocality}. It is clear that total derivatives of local observables have zero norm under it. Thus any representative of an equivalence class is defined up to addition of total derivatives. This suggests that the equivalence class of $o(x)$ can be identified with its corresponding extensive observable, its spatial integral, as indeed this spatial integral formally does not change by addition of total derivatives to $o(x)$. It is in fact convenient to {\em define the complete space of extensive observables} $\mathcal H'$ as the resulting space of Cauchy-completed equivalence classes, which we will denote by capital letters, e.g.~$O\in \mathcal H'$. Intuitively, these are observables growing extensively with the system size, but the above formal definition does not require such a finite-size analysis. The passage from local observables to equivalence classes after ``integrating out" a group action -- here that of space translations -- is dubbed ``hydrodynamic reduction" in \cite{Ben}, and is a process that is claimed to reveal the Hilbert space structures underlying the various hydrodynamic orders.

For convenience of notation, the equivalence class of a local observable $o(x)$ may be denoted\footnote{In this notation, the integral symbol has the meaning of the map $\int \dd x: o(x)\mapsto O$ from local observables $o(x)$ to their equivalence classes $O = o(x) + \mathcal N$, where $\mathcal N$ is the subspace of local observables that is null under $\aveI{\cdot,\cdot}$. This map indeed satisfies the basic expected properties of a total integral.} $O = \int \dd x\, o(x)$. Then, by definition of equivalence classes, we have $\aveI{O_1,O_2} = \aveI{o_1,o_2}$, and we extend the meaning of $\aveI{O_1,O_2}$ by continuity to all elements of $\mathcal H'$. This provides $\mathcal H'$ with a Hilbert space structure. We will also use the integral symbol in the usual way when considering correlation function of local observables, hence we write in various ways $\aveI{O_1,O_2} = \aveI{o_1,o_2} = \ave{o_1 O_2}^{\rm c} = \ave{O_1o_2}^{\rm c}$ depending on the intended emphasis.

It is shown in \cite{pseudolocality} that the space $\mathcal H'$ is in bijection with the space of pseudolocal observables, a concept first introduced by Prosen that has played such an important role in the understanding of generalised thermalisation, see the review \cite{ilievski2016quasilocal}. We can in particular recast the hydrodynamic $\mathsf B$ and $\mathsf C$ matrices, Eqs.~\eqref{eq:Bmatrix}, \eqref{eq:Cmatrix}, in terms of this inner product:
\begin{equation}\label{eq:BCinnerproduct}
    \mathsf B_{li} =
    \aveI{j_l,q_i},\quad \mathsf C_{li} = \aveI{q_l,q_i}.
\end{equation}

As proven in \cite{hydroprojectionsEuler} in the context of quantum chains, time evolution can be transferred to a one-parameter group of unitary operators $\tau_t$ acting on $\mathcal H'$ (and we will naturally denote $O(t) = \tau_t O$). Thus we may time-evolve extensive observables unitarily. Unitarity is here not related to the quantum nature of the system; it is a consequence of general properties of the state including stationarity, and holds for classical systems as well.

With unitarity, strong results are available from functional analysis, such as von Neumann's ergodic theorem \cite{Rudin1991}. For instance, consider the closed space of all time-independent extensive observables, $\mathcal H_{\rm bal} = \cap_t {\rm ker}(\tau_t)$. The index ``bal" refers to ``ballistic", as this Hilbert space will be seen to be the space of emergent, ballistic degrees of freedom at the Euler scale. In particular, projection onto $\mathcal H_{\rm bal}$ can be shown to arise under time averaging, for arbitrary extensive observables $O_1, \, O_2$ \cite{hydroprojectionsEuler}:
\begin{equation}\label{eq:zeroprojection}
    \lim_{t\to\infty} \frac1{2t} \int_{-t}^t \dd s\,\aveI{O_1(s),O_2}
    =\aveI{\mathbb P O_1,\mathbb P O_2},\quad
    \mathbb P :\mathcal H'\to\mathcal H_{\rm bal}.
\end{equation}

What are more precisely the time-independent extensive observables $\mathcal H_{\rm bal}$? Clearly, if $q(x,t)$ is a local observable that satisfies a conservation law such as in Eq.~\eqref{continuity}, then $Q(t) = \int \dd x\, q(x,t)$ is an element of $\mathcal H'$ that is independent of $t$, according to our general definitions, and thus $Q\in\mathcal H_{\rm bal}$. Thus, naturally, the conserved quantities $Q_i$ already discussed in sec. \ref{sec:hydro} are elements of $\mathcal H_{\rm bal}$. It turns out that the most accurate {\em definition of the space of extensive conserved quantities}, from the viewpoint of linearised hydrodynamics, is obtained by inverting this argument: it is exactly the Hilbert space $\mathcal H_{\rm bal}$ of time-independent extensive observables defined above. The discrete set $Q_i$ can be rigorously taken as a discrete basis for the countable-dimensional Hilbert space $\mathcal H_{\rm bal}$. The local and quasi-local conserved quantities constructed by transfer-matrix methods in integrable systems \cite{ilievski2016quasilocal} are expected to form a good basis, although there is no proof yet of this. In any case, using the basis of $Q_i$'s, the projection is explicitly written as
\begin{equation}\label{lin_proj}
    \mathbb P O =
    Q_i\mathsf C^{ij} \aveI{Q_j,O}.
\end{equation}

From this, taking into account the inner-product form \eqref{eq:BCinnerproduct}, one can easily translate the Drude weights definition \eqref{drudeKMB} as an inner product of projected currents, and obtain
\begin{equation}\label{eq:drudeexpansion}
    \mathsf D_{ab} =
    \aveI{\mathbb PJ_a,\mathbb PJ_b}=
    \mathsf B_{ai}\mathsf C^{ij}\mathsf B_{jb}.
\end{equation}
This shows \eqref{drudeAB} and \eqref{drudeAsquareC}, and further implies that the Drude matrix is positive semi-definite,
\begin{equation}
    \mathsf D\geq 0.
\end{equation}
Drude weights (as well as the Onsager matrix) take an even more suggestive form in the basis of normal mode densities $n^a$ \eqref{orto}, or their extended versions $N^a=\int \dd x\, n^a(x,0)$ \eqref{nmb}. Due to orthogonality, we have
\begin{equation}
\label{hydro_pro}
\mathsf  D_{kl}=\langle j_k, n^a\rangle_{\text{I}}\langle n_a, j_l\rangle_{\text{I}}=\langle j_k N^a\rangle^c\langle N_a j_l\rangle^c.
\end{equation}
One may not know a complete basis of conserved quantities $Q_i$, but the projection mechanism guarantees that any subset provides a {\em lower bound} for the matrix. In particular, any diagonal Drude weight is bounded by any conserved quantity $Q$ as
\begin{equation}\label{eq:drudebound}
    \mathsf D_{ii} \geq 
    \frac{|\aveI{J_i,Q}|^2}{\aveI{Q,Q}}.
\end{equation}
This is an example of the Mazur bound \cite{Mazur1969}, which has been an extremely powerful tool in the study of ballistic transport, see for instance the reviews \cite{ilievski2016quasilocal,bertini2020finite}.

Perhaps surprisingly, it is possible to extend this projection analysis to the full {\em Euler scale of linearised hydrodynamics}. This includes not only correlation functions of conserved densities, the terms of Eq.~\eqref{eq:twopointfunction} which are first-order in derivatives, but also correlation functions of arbitrary local observables:
\begin{equation}\label{eq:defSoo}
S_{o_1,o_2}(x,t) =  \ave{o_1^\dagger(x,t)o_2(0,0)}^{\rm c}.
\end{equation}
The generalisation of the projection relation \eqref{eq:zeroprojection} to such correlation functions may be referred to as the {\em Boltzmann-Gibbs principle}\footnote{This principle is widely studied in the context of interacting stochastic particle systems with few conservation laws, see for instance \cite{KipnisLandimBook}. However, it is expected to be more generally applicable to many-body systems, stochastic or deterministic, at the Euler hydrodynamic scale.}. A precise form of this principle, which is shown rigorously in the context of quantum spin chains \cite{hydroprojectionsEuler}, is expressed for the Fourier transform at wave number $k$:
\begin{equation}
    S_{o_1,o_2}(k,t) = \int \dd x\,e^{\ii kx} S_{o_1,o_2}(x,t),
\end{equation}
after time averaging at fixed $\kappa=kt$,
\begin{equation}\label{eq:Sooeulertime}
    \lim_{\rm Eul} S_{o_1,o_2}(\kappa) = \lim_{t\to\infty}\frac1{2t}\int_{-t}^t \dd s\, S_{o_1o_2}(\kappa/s,s).
\end{equation}
The Boltzmann-Gibbs principle expresses the fact that projection to the conserved quantities occur for arbitrary $\kappa\in\RaR$,
\begin{equation}\label{eq:boltzmanngibbs}
    \lim_{\rm Eul} S_{o_1,o_2}(\kappa)
    = \lim_{\rm Eul} S_{\mathbb Po_1,\mathbb Po_2}(\kappa)
    = 
    \aveI{o_1,q_i}
    \mathsf C^{ij}\,
    \lim_{\rm Eul} S_{jk}(\kappa)\ 
    \mathsf C^{kl}
    \aveI{q_l,o_2}.
\end{equation}

\begin{figure}
    \centering
    \includegraphics[width=7cm]{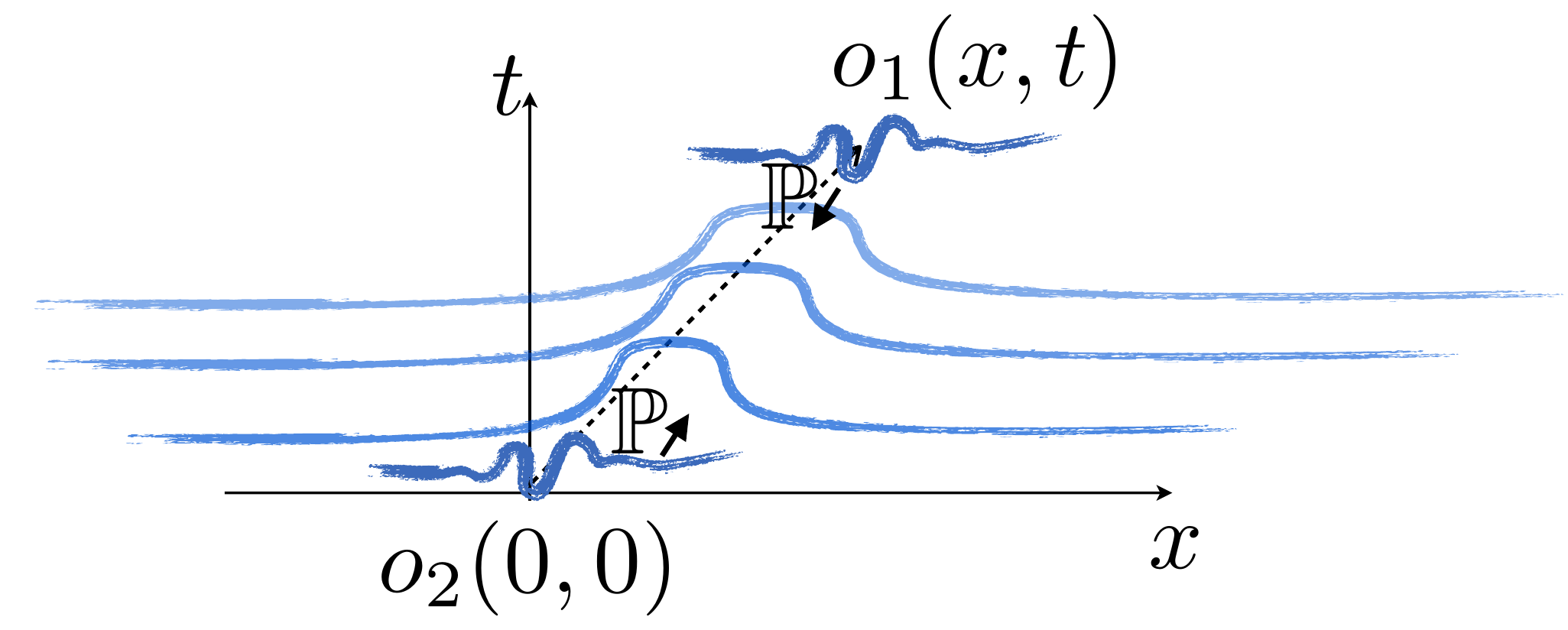}
    \caption{Cartoon representation of Eq.~\eqref{eq:boltzmanngibbs}. The observable $o_2(0,0)$ produces a generic disturbance of the state which slowly projects onto fluid modes. These propagate, and influence the observable $o_1(x,t)$ via its projection onto fluid modes. This is what produces the leading large-scale correlation between arbitrary observables.}
    \label{fig:waveprojection}
\end{figure}

Physically, this says that the leading large-wavelength, long-time correlation between local observables is due to the propagation of linear waves between these observables, and that the set of such linear waves is identified with the space of extensive conserved quantities $\mathcal H_{\rm bal}$. As this holds for arbitrary local observables, we conclude that the space $\mathcal H_{\rm bal}$ is the space of emergent dynamical degrees of freedom at large scales of space-time: the ballistic waves. Reading formula \eqref{eq:boltzmanngibbs}: a wave is created by the ``disturbance" $o_2(0,0)$, represented by the projection $\mathbb Po_2$; it propagates in space-time, represented by the dynamical structure factor $S_{ij}(\kappa)$; and it is observed by the ``probe" $o_1(x,t)$, represented by projection $\mathbb P o_1$. See fig.~\ref{fig:waveprojection}. With the exponential solution \eqref{eq:Ssolution}, the Boltzmann-Gibbs principle gives (in matrix notation)
\begin{equation}\label{eq:Seulerfull}
    \lim_{\rm Eul} S_{o_1,o_2}(\kappa) = \aveI{o_1,\boldsymbol q} \cdot \mathsf C^{-1}\exp[\ii \mathsf A \kappa]\aveI{\boldsymbol q,o_2}
\end{equation}
Using instead the space-time form of the Euler scaling limit, with \eqref{eq:Ssolutionxt}, the Boltzmann-Gibbs principle becomes
\begin{equation}\label{eq:boltzmanngibbsxt}
    \lim_{\rm Eul} S_{o_1,o_2}(x,t) =
    \aveI{o_1,\boldsymbol q} \cdot \mathsf C^{-1}\delta(x-\mathsf A t)\aveI{\boldsymbol q,o_2}.
\end{equation}
As mentioned, in general it is necessary, in the Euler scaling limit, to perform an appropriate fluid cell averaging, in order to ``wash out" possible oscillations that are not described by the Euler scale. Nevertheless, the direct asymptotic, without fluid cell averaging,
\begin{equation}
    S_{o_1,o_2}(x,t)\sim \lim_{\rm Eul} S_{o_1,o_2}(x,t)\quad (x,t\to\infty,\ x/t=\xi)
\end{equation}
is expected to hold in some cases in integrable models; in these cases, because of the delta function and the continuum of normal mode velocities, the leading decay of the correlation function is in $1/t$, times a function of $\xi$. See sec. \ref{sec:euler}.

It is also possible to extend the projection principle to non-stationary states. Here there is no general proof, but nevertheless the result is expected \cite{inprepa}. The idea is that the projection $\mathbb P$ represents a local effect, and hence should be applicable locally. Therefore, the non-stationary Euler scaling limit $\lim_{\rm Eul} S_{o_1,o_2}(x,t)$ of correlation functions of arbitrary observables $o_1,\,o_2$, defined as in \eqref{eq:Snonstateulerscale}, can be expressed solely in terms of the non-stationary Euler scaling limit $\lim_{\rm Eul} S_{ij}(x,t)$ of correlation functions of conserved densities. The projection mechanism can be written in terms of the ballistic inner product $\aveI{\cdot,\cdot}_{x,t}$ with respect to the state of the local fluid cell at $(x,t)$, and correspondingly the inverse static correlation matrix $\mathsf C_{\ x,t}^{ij}$, as
\begin{equation}\label{eq:projectionnonstat}
    \lim_{\rm Eul} S_{o_1,o_2}(x,t) =
    \aveI{o_1,q_i}_{x,t} \mathsf C_{\ x,t}^{ij}\,\lim_{\rm Eul} S_{jk}(x,t)\;\mathsf C_{\ 0,0}^{kl}\aveI{q_l,o_2}_{0,0}.
\end{equation}
Of course, this formula is not as explicit as \eqref{eq:boltzmanngibbsxt}, because the dynamical equation \eqref{eq:twopointfunctionnonstat} that $\lim_{\rm Eul} S_{jk}(x,t)$ satisfies does not in general admit, in the non-stationary case, a closed-form solution.

Three remarks are in order. First, the Boltzmann-Gibbs principle can be used to derive \eqref{eq:twopointfunction} without the need for the linear response argument made in sec. \ref{subsec:einstein}: by the microscopic equations of motion, the continuity equation $\partial S_{ij}(x,t)/\partial t + \partial  S_{j_i,q_j}(x,t)/\partial x=0$ holds for all $x,t$, and writing $S_{j_i,q_j}(x,t)$ at the Euler scale in projection form one recovers \eqref{eq:twopointfunction}. This simple argument can be used to rigorously prove $\dd S_{ij}(\kappa)/\dd\kappa = \ii \mathsf A_i^{~k}S_{kj}(\kappa)$ in quantum spin chains \cite{hydroprojectionsEuler}.

Second, as mentioned, the Boltzmann-Gibbs principle shows that the space $\mathcal H_{\rm bal}$ of conserved quantities is the correct space to describe the Euler scale. This is rigorously defined, and thus unambiguously characterises the emergent hydrodynamic degrees of freedom, at least for linearised hydrodynamics. It is a simple matter to show that $\mathcal H_{\rm bal}$ is infinite dimensional in integrable models, and one expects it to be finite dimensional in non-integrable models. The latter is in fact a crucial open problem. The recent paper \cite{Shiraishi_2019} makes progress by showing the absence of nontrivial local conservation laws in the (expectedly) non-integrable XYZ quantum chain with a magnetic field. However this does not guarantee the absence of quasi-local conserved quantities (that is, there may still be nontrivial elements in the completion $\mathcal H_{\rm bal}$).

Finally, as mentioned in sec. \ref{subsec:hydro}, one may want to characterise the absence of interactions in a model either by the fact that the current observables, in an appropriate gauge, are themselves conserved densities; or by the fact that the flux Jacobian $\mathsf A$ is independent of the state; or by the fact that diffusion matrix vanishes. In fact, a precise re-statement of the first proposition, which is perhaps the most universal characterisation as it implies the other two and holds in known free models, is that {\em the set of currents $J_i$, as elements of the space of extensive observables $\mathcal H'$, spans (a dense subspace of) the space of conserved quantities $\mathcal H_{\rm bal}$}. In this case we may choose a basis such that $\mathsf B = \mathsf C$, and the flux Jacobian is the identity. We will see below how this precise statement also implies the vanishing of the diffusion matrix.

\subsection{Projections, bounds and entropy production at the diffusive scale} \label{subsec:projectionsdiffusive}

An important recent discovery is that the notion of hydrodynamic projections can be applied to the diffusive scale in a way that parallels the Euler scale, with important consequences especially in integrable systems. In this section we discuss the main general ideas, independent of integrability.

In \cite{Ben} it was proposed that the Hilbert space formulation at the Euler scale could be extended to the diffusive scale. One defines the ``2$^{\rm nd}$ order" pre-inner product by time-integration of the $1^{\rm st}$-order inner product, thus further ``reducing" the space of spatially extensive observables to those that are also temporally extensive:
\begin{eqnarray}\label{eq:difinner1}
    \aveII{o_1,o_2}&=&
    \int_{-\infty}^{\infty} \dd t\, \aveI{O_1^-(t),O_2^-}\\
    &=& \lim_{t\to\infty}
    \int_{-t}^t \dd s\,
    \Big[\int \dd x\,
    \ave{o_1^\dagger(x,s)o_2(0,0)}^{\rm c} - \mathsf D_{o_1,o_2} \Big].
    \label{eq:difinner}
\end{eqnarray}
Here $O^{-}=(1- \mathbb{P})O$ denotes the extensive observable from which the Euler contribution is subtracted. Further, $\mathsf D_{o_1,o_2}$ is a generalisation of the Drude weight to arbitrary local observables,
\begin{equation}
    \mathsf D_{o_1,o_2}
    = \aveI{o_1,q_i}
    \mathsf C^{ij}\aveI{q_j,o_2}.
\end{equation}
The formulation \eqref{eq:difinner} clearly generalises the Onsager matrix \eqref{onsagerKMB} to arbitrary observables, so that
\begin{equation}\label{eq:Linner}
    \mathfrak L_{ij} =
    \aveII{j_i,j_j}.
\end{equation}
On the other hand, the first formulation \eqref{eq:difinner1} makes it clear that the diffusive-scale pre-inner product is positive semi-definite, hence in particular so is the Onsager matrix:
\begin{equation}
    \mathfrak L \geq 0.
\end{equation}
The Hilbert space $\mathcal H''$ is obtained by moding out the null observables under $\aveII{\cdot,\cdot}$, and Cauchy completing. This is the Hilbert space of space-time extensive observables.

Contrary to the Euler scale, there is no general statistical mechanics result guaranteeing that the diffusive inner product is actually {\em finite}. If it is infinite, then superdiffusive effects are present, which is generically the case in non-integrable systems admitting a conserved momentum according to the theory of nonlinear fluctuating hydrodynamics \cite{Spohn2014}. Here we assume that it is finite, which is typically the case in integrable systems, see sec. \ref{difhydro}.

Positivity of the Onsager matrix has an important physical consequence: positivity of the entropy production. The entropy density and its current can be defined in a completely general fashion, in any many-body system, using the free energy density and flux \eqref{eq:fg}:
\begin{equation}\label{eq:entropy}
    {\tt  s} = \beta^i {\tt q}_i - \mathsf f,\quad
    {\tt j}_{\rm s} = \beta^i {\tt j}_i - \mathsf g.
\end{equation}
From the diffusive hydrodynamic equation, \eqref{hydroequations} with \eqref{hydroexpansion}, one finds non-negativity of entropy production by positive-definiteness of the inner product (equivalently, positive-semidefinitenes of the Onsager matrix) (see e.g.~\cite{10.21468/SciPostPhysLectNotes.18}),
\begin{equation}
    \partial_t {\tt  s} + \partial_x {\tt j}_{\rm s}
    = \partial_x {\tt q}_i \mathsf C^{ij}
    \mathfrak D_j^{~k}\partial_x {\tt q}_k = \partial_x\beta^i \mathfrak L_{ij}\partial_x\beta^j
    = \aveII{\partial_x\beta^i j_i,\partial_x \beta^jj_j}
 \geq 0.
\end{equation}

The presence of a Hilbert structure $\mathcal H''$ suggests that it should be possible, by using projections, first, to express the Onsager matrix as an expansion into an appropriate basis, and, second, to bound it from below, much like for the Drude weights, Eqs.~\eqref{eq:drudeexpansion} and \eqref{eq:drudebound}. In the case of the Drude weights, the natural basis onto which to project was that of conserved quantities, $\mathcal H_{\rm bal}$. However, on $\mathcal H''$ conserved densities correspond to null elements, because of the subtraction by the projection $\mathbb P$ in \eqref{eq:difinner1} (equivalently, the subtraction by the Drude weights in \eqref{eq:difinner}). Pseudolocal conserved quantities have been ``subtracted out" in the diffusive Hilbert space.

It turns out that, with respect to the generalised times associated to higher conserved charges -- the other flows of the integrable hierarchy -- the currents $j_i$ are themselves, as elements of $\mathcal H''$, conserved quantities \cite{Ben,DurninBhaseenDoyon2020}. The subspace of invariants under all flows in $\mathcal H''$ was dubbed the diffusive subspace $\mathcal H_{\rm dif}\subset\mathcal H''$ in \cite{Ben}, and thus $j_i\in\mathcal H_{\rm dif}$. Hence for the study of hydrodynamic diffusion, this is the subspace that we must analyse. One can show \cite{Ben}, as discussed below, that {\em products of conserved charges}, which are quadratically extensive in space, are also, in fact, space-time extensive, thanks to the finite Lieb-Robinson velocity. By involution they are also invariant under all flows, so are elements of $\mathcal H_{\rm dif}$, and therefore can be used to bound the currents as element of $\mathcal H''$. Physically, the complementary space to $\mathcal H_{\rm dif}$, elements of $\mathcal H''$ which are not invariant under the integrable flows control thermalisation under integrability breaking \cite{DurninBhaseenDoyon2020}.

The first results bounding diffusion by the presence of quadratically extensive charges  was obtained in \cite{Prosen_2014}, and by the Drude weight curvature in \cite{MKP17}. These bounds can be related in a general expression in terms of fluctuations of local conservation laws \cite{10.21468/SciPostPhys.9.5.075}. By improving the bound from \cite{Prosen_2014}, the construction can be put on a more general and rigorous framework \cite{Ben}.

A quadratically extensive charge $Q^{[2]}$ can be understood as a limiting sequence of operators $Q^{[2]}_{[x_1,x_2]}$, restricted to an increasing interval $[x_1,x_2]$, such that 
\begin{equation}
    |Q^{[2]}|=\lim_{|x_1-x_2|\to\infty}\frac{\ave{Q^{[2]}_{[x_1,x_2]}Q^{[2]}_{[x_1,x_2]}}^c}{|x_1-x_2|^2}
\end{equation}
is finite\footnote{In fact, the limit does not need to exist, a quadratic bound is sufficient.}. By using the Cauchy-Schwartz inequality and assuming a maximal velocity  $v^{\text{max}}$ of correlation spreading, it turns out that\footnote{Here $o^-$ is understood by interpreting \eqref{lin_proj} as a projection on local densities of charges $q$.}
\begin{equation}
    |\ave{Q^{[2]} o^-}^{\rm c}| \leq
    \sqrt{\aveII{o,o}}
    \sqrt{2v^{\rm max} |Q^{[2]}|}.
\end{equation}
Here $v^{\text{max}}$ is a maximal velocity of correlation spreading, and is upper bounded in generic locally interacting systems by the Lieb-Robinson theorem \cite{robinson1997operator}. Correlations spread only inside of the maximal light-cone bounded by $v^{\text{max}}$, meaning that within the diamond shaped area 
\begin{equation}\label{diamond}
D_n=\left\{(x,t): -\half n+v^{\text{max}}t<x<\half n-v^{\text{max}}t,\quad |t|<\frac{ n}{2v^{\text{max}}}\right\},
\end{equation}
$Q^{[2]}_{[-n/2,n/2]}$ is time translation invariant. This allows us to consider a scaled limit of correlations within \eqref{diamond} and omit the time-dependence of the resulting quadratic charges in the definition of inner product \eqref{eq:difinner1}, resulting in the bound. Thus we have a strict lower bound on Onsager matrix elements $\mathfrak{L}_{ii}$ in the presence of a quadratically extensive charge $Q^{[2]}$:
\begin{equation}
    \mathfrak{L}_{ii}\geq \frac{\ave{j_i^- Q^{[2]}}^c\ave{ Q^{[2]} j_i^-}^c}{2 v^{\text{max}}|Q^{[2]}|}.
\end{equation}

Examples of quadratically extensive charges are the products of conserved charges
$Q_{ab}^{[2]}= \int \dd x \dd y\, q_a(x)q_b(y)$. A simple analysis using clustering shows that
\begin{equation}
    \label{dec_op}
    |Q_{ab}^{[2]}| = \aveI{Q_a,Q_a} \aveI{Q_b,Q_b} + \aveI{Q_a,Q_b}^2.
\end{equation}
In particular, using this, in isentropic fluids with a conserved number of particles $Q_0$ and momentum $Q_1$, the momentum diffusion $\mathcal L_{11}$, related to the viscosity of the hydrodynamic equation, is bounded as \cite{Ben}
\begin{equation}
    \mathcal L_{11} \geq \frac{(\chi v_{\rm s}\partial_{{\tt q}_0} v_{\rm s})^2}{v^{\rm max}}
\end{equation}
where $\chi = \mathsf C_{00}$ (the particle density susceptibility) and $v_{\rm s} = \sqrt{\mathsf A_1^{~0}}$ (the sound velocity).

A particularly useful form of the bound is obtained from the products of normal modes $N_{ij}^{[2]}=N_i N_j$. In this case the considered space-time region can be altered to
\begin{equation}\label{diamond2}
D_n=\left\{(x,t): -\half n+\max(v_i^\eff,v_j^\eff)t<x<\half n+\min(v_i^\eff, v_j^\eff)t,\quad |t|<\frac{ n}{ |v_i-v_j|}\right\},
\end{equation}
as the maximal and minimal velocity is set by the velocity of corresponding normal modes.  Due to the orthogonality of normal modes
\begin{equation}
    |N_{ij}^{[2]}|=\delta_{ik}\delta_{jl}+\delta_{il}\delta_{jk}.
\end{equation}
By considering the subspace formed by all normal modes, we obtain a simple bound on coefficients of the Onsager matrix
\begin{equation}
\label{dif_nm_bound}
\mathfrak{L}_{ii}\geq\sum_{kl}\frac{\langle j^-_i N_kN_l\rangle^c\langle N_k N_l j^-_i\rangle^c}{2|v_{k}^{\text{eff}}-v_{l}^{\text{eff}}|}.
\end{equation}

Notice that the bound diverges, if the three point function of the current and the same normal mode is non-vanishing or, more generally, between the current and two normal modes with degenerate velocities. This confirms the prediction of nonlinear fluctuating hydrodynamics, which asserts that superdiffusive behaviour emerges in such cases \cite{Spohn2014}. Note that the observed superdiffusive behaviour in integrable systems with non-abelian symmetries has different origins, see the review \cite{2103.01976} in this special issue.

In sec. \ref{sec:diffusion} we will show that in integrable systems the full Onsager matrix can be expressed by projecting onto the complete set of normal modes
\begin{equation}
\label{dif_nm_od}
\mathfrak{L}_{ij}=\sum_{kl}\frac{\langle j^-_i N_kN_l\rangle^c\langle N_k N_l j^-_j\rangle^c}{2|v_{k}^{\text{eff}}-v_{l}^{\text{eff}}|}.
\end{equation}
That is, in this case, the normal modes span $\mathcal H_{\rm dif}$. This expression for Onsager coefficients was first obtained in \cite{10.21468/SciPostPhys.9.5.075}, by considering the long lasting effects that local operators have on the state of the system within the hydrodynamic cell. A simple idea underlining the derivation is to expand generic operators on mesoscopic scales in terms of long lived excitations, by introducing hydrodynamic averages of observables within the cells of size $\Delta x=\ell$,  $o(x,t)\to \int_{x-\ell/2}^{x+\ell/2}\,\dd x\,o(x,t)$, $t\to t\ell$, $x\to x \ell$, $\mathtt{q}_i(x)\to \mathtt{q}_i(x)\ell$, and study the dynamics on the level of correlation functions on this mesoscopic lattice. Certainly these long lived excitations include conserved charges and their higher powers within cells, which motivates the expansion
\begin{equation}
\label{current_exp}
 o(x,t)=(\partial_{\mathtt{q}_i(x)} \ave{o(x)})q_i(x,t)+\frac{1}{2}(\partial_{\mathtt{q}_j(x)}\partial_{\mathtt{q}_i(x)} \ave{o(x)})q_i(x,t)q_j(x,t)+\mathcal{R},
\end{equation}
where we have chosen  $\ave{o}=0$ and $\ave{q_j}=0$. $\mathcal{R}$ includes higher order terms and other possible contributions, which we disregard. The first term corresponds exactly to the linear order hydrodynamic projection \eqref{lin_proj} within the hydrodynamic cell at point $x$. It can be shown that if Drude matrix is considered, the first term in \eqref{current_exp} is the only one that does not vanish, reproducing the result \eqref{hydro_pro}.

In order to probe the Onsager matrix, the Euler scale contributions should be moded out. This corresponds to considering $o^-(x,t)$ or, equivalently, discarding the first term in expansion \eqref{current_exp}.
The second term in expansion \eqref{current_exp} gives rise to a finite contribution $\mathfrak{L}^c_{ij}$ to the Onsager coefficient
\begin{equation}\label{eq:Lderq}
    \mathfrak{L}^c_{ij}=\frac{1}{2}(\partial_{\mathtt{q}_j(0,0)}\partial_{\mathtt{q}_i(0,0)} \ave{j_j(0,0)}) (M_{uv}^{j_i}-{\mathsf A_i^{\,\,k}}M_{uv}^{q_k}),
\end{equation}
in terms of the three-point functions
\begin{equation}\label{threep}
M_{ij}^o=\sum_{x}\int\dd t\left<q_i(0,t)q_j(0,t)o(x,0)\right>^c,
\end{equation}
which can be evaluated exactly, on the Euler scale
\begin{equation}\label{ons_bound}
\mathfrak{L}^c_{ij}=2({\mathsf R}^{-1}\tilde{ \mathsf G}^2 {\mathsf R}^{-\mathrm{T}})_{uv},
\end{equation}
in terms of Hessian $ { \mathsf H}_v^{\,\,ij}=\ell\partial_{\mathtt{q}_j(0,0)}\partial_{\mathtt{q}_i(0,0)} \ave{j_v(0,0)}$ and $G$-matrix
\begin{equation}\label{Gmat}
{ \mathsf G}_i^{\,\,jk}=\frac{1}{2}{ \mathsf R}_i^{\,\,l}\big({\mathsf R}^{-\mathrm{T}}{\mathsf H}_l{\mathsf R}^{-1}\big)^{jk},
\quad \tilde{\mathsf G}^2_{ij}=\frac{1}{|v_{i'}-v_{j'}|}{\mathsf G}_{ii'j'}{\mathsf G}_j^{\,\,i'j'}.
\end{equation}
Indeed, if we transform the result to the normal mode basis, the simple expression \eqref{dif_nm_bound} is obtained. In sec. \ref{subsec:higher} we discuss in particular how to study three-point functions of the type \eqref{threep} by nonlinear response theory.

The interpretation of these formulae is that these contributions to diffusion arise due to the non-linear effects: beyond the linear order, as characterised by higher-point functions, there is nontrivial scattering of ballistic waves, and this scattering leads to diffusive effects. This is manifested by the fact that the diffusion coefficient is bounded by the three-point correlation function. Localized fluctuations of conserved charge densities excited by the current operator at the origin, which are manifested by the second term in expansion \eqref{current_exp}, result in a finite contribution to diffusion as they reach the current operator at position $x$ \eqref{threep}. The scattering between ballistic waves is a coupling between tangent vectors to the state manifold, and hence has a geometric interpretation, see \cite{Ben}.

We see that if the flux Jacobian $\mathsf A$ is independent of the state, then by \eqref{eq:Lderq}, valid in integrable systems, the diffusion clearly vanishes; in general, this is the vanishing of the lower bound \eqref{dif_nm_bound}. If, more strongly, the total currents $J_i$ span the space of conserved quantities $\mathcal H_{\rm bal}$ (in fact, if they simply lie within it), then by \eqref{eq:Linner} it is clear that diffusion vanishes (as $\mathcal H_{\rm bal}$ is projected out in the definition of the inner product). This finally puts together the various characterisations of the absence of nontrivial interactions mentioned in sec. \ref{subsec:hydro} and \ref{subsec:projectionsEuler}.

\subsection{Non-linear response: higher-point functions} \label{subsec:higher}

As we saw in sec. \ref{subsec:einstein},  \ref{subsec:projectionsEuler} and \ref{subsec:projectionsdiffusive}, transport coefficients, the Drude weights $\mathsf D_{ij}$ or flux Jacobian $\mathsf A_i^{~j}$ at the Euler scale, and the Onsager matrix $\mathfrak L_{ij}$ or diffusion matrix $\mathfrak D_i^{~j}$ at the diffusive scale, determine the large-scale two-point dynamical correlation functions. This can be seen for instance by linear response from the hydrodynamic equations. In fact, of course, they  determine the full hydrodynamic equation, which is non-linear. Therefore, it is expected, by non-linear response, that all large-scale correlation functions be fully fixed by transport coefficients. This is a subject that requires further development, and as far as we are aware, there are no results beyond the Euler scale. However, there are some general results in the literature at the Euler scale. In this section we review two families of such results: dynamical equations for three-point correlation functions from non-linear response; and the full-counting statistics of charge transport, or more generally the moments of total line integrals of conserved 2-currents. The former helps illustrate the general idea behind how to access higher-point functions by non-linear response. The latter illustrates more advanced techniques of measure biasing. In fact, the full-counting statistics of charge transport is deeply related to two-point correlation functions of special types of local observables associated to global symmetries, referred to as twist fields. It suggests that a new theory for large-scale correlations that goes beyond the simple wave-propagation picture should be developed.

Further general results, which we will not review here, are concerned with the ``non-linear Drude weights", see for instance \cite{tanikawaExact2021}. In the more special context of GHD, a diagrammatic theory for nonlinear response to external fields was proposed recently \cite{favaHydrodynamic2021}, which we also do not review.

\subsubsection{Three-point functions} The first set of results is concerned with the Euler scaling limit of certain three-point functions, and serves to illustrate the idea that higher correlation functions can be obtained by a simple non-linear response theory based on formal functional differentiation. 

Consider the three-point function $\ave{o_1(x,t)o_2(y,0)o_3(z,0)}^{\rm c}$ in a maximal entropy state. Its Euler scaling limit may be defined as usual by fluid-cell averaging,
\begin{eqnarray}
    \hspace{-2cm}\lefteqn{\lim_{\rm Eul}
    \ave{o_1(x,t)o_2(y,0)o_3(z,0)}^{\rm c}}\\
    &\hspace{-0.8cm}=&\hspace{-0.3cm}
    \lim_{a\to 0^+}\lim_{\ell\to\infty} \ell^2
    \int_{[-a\ell,a\ell]^4} \frac{\dd x'\dd y' \dd t' \dd s'}{(2a\ell)^4}\,\ave{o_1(\ell x+x',\ell t+t')o_2(\ell y + y',s')o_3(\ell z,0)}^{\rm c}.\nonumber
\end{eqnarray}
Note that, for Euler-scale $n$-point functions, the scaling factor is $\ell^{n-1}$. In particular, if no fluid-cell averaging is required, then the connected three-point function behaves as
\begin{equation}
    \ave{o_1(\ell x,\ell t)o_2(\ell y,0)o_3(\ell z,0)}^{\rm c}
    \sim \ell^{-2} \ave{o_1(x,t)o_2(y,0)o_3(z,0)}^{\rm Eul}.
\end{equation}

Re-interpreting the discussion of linear response made in sec. \ref{subsec:einstein}, the insertion of a local observable $o(0,0)$ is seen as being the result of functional differentiation with respect to an appropriate (but here formally defined) conjugate potential $\beta^o(0,0)$. For instance, we may write, formally,
\begin{equation}
    \lim_{\rm Eul}
    \ave{o_1(x,t)o_2(y,0)o_3(z,0)}^{\rm c}
    = \frac{\partial^2\ave{o_1(x,t)}_{\rm ns}}{\partial \beta^{o_3}(z,0)\beta^{o_2}(y,0)}\Big|_{\rm stat}\\
\end{equation}
where the average $\ave{\cdots}_{\rm ns}$ is within an Euler-scale non-stationary state (of the form Eq.~\eqref{eq:nonstatdensity} but with extra conjugate potentials for $o_2$ and $o_3$), the functional derivatives affect its initial condition by perturbing the Euler-scale conjugate potentials, and the final result is evaluated in the stationary limit. This then takes the form, using the non-stationary projection formula \eqref{eq:projectionnonstat},
\begin{eqnarray}
    \hspace{-0.5cm}\lefteqn{\lim_{\rm Eul}
    \ave{o_1(x,t)o_2(y,0)o_3(z,0)}^{\rm c}} && \\
    &=& \frac{\partial}{\partial \beta^{o_3}(z,0)}\lim_{\rm Eul}
    \ave{o_1(x,t)o_2(y,0)}^{\rm c}_{\rm ns}\Big|_{\rm stat}\nonumber \\
    &=& \frac{\partial}{\partial \beta^{o_3}(z,0)}\Big[\aveI{o_1(x,t),q_i}_{x,t}\mathsf C^{ij}_{\ x,t} \lim_{\rm Eul} \ave{q_j(x,t)o_2(0,0)}^{c}_{\rm ns}\Big]\Big|_{\rm stat}.\nonumber
\end{eqnarray}
By using the Leibniz rule of differentiation, we obtain non-trivial relations between three-point functions. The result, as derived in \cite{Ben}, can be expressed as two statements. Let us denote, as in sec. \ref{subsec:projectionsdiffusive}, $o^-(x,t) = (1-\mathbb P)o(x,t)$, the observable $o(x,t)$ from which its projection onto  ballistic waves is subtracted. Then:
\begin{equation} \label{hydro_3point}
    \lim_{\rm Eul}\ave{o_1^-(x,t)o_2(y,0)o_3(z,0)}^{\rm c}
    = \lim_{\rm Eul}\ave{o_1^-(x,t)\mathbb Po_2(y,0)\mathbb Po_3(z,0)}^{\rm c}
\end{equation}
and
\begin{eqnarray}\label{hydro_3point_2}
    \lefteqn{\lim_{\rm Eul}\ave{o^-(x,t)q_a(y,0)q_b(z,0)}^{\rm c}} \qquad && \\
    &=& \aveI{o^-,q_i,q_j}\,\mathsf C^{ik}
    \mathsf C^{jl}\, \lim_{\rm Eul}S_{ka}(x-y,t)\, \lim_{\rm Eul}S_{lb}(x-z,t)\nonumber
\end{eqnarray}
where the ballistic ``three-point coupling" is
\begin{equation}\label{eq:3pointcoupling}
    \aveI{o,q_i,q_j} =
    \int \dd x\dd y \,\ave{o(0)q_i(x)q_j(y)}^{\rm c}
    = \frac{\partial^2}{\partial\beta^i\partial\beta^j} \ave{o}.
\end{equation}

\begin{figure}
    \centering
    \includegraphics[width=8cm]{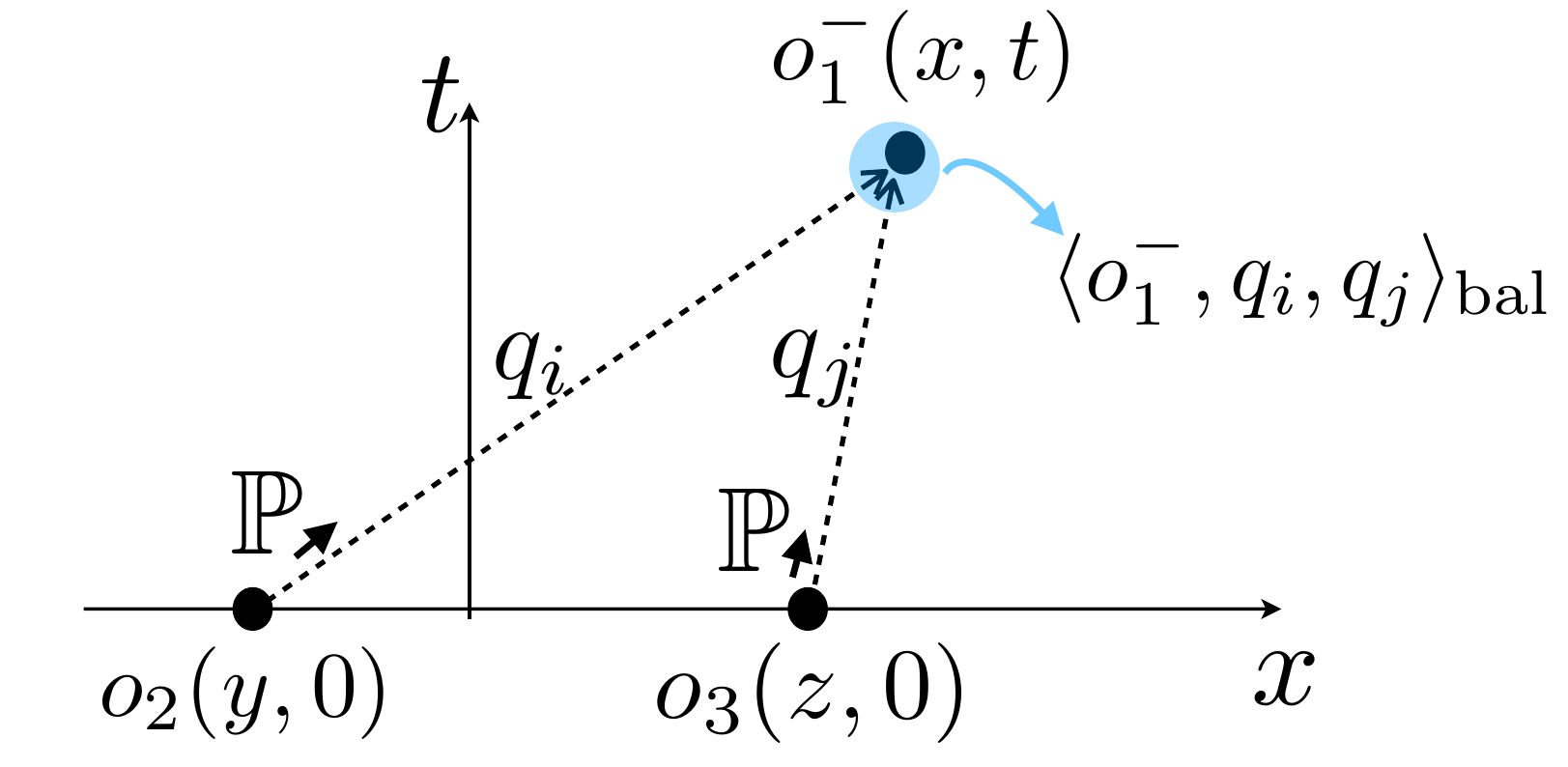}
    \caption{The connected three-point function at the Euler scale is controled by linear waves propagating from the observables at time 0, and an appropriate three-point coupling to the observable at time $t$.}
    \label{fig:threepoint}
\end{figure}

The first statement indicates that, after subtracting the part of the observable $o_1(x,t)$ that projects onto the conserved densities in the three-point function with $o_2(y,0)$ and $o_3(z,0)$ (that is, taking only the orthogonal component), the observables $o_2$ and $o_3$, both at time 0, may be projected onto the conserved densities (that is, their orthogonal components give no contribution). Further, the second statement means that the full space-time dependence is then given by a product of dynamical structure factors. These are the amplitudes for the propagation of waves from $(y,0)$ to $(x,t)$, and from $(z,0)$ to $(z,t)$. The result involves an appropriate three-point coupling between these amplitudes, which is the only part with information about the observable $o_1$. See fig.~\ref{fig:threepoint}. It is perhaps surprising that dynamical structure factors control three-point functions in such an intuitively simple way. It is this structure which ultimately allows the evaluation of \eqref{ons_bound}. Note that there is a possible ambiguity in these formulae: the projection $\mathbb P$ acts on the ballistic Hilbert space, not necessarily on the space of local observables. In particular, there is ambiguity by addition of total derivatives. However, these are not expected to affect the Euler scaling limit of higher-point functions, and thus the formula is still expected to make sense. The mathematical theory underlying this is still not developed.

\subsubsection{Full counting statistics} 
\begin{figure}
    \centering
    \includegraphics[width=5cm]{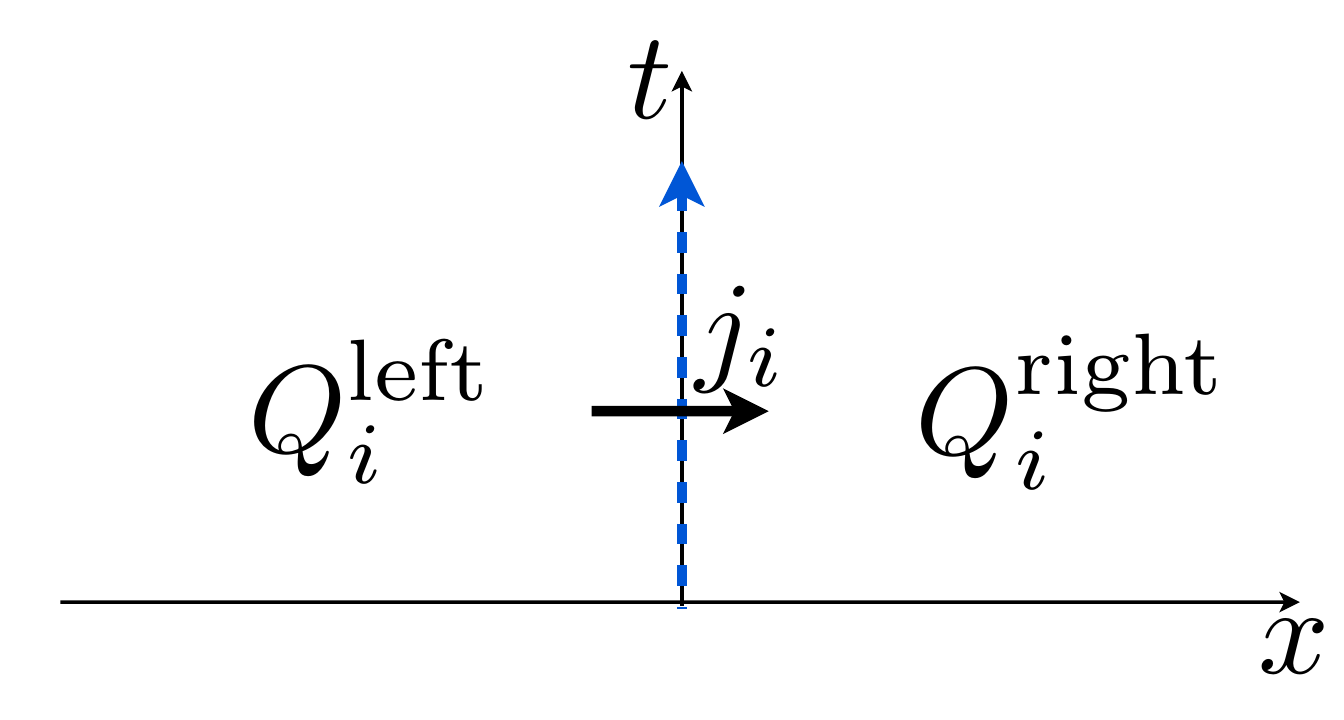}
    \caption{The total amount of charge transferred from the left to the right halves of the system in a time $t$ is the time integral of the current $j_i$. This fluctuates because of the fluctuations in the state, and at long time $t$, these fluctuations are accessible by Euler-scale hydrodynamics.}
    \label{fig:fcs}
\end{figure}

The second set of results is concerned with total line integrals of 2-currents, and their moments, in a maximal entropy state $\ave{\cdots}$. Consider the total current $J_i(t) = \int_{-t/2}^{t/2}\dd s\,j_i(0,s)$ over a period of time $t$. This is the total quantity of charge $Q_i$ that has been transferred, within the stationary state, from the left to the right of the point $x=0$, during a period of time $t$; see fig.~\ref{fig:fcs}. If the state carries currents for this quantity -- for instance, it is a non-equilibrium steady state --, then $J_i(t)$ has nonzero average. In general, this quantity is {\em fluctuating}, and characterises the non-equilibrium fluctuations of transport of $Q_i$. We are interested in describing these fluctuations. In particular, we wish to analyse the cumulants (connected correlation functions) $\ave{J_i(t)^n}^{\rm c}$. Clearly the first cumulant, $\ave{J_i(t)} = t\ave{j_i}$, grows like the length of the interval $t$. But also, if clustering is strong enough in time, then linear growth holds for every moment,
\begin{equation}
    \ave{J_i(t)^n}^{\rm c}\sim t c_n\quad (t\to\infty).
\end{equation}
The values of $c_n$ are referred to as the scaled cumulants of the total current, and take the form
\begin{equation}
    c_n = \Big\langle\int \dd s_1\cdots\dd s_{n-1}\, j_i(0,0) j_i(0,s_1)\cdots j_i(0,s_{n-1})\Big\rangle^{\rm c}.
\end{equation}
The growing asymptotic form $tc_n$ of the cumulants describe the large fluctuations of the quantity of transferred charge $Q_i$ during a long period of time $t$. As these are large scale quantities, it turns out that they are accessible by non-linear response from Euler hydrodynamics: the fluctuations, at large time, are due to ballistic transport. The theory is referred to as the ballistic fluctuation theory (BFF).

The same question can be asked for total line integrals along other rays $\alpha\in[0,\pi/2]$ in space-time
\begin{equation}
J_i(\ell,\alpha) = \int_{-\ell z_\alpha/2}^{\ell z_\alpha/2}\dd x^\mu \epsilon_{\mu\nu} j_i^\mu
\end{equation}
with the usual 2-current notation $j^0_i = q_i$ (time component), $j^1_i = j_i$ (space component). Here the space-time point $z_\alpha = (z_\alpha^1,z_\alpha^0) = (\sin\alpha ,\cos\alpha)$ determines the ray, and the case $\alpha=0$ is the time-ray case discussed above. Again, linear scaling holds if correlations decay sufficiently fast along these rays,
\begin{equation}
    \ave{J_i(\ell,\alpha)^n}^{\rm c}\sim \ell c_n(\alpha)\quad (\ell\to\infty),
\end{equation}
and the cumulants $c_n(\alpha)$ are controlled by Euler-scale hydrodynamics. The case $\alpha=\pi/2$ is the case of the space-integral of the conserved density,  $J_i(\ell,\pi/2) = -\int_{-\ell/2}^{\ell/2} \dd x\,q_i(x)$. In this case, the scaled cumulants are obtained simply from the specific free energy $\mathsf f$,
\begin{equation}
    c_n(\pi/2) = -\Big(\frac{\partial}{\partial\beta^i}\Big)^n\mathsf f.
\end{equation}
That is, they describe the well understood large-volume thermodynamic fluctuations of conserved quantities. The question of transport is a generalisation of this question to the time domain. The main object is the function
\begin{equation}
\mathsf f_{\lambda,\alpha} = \lim_{\ell\to\infty} \ell^{-1} \log\big(\ave{\exp[\lambda J_i(\ell,\alpha)]}\big)
= \sum_{n=1}^\infty
    c_n(\alpha) \frac{\lambda^n}{n!}
\end{equation}
which generalises the specific free energy (up to a change of sign). In particular, it controls the leading exponential behaviour as
\begin{equation}
    \ave{\exp\Big[\lambda 
    \int_{-\ell z_\alpha/2}^{\ell z_\alpha/2}\dd x^\mu \epsilon_{\mu\nu} j_i^\mu
    \Big]}
    \asymp \exp\big[\ell\, \mathsf f_{\lambda,\alpha}\big].
\end{equation}

Using the direct non-linear response techniques described above would be very tedious. Instead, the techniques introduced in the context of conformal field theory \cite{Bernard_2016} have been extended to arbitrary multi-component hydrodynamics \cite{doyonFluctuations2020}. The main idea is to use the generating function $\ave{\exp[\lambda J_i(\ell,\alpha)]}$. The argument is as follows. By conservation of currents, the object $\exp[\lambda J_i(\ell,\alpha)]$ is, at large $\ell$, invariant under space-time translations except for boundary terms far in space-time, and therefore the new state
\begin{equation}
    \ave{\cdots}_{\lambda,\alpha} = \lim_{\ell\to\infty} \frac{\ave{\exp[\lambda J_i(\ell,\alpha)]\cdots}}{
    \ave{\exp[\lambda J_i(\ell,\alpha)]}}
\end{equation}
is stationary and homogeneous. One can also argue that it is clustering, and as a consequence, by the general ergodicity principle, it must be a maximal entropy state, characterised by some $\beta^i_{\lambda,\alpha}$. Differentiation of $\mathsf f_{\lambda,\alpha}$ with respect to $\lambda$ gives the average of the $\alpha$-direction current $j_i^\alpha = \cos\alpha\, j_i-\sin\alpha\, q_i$ within this $\lambda$-dependent maximal entropy state. One concludes that
\begin{equation}\label{eq:generating}
    \mathsf f_{\lambda,\alpha} 
    = \int_0^\lambda \dd\lambda'\,
    \ave{j_i^\alpha}_{\lambda',\alpha}.
\end{equation}
But also, differentiation of any average $\ave{o}_{\lambda,\alpha}$ with respect to $\lambda$, which gives $\lim_{\ell\to\infty} \ave{J_i(\ell,\alpha) o}_{\lambda,\alpha}^{\rm c}$, can be re-written as differentiation with respect to $\beta^j_{\lambda,\alpha}$ by the chain rule, which involves instead $\ave{Q_jo}^{\rm c}_{\lambda,\alpha}$'s. Relating these to the hydrodynamic matrices \eqref{eq:Bmatrix}, \eqref{eq:Cmatrix}, leads to a {\em flow equation} for the $\beta^j_{\lambda,\alpha}$\footnote{Note that there is small mistake in the flow equation written in \cite[Eq 3.25]{doyonFluctuations2020}}:
\begin{equation}\label{eq:flow}
    \frac{\dd\beta^j_{\lambda,\alpha}}{\dd\lambda}=
    {\rm sgn}\big(\sin\alpha\,{\bf 1}-\cos\alpha\,\mathsf A_{\lambda,\alpha}\big)_i^{~j}.
\end{equation}

Again, the special case $\alpha=\pi/2$ is simple: $\beta^j_{\lambda,\pi/2} = \beta^j + \lambda\delta_i^j$ and $\mathsf f_{\lambda,\pi/2} = -(\mathsf f|_{\boldsymbol\beta + \lambda \boldsymbol\delta_i} - \mathsf f|_{\boldsymbol\beta})$ is the change of free energy due to the shift, that is\footnote{As usual in large-deviation theory, the symbol $\asymp$ means equality of the leading asymptotic behaviour of the logarithm of each side; hence this does not imply equality of possible algebraic pre-factors, for instance.}
\begin{equation}
    \ave{\exp\Big[-\lambda \int_{0}^x \dd y\,q_i(x,0)\Big]}
    \asymp \exp\Big[-x\big(\mathsf f|_{\boldsymbol\beta + \lambda \boldsymbol\delta_i} - \mathsf f|_{\boldsymbol\beta}\big)\Big].
\end{equation}
It is of course natural that the free-energy shift due to the insertion of the charge $Q_i$ be the quantity that controls the large-$x$ exponential behaviour, as $x$ may be interpreted as the volume of the subsystem on which the extra charge $Q_i$ is applied; this idea was exploited for instance in \cite{doyonFinite2007,Chen_2014}, see the discussion in \cite{doyonFluctuations2020}.

Interestingly, the general formalism extends this simple formula to arbitrary rays. In particular, along the time direction,
\begin{equation}\label{eq:FCS}
    \ave{\exp\Big[\lambda \int_{0}^t \dd s\,j_i(0,s)\Big]}
    \asymp \exp\Big[t\int_0^\lambda \dd \lambda'\,\ave{j_i}_{\lambda',0}\Big].
\end{equation}
Note how the role of $\lambda$ and $t$ are exchanged, and how the ensemble average is brought inside the exponential.

The BFF essentially applies ideas of change-of-measure and counting-field used in classial large-deviation theory (see e.g.~\cite{chetriteNonequilibrium2015}) to the realm of hydrodynamics. The quantity $\mathsf f_{\lambda,0}$ is often referred to as the {\em full counting statistics} for the transport of the charge $Q_i$; or, in the large-deviation literatyre, the {\em scaled cumulant generating function}. It is an important property of non-equilibrium steady states, studied for instance in mesoscopic physics for electric charge transport. The celebrated Levitov-Lesovik formula \cite{levitovlesovik,avronFredholm2008} in free-fermion models is one of the first important results. Beyond free theories, there are results in Luttinger liquids and conformal field theory, see the review \cite{Bernard_2016} and rigorous results in \cite{gawedzkikozlowski1,gawedzkikozlowski2}. The full counting statistics satisfies important fluctuation relations characterising non-equilibrium steady states (see e.g.~the review \cite{RevModPhys.81.1665}), proving such relations directly within the BFF is an open problem. In the case without interactions, where $\mathsf A$ is independent of the state, then the stronger extended fluctuation relations \cite{Bernard_2013} are immediate from the BFF \cite{doyonFluctuations2020}. As argued in \cite{doyonFluctuations2020}, when considering an arbitrary ray $\alpha$, the BFF gives predictions for correlation functions of certain types of local fields called ``twist fields". This in principle extends the applicability of hydrodynamics to fields which do not necessarily couple with linear waves; although much needs to be developed still in this direction. The assumption of strong enough clustering along the ray $\alpha$ is crucial, and its breaking leads to dynamical phase transitions, here interpreted using hydrodynamics \cite{doyonFluctuations2020}.

\section{Correlation functions in integrable systems}\label{sec:corrFF}

In this section we change the perspective. We move from a hydrodynamic theory, applicable to vast class of theories, to fine-tuned quantum integrable many-body models. Our main aim here is to develop methods to study dynamic correlation functions in homogeneous, stationary, states. Notable examples of such states are the ground states and finite temperature states. As explained in previous sections the non-equilibrium physics and hydrodynamics is aimed at identifying and studying dynamics of any maximum entropy states. We will see that quantum integrable models provide techniques to approach dynamic correlation function in such cases. Whereas tools developed with integrability, in principle should allow to determine correlation functions with their full space-time dependence, in this review we focus on the regime relevant for the hydrodynamics, i.e. the long-distance, large time correlations akin to considering the low-frequency and small momentum. We start with providing a broader perspective on the problem of determination of correlation function within integrable models. Then, we describe two approaches that build upon the concept of thermodynamic form-factors. We conclude this part by providing the thermodynamic form-factors for conserved densities and currents which play a central role in the hydrodynamic considerations in following sections.

Recall that one way of studying a many-body model is by determining its space of asymptotic states: these are the states obtained after evolving, for a long (positive or negative) time, any initial state with finite energy support in an otherwise vacuous environment. Asymptotic states are extremely useful for studying vacuum correlation functions in integrable models. This is because thanks to factorised scattering, the general axiomatic, analytic framework for the so-called form factors -- matrix elements of local observables in asymptotic states -- leads to exact solutions. Therefore, K\"all\'en–Lehmann spectral representation is explicit, and often rapidly convergent. But standard asymptotic states are properties of the vacuum. Based on the Bethe ansatz formulation of quantum integrability, similar concepts can be developed with respect to states of finite energy density. In such cases we have thermodynamic asymptotic states, thermodynamic form factors, and an axiomatic, analytic framework.  Scattering theory on top of finite-density states should exist beyond integrability, but we are not currently aware of useful results in this direction. 

Studies of correlation functions in quantum integrable models have a long history, starting with a work on the quantum 1d Ising~\cite{McCoyWuBOOK} and XY~\cite{PhysRevA.4.2331} models. The contemporary developments follow roughly along two different lines. One involve setup of relativistic quantum filed theories with purely elastic S-matrices, the Integrable Quantum Field Theories~\cite{MussardoBOOK}. The second, is based on the Bethe Ansatz for a wave function of quantum mechanical many-body system~\cite{KorepinBOOK}. In both setups computations of correlation functions rely on 3 ingredients: 
\begin{enumerate}
\item knowledge of the spectrum
\item knowledge of the form-factors,
\item performing the spectral sum of the K\"all\'en–Lehmann spectral representation.
\end{enumerate}
Whereas in solvable systems, i) is guaranteed almost from the definition, our access to ii) and iii) is not immediate. In the following, we will review the current state of art in determining form-factors and computing correlation functions. We shall be mainly focused on results important for the hydrodynamics theory, where we are interested in correlation functions in finite energy density states. Such states could correspond to the thermal equilibrium, GGE states, or some other non-equilibrium states of the system. In the passing let us note that determining the spectrum of integrable models is an active area of research with new results for various models including spin chains based on higher rank quantum groups~\cite{10.21468/SciPostPhysLectNotes.6,GERRARD2020115021}, quantum Kadomtsev-Petviashvili model~\cite{Kozlowski_2017} or new approach based on the Seperation of Variables, see~\cite{Maillet_2018} and references therein.

In the realm of the Bethe Ansatz solvable models determination of the form-factors is based on the Quantum Inverse Scattering method~\cite{1984_Izergin_CMP_94} mixed with the Slavnov formula for the scalar product~\cite{1989_Slavnov_TMP_79}. This way we can access form-factors of the field~\cite{Kojima_1997} and density~\cite{1990_Slavnov_TMP_82} operators in the Lieb-Liniger model, or form-factors of local spin operators in the XXZ spin chain~\cite{1999_Kitanine_NPB_554,Belliard2013,Fuksa2017,Pakuliak2014,Hutsalyuk2016}. The resulting form factors are for arbitrary states in finite systems. Given their explicit form, the spectral sum can be evaluated numerically as materialized by the ABACUS method~\cite{2009_Caux_JMP_50,Schlappa2012,PhysRevA.89.033605}. 
Recently new techniques, allowing to perform the spectral sum in the low density limit, were developed for the quantum Ising model~\cite{10.21468/SciPostPhys.9.3.033} and Lieb-Liniger gas~\cite{granet2020lowdensity}.

On the other hand, in the Integrable Quantum Field Theories, the vacuum form-factors can be determined from the bootstrap program~\cite{KAROWSKI1978455,SmirnovBOOK}. These form-factors are for infinite system and are not directly usable for two (and more)-point functions in finite energy density states. However, as per LeClair-Mussardo formula they allow for the computation of one-point functions~\cite{Leclair1996, Saleur:1999hq, Pozsgay:2010xd}. There were various attempts~\cite{Leclair_1996,Castro_Alvaredo_2002,Doyon_2005,Doyon_2007} in generalizing the vacuum form-factors to finite density states albeit in the end limited to free theories. Relying on the vacuum form-factors one can, however, access the regime of low temperatures (or in general low densities) as shown in~\cite{Essler:2009zz,Pozsgay:2010cr,Pozsgay2018}.

Irrespective of the origins of the form-factors, their important feature is the kinematic singularity. Form-factors have (at least) a simple pole whenever a particle in bra and ket state coincide. This fine structure causes difficulties when evaluating the correlation function in infinite volume. Indeed, one way of circumventing this problem in IQFT is a finite volume regularization~\cite{Pozsgay:2007kn,POZSGAY2008209,Essler:2009zz}. Another, still in the IQFT context, is a "point-splitting" regularization~\cite{Essler:2009zz}. 

The presence of kinematic singularities creates a fundamental obstacle in performing the spectral sum in the thermodynamic limit. To better address this issue it is convenient to change our perspective. Instead of considering the form-factors of arbitrary two states we can understand one state as an excited state formed by creating a certain number of particles and holes on top of the other state. This picture is especially convenient when considering correlation functions of local operator, which do not couple the states that are thermodynamically different. From this point of view the kinematic singularities occur when a particle in an excited state coincides with a hole, that is when an excited state contains small particle-hole excitations.

Another, however related, feature of the finite system form-factors is their irrational dependence on the volume. This signals that the finite-size system excitations need to be reorganized and resummed when taking the thermodynamic limit. The standard picture involves then a resummation of the form-factors over small particle-hole excitations, referred to as soft-modes~\cite{Caux_2016}. This resummation, due to presence of the kinematic singularities, is technically very involved. The soft-modes resummation was however successfully implemented in few cases. An important result in that aspect was derivation of the large-time and long-distance asymptotics of the two-point functions in the Lieb-Liniger model~\cite{1742-5468-2012-09-P09001} and XXZ spin chain in the massless regime~\cite{doi:10.1063/1.5094332}.  Traces of the resummation can be also identified in the perturbative expansion of the Lieb-Liniger gas in the Tonks-Girardeau regime~\cite{10.21468/SciPostPhys.9.6.082} and in the form-factor approach to correlation functions in the XY model~\cite{Gamayun_2021}.

Finally, we mention the quantum transfer matrix (QTM) approach to the dynamic correlation functions at finite temperature~\cite{Dugave_2013,G_hmann_2017,2011.12752}. Contrary to the previous approaches, that were based on form-factors between eigenstates of the Hamiltonian, the spectral sum in this approach is spanned over the eigenstates of the time-dependent transfer matrix. Recently, closed-form expressions for a real-time correlation function in the massive regime of the XXZ model were derived~\cite{babenko2020exact}.

The approach that we review here, and which was successful in addressing the hydrodynamic regime, is based on the concept of thermodynamic form-factors. These are form-factors for finite energy density states in the infinite system, which enter directly the spectral sum. The latter takes a form of an integral over allowed particles and holes in an excited state. The thermodynamic form-factors still contain kinematic singularity and a regularization scheme has to be adopted to define the integrations. These take a form of the Hadamard integrals. The thermodynamic form-factors can be computed either from thermodynamic limit of the finite system form-factors~\cite{1742-5468-2015-2-P02019,SciPostPhys.1.2.015,DeNardisP2018} or by the thermodynamic bootstrap program~\cite{Bootstrap_JHEP,Bootstrap_SciPost}. In the following we review both approaches. The thermodynamic form-factors, combined with the quench action, provide also an alternative derivation of the Euler scale GHD equations~\cite{cubero2020generalized}, as reviewed in~\cite{TakatoReview}.

The quantum integrable systems are characterized by a stable (quasi-)particle content which, in a finite system, uniquely specifies the eigenstates. In a system with the periodic boundary conditions, the eigenstates are then described by quantum numbers. For simplicity of the presentation we focus on the theories with a single particle type. We remark on the generalization to theories with multiple particle types later. We denote $|\bfI_N \rangle$ an eigenstate described by quantum numbers $\bfI_N = \{I_j\}_{j=1}^N$. Then, the dynamic two-point function, 
\begin{equation}
    \langle \Omega| o_1(x, t) o_2(0, 0) |\Omega \rangle = \sum_{\bfI} e^{ix (P_\bfI - P_{\Omega}) - it (E_\bfI - E_{\Omega})} \langle \Omega| o_1(0) | \bfI \rangle \langle \bfI | o_2(0) |\Omega \rangle, \label{spectral_finite}
\end{equation}
where $|\Omega\rangle$ is a generic eigenstate of the system and the summation extends over all allowed sets of quantum numbers. The momentum $P_\bfI$ and energy $E_\bfI$ of any eigenstate are explicitly known for a generic integrable model.  Instead, the matrix elements, form-factors, $\langle \bfI | o_2(0) |\Omega \rangle$ are known only for some operators and only some specific models. The prominent examples are form-factors of local spin operators in the XXZ spin chain~\cite{1999_Kitanine_NPB_554}, form-factors of the field and density operators in the Lieb-Liniger model~\cite{1989_Slavnov_TMP_79,KorepinBOOK}. To simplify the notation we will consider in details the two-point functions of the same operator. Generalization to the case of two different operators is then straightforward

We consider two-point functions of operators conserving number of particles, such as local conserved density or current. In such cases, the states $\{I_j\}$ can be understood as consisting of a number of particle-hole excitations with respect to the reference state $|\Omega\rangle$. This centers the spectral sum in~\eqref{spectral_finite} around the quantum numbers of $|\Omega\rangle$. We denote by $\bfI^+$ and $\bfI^-$ the quantum numbers of particles and holes respectively. Then
\begin{equation}
    \langle \Omega| o(x, t) o(0, 0) |\Omega \rangle = \sum_{m=1}^{\infty}\frac{1}{(m!)^2} \sum_{\bfI_m^+, \bfI_m^-}  e^{ix p_{\Omega}(\bfI^+, \bfI^-) - it e_{\Omega}(\bfI^+, \bfI^-)} |\langle \Omega| o(0) | \Omega; \bfI^+, \bfI^- \rangle|^2, \label{spectral_finite_ph}
\end{equation}
with $p_{\Omega}(\bfI^+, \bfI^-)$ and $e_{\Omega}(\bfI^+, \bfI^-)$ the momentum and energy of the excited state $| \Omega; \bfI^+, \bfI^- \rangle$ with respect to $|\Omega\rangle$. 

Whereas quantum numbers are convenient when spanning the Hilbert space, the momentum, energy and form-factors are explicit functions of rapidities (quasi-momenta) $\{\theta_j\}$. The two sets are related by the Bethe equations  
\begin{equation}
    L p(\theta_j) = 2\pi I_j + i \sum_{k \neq j} \log S(\theta_j - \theta_k), \qquad j = 1, \dots, N,
\end{equation}
stated for the system of length $L$ with $N$ particles. In this expression $p(\theta)$ is the (bare) momentum and $S(\theta)$ is the $2$-body scattering matrix. In writing Bethe equations, we have assumed that the scattering matrix depends only on the difference of the rapidities. The map defined by the Bethe equations is one-to-one and therefore the sum in~\eqref{spectral_finite_ph} can be equivalently expressed in terms of the rapidities
\begin{equation}
    \langle \Omega| o_1(x, t) o_2(0, 0) |\Omega \rangle = \sum_{m=1}^{\infty} \frac{1}{(m!)^2} \sum_{\bft_m^+, \bft_m^-} e^{ix k_{\Omega}(\bft^+, \bft^-) - it \varepsilon_{\Omega}(\bft^+, \bft^-)} |\langle \Omega| o(0) | \Omega; \bft^+, \bft^- \rangle|^2, \label{spectral_finite_rapidities}
\end{equation}
The finite system representations are useful for numerical evaluation of the correlation function \cite{2009_Caux_JMP_50,Schlappa2012,PhysRevA.89.033605}. To pursue an analytic approach further it is practical to consider the thermodynamic limit in which $L \rightarrow \infty$ with $N/L$ fixed. In the thermodynamic limit, rapidities follow a distribution $\rho_{\rm p}(\theta)$, such that $L \rho_{\rm p}(\theta) {\rm d}\theta$ is the number of rapidities in the range $[\theta, \theta + {\rm d}\theta]$.
In the thermodynamic limit the summations over particle and hole excitations turn into integrals weighted by appropriate density functions
\begin{equation}
    \langle \rho_{\rm p}| o_1(x, t) o_2(0, 0) |\rho_{\rm p} \rangle = \sum_{m=1}^{\infty} \frac{1}{(m!)^2} \fint {\rm d}\bfp_m {\rm d}\bfh_m e^{ix k(\bfp, \bfh) - it \varepsilon(\bfp, \bfh)} |\langle \rho_{\rm p}| o(0) |  \rho_{\rm p} ;  \bfp, \bfh \rangle|^2, \label{spectral_th}
\end{equation}
with
\begin{equation}
    {\rm d}\bfp_m {\rm d}\bfh_m = \prod_{j=1}^m {\rm d} \theta^+_j {\rm d} \theta^-_j \rho_{\rm p}(\theta^-_j) \rho_{\rm h}(\theta^+_j).
\end{equation}
Here $|\rho_{\rm p}\rangle$ denotes a thermodynamic state of the system, the finite-size state $|\Omega\rangle$ can be viewed as its discretization.
The thermodynamic form-factors $\langle \rho_{\rm p}| o(0) |  \rho_{\rm p} ;  \bfp, \bfh \rangle$ are defined through a resummation of the finite-system form-factors $\langle \Omega| o(0) | \Omega; \bft^+, \bft^- \rangle$. 
One should also notice that kinematic properties of excitations are now also taken in the thermodynamic limit. In particular we defined momentum and energy of excitations, which are now dressed by the finite density background 
\begin{align}\label{eq:PE}
k(\bfp, \bfh) = \sum_{j=1}^m \left( k(\theta^+_j) - k(\theta^-_j) \right)\\
\varepsilon(\bfp, \bfh) = \sum_{j=1}^m \left( \varepsilon(\theta^+_j) - \varepsilon(\theta^-_j)   \right).\no
\end{align}
These functions depend on the reference state via the so-called back-flow function. Namely, given the single-particle energy $e(\theta)$ and the single-particle momentum $p(\theta)$, we have 
\begin{align}
\varepsilon(\theta) = e(\theta) + \int  \dd \alpha\,  F( \theta, \alpha) e'(\alpha ) n(\alpha) \\
k(\theta) = p(\theta) + \int  \dd \alpha\, F( \theta, \alpha)   p'(\alpha ) n(\alpha) 
\end{align}
with the back-flow $F(\theta,\alpha)$ being the amplitude of the global $1/L$ shift of the rapidities close to $\theta$ in the presence of the excitations $\alpha$, \cite{KorepinBOOK}. The back-flow is written in terms of the dressed scattering phase shift, more precisely
\begin{equation}
F  =   \frac{\log S }{2\pi i} (1 - nT)^{-1}.
\end{equation}
where we defined the linear operator 
\begin{equation}
    (1 - nT)(\theta,\alpha) \equiv \delta(\theta-\alpha) - n(\theta) T(\theta,\alpha)
\end{equation}
\begin{equation}
    (1 - Tn)(\theta,\alpha) \equiv \delta(\theta-\alpha) - n(\alpha) T(\theta,\alpha)
\end{equation}
 with the fermonic occupation numbers $n(\theta)= \rho_{\rm p}(\theta)/\rho_{\rm tot} (\theta)$, and the total density of particles given by $\rho_{\rm tot} = \rho_{\rm p} + \rho_{\rm h}$. The differential scattering shift $T$ is defined in terms of the scattering shift $\log S$ as its logarithmic derivative,
 \begin{equation}
    T(\theta, \alpha) = \frac{\dd}{\dd\theta} \frac{\log S(\theta, \alpha)}{2\pi \ii}.
 \end{equation}
This will be essential in all hydrodynamic equations, in particular for the definition of dressing of charges. Given a function $f(\theta)$ we shall indeed define its dressed version as 
 \begin{equation}\label{eq:dress}
    f^{\rm dr} = (1- T n)^{-1} f
\end{equation}
We shall also define the effective operation, in particular the effective velocity is the derivative of the dispersion relation respect to momentum
\begin{equation}\label{eq:effective-velo}
    v^{\rm eff} = \frac{ \partial \varepsilon }{\partial k } = \frac{ (e')^{\rm dr} }{ (p')^{\rm dr} },
\end{equation}
The thermodynamic form-factors have simple poles (kinetic singularities) when $\theta_j^+ \rightarrow \theta_k^-$. The integrals in~\eqref{spectral_th} are regularized by adopting the Hadamard prescription for integrals with double poles
\begin{equation}
    \fint {\rm d}x \frac{f(x)}{x^2} = \lim_{\epsilon\rightarrow 0} \left( \int {\rm d}x \,\Theta(|x| - \epsilon) \frac{f(x)}{x^2} + \frac{2f(0)}{\epsilon}\right).
\end{equation}
To evaluate Eq.~\eqref{spectral_th} we should first use Hadamard prescription to perform the integrals over particles (holes). The remaining integrations over holes (particles) are then regular. 

In general, the spectral sum~\eqref{spectral_th} is difficult to evaluate given the complicated expressions for the thermodynamic form-factors. The exception is the 1ph contribution which is straightforward to compute due to the energy-momentum conservation. Higher contributions require performing actual integrations and as such dealing with the Hadamard regularization. Known results in this direction concern so far only small momentum and energy contribution from the 2ph excitations to the two-point functions of conserved densities and currents~\cite{Panfil_2021}. 

The thermodynamic form-factors can be obtained either by relating them to finite-size form-factor or from the Thermodynamic Bootstrap program. In the following we recall the two approaches.

\subsection{Thermodynamic form-factors from finite-size form-factors}

To define thermodynamic form-factor we first define the thermodynamic state $|\rho_{\rm p}\rangle$, as the thermodynamic limit of the sum over all finite-size states such that their rapidities $\bft$ are a discretization of~$\rho_{\rm p}(\theta)$ (a precise construction of the set of allowed discretization can be found in~\cite{Caux_2016,granet2020lowdensity})
\begin{equation}
    |\rho_{\rm p}\rangle = \lim_{\rm th} \frac{1}{\mathcal{N}_{\rho}} \sum_{\bft} |\bft\rangle.
\end{equation}
The normalization factor, assuming normalization of the finite-size states, is equal to the square root of the cardinality of the set of allowed discretizations. This in turn is given by the exponential of the Yang-Yang entropy,
\begin{equation}
    \mathcal{N}_{\rho_{\rm p}} = e^{-\frac{1}{2} S_{\rm YY}[\rho_p]},
\end{equation}
where Yang-Yang entropy $S_{\rm YY}[\rho]$ is a functional of the particles distribution
\begin{equation}
    S_{\rm YY}[\rho_{\rm p}] = L \int {\rm d}\theta s[\rho_{\rm p}; \theta], \quad s[\rho_{\rm p}; \theta] = \rho_{\rm tot}(\theta) \ln \rho_{\rm tot}(\theta) - \rho_{\rm p}(\theta) \ln \rho_{\rm p}(\theta) - \rho_{\rm h}(\theta) \ln \rho_{\rm h}(\theta). \label{S_YY}
\end{equation}
The thermodynamic form factor is then defined as 
\begin{equation}
    \langle \rho_{\rm p}| o(0) |  \rho_{\rm p} ;  \bfp, \bfh \rangle = \lim_{\rm th} \frac{1}{\mathcal{N}_{\rho_{\rm p}}\, \mathcal{N}_{\rho_{\rm p}; \bfp, \bfh}} \sum_{\bft, \bft'} \langle \bft | o(0) | \bft'\rangle, \label{TFF_th}
\end{equation}
where $|\bft'\rangle$ is a discretization of the excited state $|\rho_{\rm p} ;  \bfp, \bfh \rangle$. The thermodynamic form-factors of the density operator in the Lieb-Liniger model were computed in~\cite{1742-5468-2015-2-P02019}, see also \cite{SciPostPhys.1.2.015, DeNardisP2018}, following this route. The computations rely on two further assumptions.

First one implies that in the double summation in~\eqref{TFF_th} one can fix one of the finite-size states, thus setting a reference state. We denote the chosen set of rapidities $\bft_{\rm ref}$. Then for a given set of particle-hole excitations $(\bfp, \bfh)$,
\begin{equation}
    \langle \rho_{\rm p}| o(0) |  \rho_{\rm p} ;  \bfp, \bfh \rangle = \exp\left( - \frac{1}{2} \delta S[\rho_{\rm p}] \right) \lim_{\rm th} \sum_{\bft'} \langle \bft_{\rm ref} | o(0) | \bft'\rangle, \label{TFF_gauge}
\end{equation}
 where we used that
\begin{equation}
     \frac{\mathcal{N}_{\rho_{\rm p}}}{\mathcal{N}_{\rho_{\rm p}; \bfp, \bfh}} = \exp\left( - \frac{1}{2} \delta S[\rho_{\rm p}] \right),
\end{equation}
is an intensive quantity, where the differential entropy $\delta S[\rho_{\rm p}]$ is defined as~\cite{KorepinBOOK}
\begin{equation}
    \delta S[\rho_{\rm p}] = \int {\rm d}\theta\, s[\rho_{\rm p}, \theta] \frac{\partial}{\partial \theta} \left(\frac{F(\theta)}{\rho_{\rm p}(\theta)} \right).
\end{equation}
To analyze the remaining sum we explicitly pull out of the set of $\bft'$ the set of rapidities corresponding to the particle-hole excitations $(\bfp, \bfh)$. The remaining rapidities form the background state which can be understood as the reference state with small (namely whose amplitudes vanish in the limit $L \to \infty$) modifications, the soft-modes, excitations that carry sub-intensive momentum and energy in the thermodynamic limit. The summation is over these soft-modes,
\begin{equation}
    \sum_{\bft'} \langle \bft_{\rm ref} | o(0) | \bft'\rangle = \sum_{{\rm soft-modes}} \langle \bft_{\rm ref} | o(0) | \bft_{\rm ref} + (\mathrm{soft-modes}); \bfp, \bfh \rangle.
\end{equation}
The soft-mode summation is technically difficult to perform. Certain progress was achieved by considering only specific classes of excited states. This way it was possible, for example, to obtain the Luttinger liquid predictions for the large-time, long-distance correlation functions from microscopic models, see~\cite{1742-5468-2012-09-P09001} and references therein. Conceptually similar problems were also addressed within the Quantum Transfer Matrix approach~\cite{Dugave_2013,G_hmann_2017,doi:10.1063/1.5094332}. Techniques pertaining to the problem of soft-modes summation were recently developed in~\cite{10.21468/SciPostPhys.9.6.082}. 

In the hydrodynamic context, where we are interested in small particle-hole excitations one can expect an additional simplification. Heuristically, the soft-modes are caused by the perturbation of the reference state caused by creating excitations. When these excitations are small, the resulting perturbation should also be small. This way, the sum can be further approximated by choosing a representative excited state (we will simply choose $\bft_{\rm ref}$) and multiplying by an appropriately chosen weight factor. The weight factor corresponds to the phase-space given to the soft-modes. These phase space can be estimated as the number of discretizations of the excited state minus the number of the discretizations of the representative state. This yields
\begin{equation}
    \sum_{\bft'} \langle \bft_{\rm ref} | o(0) | \bft'\rangle = \exp\left( \delta S[\rho_{\rm p}] \right)\langle \bft_{\rm ref} | o(0) | \bft_{\rm ref}; \bfp, \bfh \rangle.
\end{equation}
Within these approximations, the thermodynamic form-factors are given by
\begin{equation}
    \langle \rho_{\rm p}| o(0) |  \rho_{\rm p} ;  \bfp, \bfh \rangle = \exp\left( \frac{1}{2}\delta S[\rho_{\rm p}] \right) \lim_{\rm th}\langle \bft_{\rm ref} | o(0) | \bft_{\rm ref}; \bfp, \bfh \rangle.
\end{equation}
In practice, to evaluate the right hand side, one has to choose a reference state: a discretization of the representative state $|\rho_{\rm p}\rangle$. In \cite{DeNardisP2018} a uniform discretization was chosen. The resulting form-factors are quite involved, therefore we won't display here their full formulas. Instead, in sec.~\ref{subsec:small_ff}, we show just the limiting expressions relevant for the theme of this review.

\subsection{Thermodynamic Bootstrap Program}\label{subsec:bootstrap}

The Thermodynamic Bootstrap Program~\cite{Bootstrap_JHEP} provides an alternative and direct route to the thermodynamic form-factors. It consists of a number of axiom that the thermodynamic form-factor should obey. From them, one can hope to bootstrap the formulas desribing them. So far the full thermodynamic form-factors haven't been obtained in this way. However it was possible to fix the minimal form-factors and obtain limiting expressions, valid in the regime relevant for the hydrodynamics. In the following we recall the axioms. They are formulated for theories without bound states in the spectrum.

The Thermodynamic Bootstrap Program is formulated for Integrable Quantum Field Theories. These are relativistic theories characterized by purely elastic scattering matrix $S(\theta)$~\cite{MussardoBOOK}. We introduce the following notation
\begin{equation}
    f_{\rho_{\rm p}}^o(\theta_1, \dots, \theta_m) = \langle \rho_{\rm p}| o(0) | \rho_{\rm p}; \theta_1, \dots, \theta_m), \label{ff_TBP}
\end{equation}
for a form-factor with $m$ particle excitations. The relativistic invariance of IQFT’s implies that a hole-excitation in the ket, is equivalent, by crossing
symmetry, to a particle excitation in the bra. The crossing of a particle from bra to ket amounts to the transformation of its rapidity, $\theta \rightarrow \theta + i \pi$. Therefore the notation~\eqref{ff_TBP} includes the insertion of both, particles and holes, as long as we allow $i \pi$ shifts in rapidity. Of main interest for us are the $m$ particle-hole form factors. They are expressed through $2m$ particles form-factor in the following way
\begin{align}\label{eq:secFFCorr}
    \langle \rho_{\rm p}| o(0)| \rho_{\rm p}, \bft^+, \bft^- \rangle = f_{\rho_{\rm p}}(\bft^+, \bft^- + i \pi).
\end{align}   
In the case of the operator acting on some position $x$ and time $t$, the space-time dependence of the corresponding form-factor factorizes such that
\begin{equation}
        \langle \rho_{\rm p}| o(0) | \rho_{\rm p}; \theta_1, \dots, \theta_m) = e^{\ii x k(\bft) - \ii t \varepsilon(\bft)} f_{\rho_{\rm p}}^o(\theta_1, \dots, \theta_m).
\end{equation}

The Thermodynamic Bootstrap Program consists of the following axioms for $f_n^o(\theta_1, \dots, \theta_m)$:
\begin{itemize}
    \item the scattering axiom
    \begin{equation}
        f_{\rho_{\rm p}}^o(\theta_1, \dots, \theta_i, \theta_{i+1}, \dots, \theta_m) = S(\theta_i - \theta_{i+1}) f_{\rho_{\rm p}}^o(\theta_1, \dots, \theta_{i+1}, \theta_i, \dots, \theta_m),
    \end{equation}
    \item the periodicity axiom
    \begin{equation}
        f_{\rho_{\rm p}}^o(\theta_1, \dots, \theta_m) = R_{\rho_{\rm p}}(\theta_m| \theta_1, \dots, \theta_m) f_{\rho_{\rm p}}^o(\theta_n + 2\pi i, \theta_1, \dots, \theta_{m-1}), 
    \end{equation}
    \item the annihilation axiom (kinematic singularity)
    \begin{align}
        &-i {\rm res}_{\theta_1 \rightarrow \theta_2} f_{\rho_{\rm p}}^o(\theta_1 + \pi \ii, \theta_2, \dots, \theta_n) = \no \\
        &\left( 1 - R_{\rho_{\rm p}}(\theta_2| \theta_3, \dots, \theta_n) \prod_{j=3}^n S(\theta_2 - \theta_j) \right) f_{\rho_{\rm p}}^o(\theta_3, \dots, \theta_n),
    \end{align}
    \item normalization of the $2$-particle form-factor
    \begin{equation}
        f_{\rho_{\rm p}}^o(\theta + \ii \pi, \theta) = V^o(\theta). \label{axiom_normalization}
    \end{equation}
    \item cluster property
    \begin{equation}
        \lim_{\alpha \rightarrow\infty} f_{\rho_{\rm p}}^{o_1}(\theta_1, \dots, \theta_i, \theta_{i+1} - \alpha, \dots, \theta_m - \alpha) = \mathcal{N}_{\rm cl}^{o_1} f_{\rho_{\rm p}}^{o_2}(\theta_1, \dots, \theta_i) f^{o_3}(\theta_{i+1}, \dots, \theta_m), 
    \end{equation}
    \item bounds on growth
    \begin{equation}
        f_{\rho_{\rm p}}^o(\theta_1, \dots, \theta_n) \sim e^{y_o |\theta_i|}, \quad {\rm when} \quad  |\theta_i| \rightarrow \infty, 
    \end{equation}
\end{itemize}
Here ${\rho_{\rm p}}$ specifies the representative (background) state, $S(\theta)$ is the standard $2$-body scattering matrix, and $R_{\rho_{\rm p}}(\theta| \theta_1, \dots, \theta_n)$ is related to the back-flow function
\begin{align}
    R_{\rho_{\rm p}}(\theta| \theta_1, \dots, \theta_m) &= \prod_{j=1} R_{\rho_{\rm p}}(\theta|\theta_j), \\
    R_{\rho_{\rm p}}(\theta|\theta_j) &= \exp\left( 2\pi \ii F(\theta|\theta_j) - \ii \delta(\theta - \theta_j)\right).
\end{align}

In the zero-density limit, the axioms reduce to the standard vacuum form-factors axioms~\cite{SmirnovBOOK}. Specifically, in this limit $R_{\rho_{\rm p}}$ becomes one and $k(\theta)$ and $\varepsilon(\theta)$ take the standard, bare form.

The scattering and periodicity axioms combined predict a universal structure of the form factors
\begin{equation}
    f_{\rho_{\rm p}}^o(\theta_1, \dots, \theta_m) = K_{\rho_{\rm p}}^o(\theta_1, \dots, \theta_m) \prod_i A_{\rho_{\rm p}}^{\rm min}(\theta_i; \theta_1, \dots, \theta_m) \prod_{i < j} f^{\rm min}(\theta_i - \theta_j),
\end{equation}
where $f^{\rm min}(\theta)$ is the minimal vacuum form-factor and $A_{\rho_{\rm p}}^{\rm min}(\theta_i; \theta_1, \dots, \theta_n)$ is a known function 
expressed in the terms of the back-flow with its precise form given in~\cite{Bootstrap_JHEP}.
The still unknown are functions $K_{\rho_{\rm p}}^o(\theta_1, \dots, \theta_m)$. These are symmetric and $2\pi i$ periodic functions in all the rapidities. The annihilation axiom can be used to derive further recursion relations between $K_{\rho_{\rm p}}^o(\theta_1, \dots, \theta_m)$ and $K_{\rho_{\rm p}}^o(\theta_3, \dots, \theta_m)$. 

The normalization factor in~\eqref{axiom_normalization} is given by
\begin{equation}
    V^o(\theta) = \frac{1}{2\pi \rho_{\rm tot}(\theta)} \sum_{k=0}^{\infty} \int \prod_{j=1}^k \left( \frac{{\rm d}\theta_j}{2\pi} n(\theta_j) \right) f_c^o(\theta_1, ..., \theta_k, \theta), \label{V_o}
\end{equation}
where $f_c^o(\theta_1, \dots, \theta_n)$ is the connected form-factor extracted from the vacuum form-factor through the following procedure~\cite{POZSGAY2008209}. We consider the following, almost diagonal, form-factor
\begin{equation}
    f^o(\theta_n + \pi \ii, \dots, \theta_1 + \pi \ii, \theta_1 + \kappa_i, \dots, \theta_n + \kappa_n) = \frac{1}{\kappa_1 \cdots \kappa_n} \sum_{i_1=1}^n \dots \sum_{i_n=1}^n a_{i_1, \dots, i_n} \kappa_{i_1} \cdots \kappa_{i_n},
\end{equation}
where the form of the expansion on the right hand-side, for small $\kappa_i$'s follows from the annihilation axiom. Coefficients $a_{i_1, \dots, i_n}$ are finite and symmetric in the indices. The connected form-factor is then defined as the finite part of these form-factor when any of the $\kappa_i$ approaches zero
\begin{equation}
    f_c^o(\theta_1, \theta_n) = n!\, a_{1,\dots, n}.
\end{equation}
For the derivation of the normalization of the $2$-particle form-factor we refer to~\cite{Bootstrap_SciPost}. 

In the context of this review we are interested in the small particle-hole excitations limit of the form-factors. In the following we will show how the bootstrap program yields the $1$ and $2$ particle-hole form factors in such limit.

For $1$ particle-hole form-factor the result follows immediately from the normalization axiom,
\begin{equation}
    \langle \rho_{\rm p}| o(0)| \rho_{\rm p}, \theta^+, \theta^- \rangle = V^o(\theta^-) + \mathcal{O}(\theta^+ - \theta^-), \label{1ph_TBP}
\end{equation}
with $V^o(\theta)$ given above in~\eqref{V_o}. The same formula, using the hydrodynamic picture, was derived in~\cite{10.21468/SciPostPhys.5.5.054}.

The small particle-hole limit of the $2$ particle-hole form-factor is fully captured by the annihilation pole axiom together with the small excitation limit of the $1$ particle-hole form-factor. We consider $f_n^o(\theta_1 + \pi \ii + \kappa_1, \theta_1, \theta_2 + \pi \ii + \kappa_2, \theta_2)$ in the limit when $\kappa_1$ and $\kappa_2$ are small. The results depends on the order of the limits. Taking first $\kappa_1$ to zero and using the annihilation axiom, in the leading order in $\kappa_2$, we find
\begin{equation}
    f_{\rho_{\rm p}}^o(\theta_1 + \pi \ii + \kappa_1, \theta_1, \theta_2 + \pi \ii + \kappa_2, \theta_2) \sim - 2 \pi \frac{\kappa_2}{\kappa_1} T^{\rm dr}(\theta_1, \theta_2) f_{\rho_{\rm p}}^o(\theta_2 + \pi \ii, \theta_2).
\end{equation}
Working out the limits in the other order we find an analogous expression with indices $1$ and $2$ switched. One can also change the position of the particles by employing the scattering axiom. Together, this allows to write a general formula for the non-analytic part of the $4$ particles form-factor in the small excitations limit
\begin{align}
    f_{\rho_{\rm p}}^o(\theta_1, \theta_2, \theta_3 , \theta_4) =& - 2\pi \sum_{\sigma \in P_4} S_\sigma(\theta_1, \dots, \theta_4) \frac{\theta_{\sigma_3} - \theta_{\sigma_4} - \ii \pi}{\theta_{\sigma_1} - \theta_{\sigma_2} - \ii \pi} T^{\rm dr} (\theta_{\sigma_2}, \theta_{\sigma_4}) f_{\rho_{\rm p}}^o(\theta_{\sigma_3}, \theta_{\sigma_4}) \no \\
    &+ (\dots), \label{2ph_TBP}
\end{align}
where $(\dots)$ denotes terms that are finite whenever $\theta_i = \theta_j + \ii\pi$ for any pair of $i,j = 1,\dots, 4$. 

In the following section we show how these general results for one particle-hole~\eqref{1ph_TBP} and two particle-hole~\eqref{2ph_TBP} form-factors can be applied to compute form-factors of conserved densities and currents.

\subsection{Small energy and momentum limit of one and two particle-hole form factors} \label{subsec:small_ff}

In general obtaining a full expression for thermodynamic form factors, either from the thermodynamic limit of their finite-size expressions or from thermodynamic bootstrap, is complicated and today still unsolved. However there are some special limits where exact and reasonably simple results can be found. These are the regimes where the rapidity of the particle excitations are close to the ones of the holes, namely when the energy and the momentum of the full excitation is small, but still of order one in the system size. This limit is far from being uninteresting as it provides the building blocks for a hydrodynamic theory. Focusing on the latter in particular we are interested in form factors of local conserved charges $q_i(x)$ and their currents $j_i(x)$. First of all continuity equation imposes a relation between their form factors, reading (we remind the correspondence between particle-hole form factors and relativistic invariant ones in \eqref{eq:secFFCorr})
\begin{equation}
   k (\bfp, \bfh ) \langle \rho_{\rm p}| j_i  |  \rho_{\rm p} ;  \bfp, \bfh \rangle = {\varepsilon(\bfp, \bfh )}{}  \langle \rho_{\rm p}| q_i  |  \rho_{\rm p} ;  \bfp, \bfh \rangle 
\end{equation}
The above equation can be written as following, namely form factors of charges and of currents are given by the same function of particle-hole modulus their different proportionality factor 
\begin{equation}
    \langle \rho_{\rm p}| q_i  |  \rho_{\rm p} ;  \bfp, \bfh \rangle = k(\bfp, \bfh )   f_i[\bfp, \bfh ] 
\end{equation}
\begin{equation}
   \langle \rho_{\rm p}| j_i  |  \rho_{\rm p} ;  \bfp, \bfh \rangle = {\varepsilon(\bfp, \bfh )}{}   f_i[\bfp, \bfh ] 
\end{equation}
This clearly does not simplify the problem, which is now recast into the question of determining the functions $f_i[\bfp, \bfh ]$, however it is useful to factorise the proportionality to momentum or energy for later calculations within hydrodynamic theory. 
Another important form factor propriety is their symmetry under exchange of particles and hole, namely, for any local (Hermitian) operator $O$ we have 
\begin{equation}
    \langle \rho_{\rm p}| O  |  \rho_{\rm p} ;  \bfp, \bfh \rangle =   \langle \rho_{\rm p}| O  |  \rho_{\rm p} ;  \bfh, \bfp \rangle 
\end{equation}
All these properties combined, together with the susceptibility sum rule, allow to determine the small momentum and energy of the one particle-hole form factor. Alternatively, we can combine~\eqref{1ph_TBP} with the particle-hole symmetry of the density form-factor and use that $V^{q_j}(\theta) = h_j^{\rm dr}(\theta)$~\cite{10.21468/SciPostPhys.5.5.054}. The one particle-hole form-factor of the density operator, in the limit when $\theta^+ \rightarrow \theta^-$, is then
\begin{equation}
    \langle \rho_{\rm p}| q_i |  \rho_{\rm p} ;  \theta^+, \theta^- \rangle = h_i^{\rm dr}(\theta^-) + \frac{\partial_{\theta^-} h_i^{\rm dr}(\theta^-)}{2} (\theta^+ - \theta^-) +\mathcal{O}((\theta^+ - \theta^-)^2). \label{LL_1ph_limit-charges}
\end{equation}
Analogously for the current operators we have 
\begin{equation}
    \langle \rho_{\rm p}| j_i |  \rho_{\rm p} ;  \theta^+, \theta^- \rangle = v^{\rm eff}(\theta^-) h_i^{\rm dr}(\theta^-) + \frac{\partial_{\theta^-} (h_i^{\rm dr}(\theta^-) v^{\rm eff}(\theta^-)) }{2} (\theta^+ - \theta^-) +\mathcal{O}((\theta^+-\theta^-)^2). \label{LL_1ph_limit-currents}
\end{equation}
These two results are the building blocks to obtain the small energy and momentum of any higher particle-hole ones, via the kinematic pole relation. The two particle-hole form factors follow then the structure shown in~\eqref{2ph_TBP}. The density operator, in the limit when $\theta^+_i \rightarrow \theta^-_i$, is given by
\begin{align}\label{eq:FFcharges}
    \langle \rho_{\rm p}| q_i |  \rho_{\rm p} ;  \bfp, \bfh \rangle & = 2\pi  k(\bfp, \bfh) \times  
	\left(\frac{T^{\rm dr}(\theta^-_2, \theta^-_1) h_i^{\rm dr}(\theta^-_2) }{k'(\theta^-_1) k'(\theta^-_2) (\theta^+_1 - \theta^-_1)}+ \frac{T^{\rm dr}(\theta^-_1, \theta^-_2) h_i^{\rm dr}(\theta^-_1) }{k'(\theta^-_2) k'(\theta^-_1) (\theta^+_2 - \theta^-_2)} \right. \nonumber\\
	&\left. + \frac{T^{\rm dr}(\theta^-_2, \theta^-_1) h_i^{\rm dr}(\theta^-_2) }{k'(\theta^-_1) k'(\theta^-_2) (\theta^+_2 - \theta^-_1)} + \frac{T^{\rm dr}(\theta^-_1, \theta^-_2) h_i^{\rm dr}(\theta^-_1) }{k'(\theta^-_2) k'(\theta^-_1) (\theta^+_1 - \theta^-_2)} + (\dots) \right). 
\end{align}
where the notation $(\dots)$ includes terms that are order one or higher when one of the particle rapidity approach one of the holes. Analogously for the current operators we have 
\begin{align}\label{eq:FFcurrents}
    \langle \rho_{\rm p}| j_i |  \rho_{\rm p} ;  \bfp, \bfh \rangle & = 2\pi  \varepsilon(\bfp, \bfh) \times  
	\left(\frac{T^{\rm dr}(\theta^-_2, \theta^-_1) h_i^{\rm dr}(\theta^-_2) }{k'(\theta^-_1) k'(\theta^-_2) (\theta^+_1 - \theta^-_1)}+ \frac{T^{\rm dr}(\theta^-_1, \theta^-_2) h_i^{\rm dr}(\theta^-_1) }{k'(\theta^-_2) k'(\theta^-_1) (\theta^+_2 - \theta^-_2)} \right. \nonumber\\
	&\left. + \frac{T^{\rm dr}(\theta^-_2, \theta^-_1) h_i^{\rm dr}(\theta^-_2) }{k'(\theta^-_1) k'(\theta^-_2) (\theta^+_2 - \theta^-_1)} + \frac{T^{\rm dr}(\theta^-_1, \theta^-_2) h_i^{\rm dr}(\theta^-_1) }{k'(\theta^-_2) k'(\theta^-_1) (\theta^+_1 - \theta^-_2)} + (\dots) \right). 
\end{align}
These results can be generalised to integrable theories with more than one particle species, as spin chains and nested systems. In such cases quasiparticles are labelled by their rapidities $\theta$ and a set of integer numbers $s_1,s_2,\ldots \equiv s$, where $s$ is a collective discrete index. Thermodynamic quantities as such as scattering shift and dressed functions acquires then an extra dependence on the quasiparticle labels
\begin{align}\label{eq:multi}
     & T(\theta,\alpha) \to T_{s,s'}(\theta,\alpha) \nonumber \\
   &   n(\theta) \to n_s(\theta)
     \nonumber \\
    & h(\theta) \to h_{s}(\theta)
\end{align}
where $h(\theta)$ is a generic single particle eigenvalue.  Clearly there are now two different options, either one particle and one hole of the same species share the same rapidity or particle and holes of different species approach the same value. The latter is not expected to generate a kinematic pole in the form factors (although there may be cases where this is not true, which so far have not been studied in detail) while the first gives kinematic poles as in \eqref{eq:FFcurrents}, \eqref{eq:FFcharges}, with the extension \eqref{eq:multi}.

\section{Euler scale}\label{sec:euler}

In this section, we describe the exact results for correlation functions available in integrable systems at the Euler scale. These results can be obtained by two different sets of techniques: those coming from the hydrodynamic perspective, as described in sec. \ref{sec:hydroCorr}, and those coming from the microscopic perspective and form factor expansions, as described in sec. \ref{sec:corrFF}. One of the beautiful discoveries in the development of generalised hydrodynamics is how these two perspectives connect, both in the actual formulae that emerge, and in the ideas behind these formulae. This makes the equivalence of the hydrodynamic and kinetic viewpoints in integrable systems quite apparent.

\begin{figure}
    \centering
    \includegraphics[width=5cm]{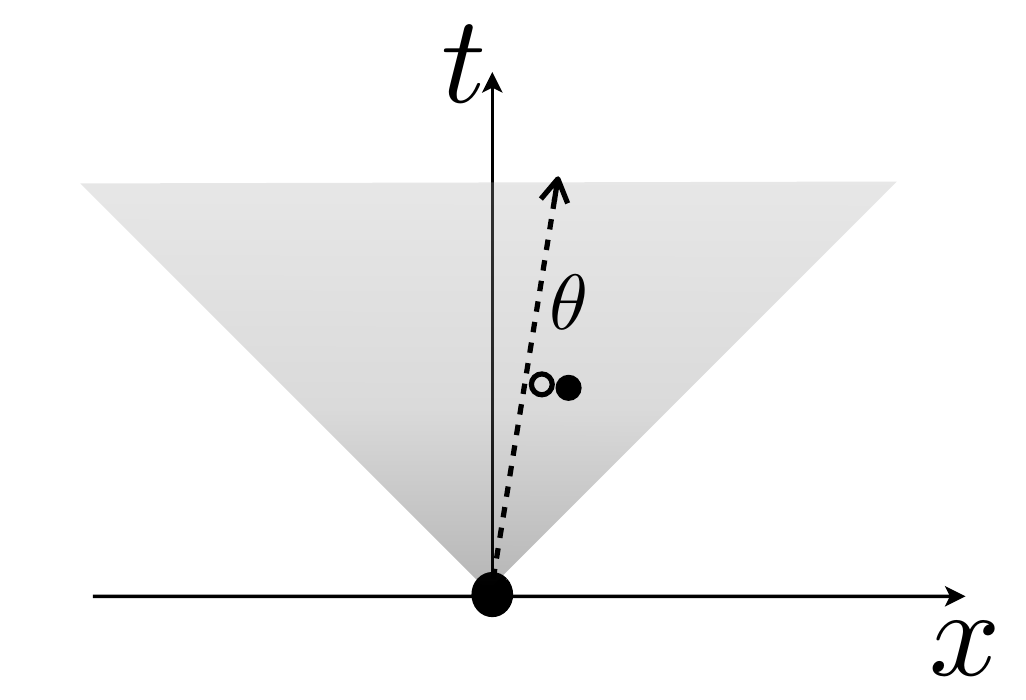}
    \caption{The decay of correlation functions in integrable models is  dominated by Euler spreading if the observable overlaps with particle-hole pairs. Co-propagating particle-hole pairs form a continuum of ballistic modes.}
    \label{fig:particlehole}
\end{figure}

In fact, one of the main ideas behind GHD is that the asymptotic states of an integrable many-body system -- technically described by the Bethe roots in the Bethe ansatz formulation -- morph, within finite-density states, into the emergent hydrodynamic modes -- technically implemented in terms of the finite-density form factors. Results for Euler-scale correlation function make this correspondence rather explicit. In particular, reviewing the thermodynamic Bethe ansatz structure of integrable systems in sec. \ref{subsec:TBA_hydro}, one finds, as explained in sec. \ref{subsec:spaceballistic}, that the space $\mathcal H_{\rm bal}$ of conserved quantities, the ballistic waves that propagate correlations between local observables at large scales, has the form of a weighted $L^2$ space. Although a full mathematical characterisation is not completely understood, its physical meaning is clarified by the agreement between the results of sec. \ref{subsec:drudedynamical}, which applies the hydrodynamic principles to integrable models, and \ref{subsec:ffEuler}, which uses the finite-density form factor description. These suggest that in integrable systems, $\mathcal H_{\rm bal}$ can generically be described, in the sense of spectral theory, in terms of un-normalisable scattering states, formed of co-propagating particle-hole pairs of all available momenta (corresponding to all available rapidities), travelling within the maximal entropy state with respect to which $\mathcal H_{\rm bal}$ is defined. It is these particle-hole pairs that carry correlations at large scales. They form a continuum of ballistic modes, and thus the decay is generically seen to be in $1/t$, see fig.~\ref{fig:particlehole}. In certain models, in addition to this continuum, there may be additional isolated modes, and diffusive decay is observed for observables that overlap only with such modes (an example is the XXZ quantum chain).

It is important to emphasise that the hydrodynamic approach is, currently, {\em more general} than the form factor analysis of sec. \ref{sec:corrFF}. Because of the use of general hydrodynamic principles, expected to enjoy a very high universality, all results are expected to hold not only in Bethe-ansatz quantum integrable models, but also in classical integrable field theories and chains, in classical soliton gases, and in cellular automata. Nevertheless, the interpretation in terms of particle-hole pairs is still expected to make sense beyond Bethe-ansatz quantum integrable models.

\subsection{Thermodynamics of integrable models and hydrodynamic matrices} \label{subsec:TBA_hydro}

The starting point of the hydrodynamic approach are the specific free energy $\mathsf f$ and free energy flux $\mathsf g$, defined by \eqref{eq:fg}. Physically, these are the generating functions for the extensive part of the total conserved quantities and currents and their static fluctuations at large system size. This means that once $\mathsf f,\,\mathsf g$ are known in arbitrary maximal entropy states, the average conserved densities and currents are known, and hence the $\mathsf B$, $\mathsf C$ and $\mathsf A$ matrices, \eqref{eq:Bmatrix}, \eqref{eq:Cmatrix}, \eqref{eq:Amatrix}, which are the building blocks for all hydrodynamic correlations at the Euler scale. The $\mathsf A$ matrix for integrable models was first found in \cite{SciPostPhys.3.6.039}, and all of the material presented here has been reviewed in \cite[sec 3.3 - 3.5]{10.21468/SciPostPhysLectNotes.18}.

In integrable models, expressions for $\mathsf f$ and $\mathsf g$ are known, at least as conjectures, in generality, for quantum and classical models of many types, and derived from the microscopic model in a number of cases using the thermodynamic Bethe ansatz (TBA) or other methods. We will refer to the resulting structure as the TBA. The specific free energy $\mathsf f$ has been widely discussed \cite{Yang1969,Takahashi1999,Mossel_2012}, and the free energy flux $\mathsf g$ was first obtained in \cite{PhysRevX.6.041065}, see the discussion in \cite{10.21468/SciPostPhysLectNotes.18}. They can be expressed in terms of the notion of quasiparticles introduced in sec. \ref{sec:corrFF}. Assuming that the model admits a single quasiparticle type (such as the Lieb-Liniger model), they take the form
\begin{equation}\label{eq:fgintegrable}
    \mathsf f = \int \dd \theta\,p'(\theta) \mathsf F(\epsilon(\theta)),\quad
    \mathsf g = \int \dd \theta\,e'(\theta) \mathsf F(\epsilon(\theta))
\end{equation}
where we recall from sec. \ref{sec:corrFF} that $p(\theta)$ and $e(\theta)$ are the bare momentum and energy of a quasiparticle with rapidity  $\theta$. This form is in fact very universal in integrable systems, and not just valid in the quantum integrable context of sec. \ref{sec:corrFF}. In general, a quasiparticle is an asymptotic excitation of the many-body model in the sense of scattering theory (for instance, a soliton in classical integrable field theory), and its bare momentum and energy are its asymptotic momentum and energy.

In \eqref{eq:fgintegrable}, the function $\mathsf F(\epsilon)$ characterises the {\em statistics} of the quasiparticles, and is interpreted as the contribution to the free energy due to a single quasiparticle of ``pseudoenergy" $\epsilon$. In quantum integrable models it often takes the fermionic form $\mathsf F(\epsilon) = -\log(1+e^{-\epsilon})$, and more generally:
\begin{equation}\label{eq:qpStatistics}
    \mathsf F(\epsilon) =
\begin{cases}
-\log(1+e^{-\epsilon}), \ \text{ fermions}\\
 \log(1-e^{-\epsilon}), \  \text{  bosons}\\
 -e^{\epsilon}, \ \text{ classical particles}\\
 \log\epsilon, \ \text{ radiative   modes.}\\
\end{cases}
\end{equation}
See \cite{10.21468/SciPostPhys.5.5.054,10.21468/SciPostPhysLectNotes.18}. The pseudoenergy $\epsilon(\theta)$ encodes the interaction, and satisfies
\begin{equation}\label{eq:pseudoenergy}
    \epsilon(\theta) = w(\theta)
    + \int \dd \alpha\, \varphi(\alpha-\theta) \mathsf F(\epsilon(\alpha))
\end{equation}
where $\varphi(\theta) = -\ii \,\dd \log S(\theta)/\dd \theta$ is the differential scattering phase\footnote{Here we assume that it depends on the difference of rapidities, and below we will assume that it is a symmetric function. See the discussion in \cite{10.21468/SciPostPhys.5.5.054,De_Nardis_2019} concerning how these properties often occur and how they are related to the choice of parametrisation of the physical bare momentum $p(\theta)$.}. As described in sec. \ref{sec:corrFF}, the differential scattering phase describes two-body scattering. Here it is expressed using the quantum notion of the two-body S-matrix $S(\theta)$, but it also has a clear meaning in the classical context, in agreement with a semi-classical interpretation of the quantum object. Indeed, it is (simply related to) the space shift incurred by the quasiparticles upon two-body scattering, see \cite{PhysRevLett.95.204101,PhysRevLett.120.045301,10.21468/SciPostPhysLectNotes.18} and especially the reviews \cite{TakatoReview,GennadyEl} in this special issue. Finally, in Eq.~\eqref{eq:pseudoenergy}, the function $w(\theta)$ characterises the maximal entropy state. If the state is written in the GGE form \eqref{eq:gge}, then (recall that summation is implied)
\begin{equation}\label{eq:wgge}
    w(\theta) = \beta^ih_i(\theta).
\end{equation}
More generally, the manifold of maximal entropy states can be seen as a certain space of functions $w(\theta)$, which may or may not be expressible as a convergent series in any given basis of functions $h_i(\theta)$. It is not known, in general, how to fully specify the precise space of functions $w(\theta)$ which result in clustering states.

The interpretation of the expression for $\mathsf f$ in \eqref{eq:fgintegrable}, with \eqref{eq:pseudoenergy}, is simple. It is based on the geometric interpretation of quasiparticle scattering first proposed in \cite{Doyon2018}. In this geometric interpretation, a quasiparticle of rapidity $\theta$ adds, as perceived by a quasiparticle of rapidity $\alpha$, a new region of effective physical space around which the $\theta$ quasiparticle lies, of length $\varphi(\alpha-\theta)/p'(\alpha)$. Thus, upon inserting, in a gas, a quasiparticle of rapidity $\theta$, the free energy is not only modified by the extra Boltzmann weight $e^{-w(\theta)}$, but also by the modification of the partition function of the rest of the gas due to the perceived change of system size by each quasiparticle in the gas, $e^{-\int \dd \alpha\, \varphi(\alpha-\theta)\mathsf F(\epsilon(\alpha))}$. This gives the self-consistent condition \eqref{eq:pseudoenergy} for the combined Boltzmann weight $e^{-\epsilon(\theta)}$, which itself gives a contribution $\dd p(\theta) \mathsf F(\epsilon(\theta))$ to the free energy $\mathsf f$.

With $\mathsf f,\,\mathsf g$ determined in an arbitrary maximal entropy state characterised by $w(\theta)=\sum_i\beta^i h_i(\theta)$, the hydrodynamic matrices are obtained by simple differentiation,
\begin{equation}
    \mathsf C_{ij} = \frac{\partial^2 \mathsf f}{\partial \beta^i\partial\beta^j},\quad
    \mathsf B_{ij} = \frac{\partial^2 \mathsf g}{\partial \beta^i\partial\beta^j}.
\end{equation}
Using in particular $\partial \epsilon / \partial \beta^i = h_i^{\rm dr}$, the results can be expressed in terms of the dressing operation \eqref{eq:dress} and the effective velocity \eqref{eq:effective-velo}:
\begin{equation}\label{eq:BCintegrals}
    \mathsf C_{ij} = \int \dd \theta\,\rho_{\rm p}(\theta) f(\theta)
    h_i^{\rm dr}(\theta)h_j^{\rm dr}(\theta),\quad
    \mathsf B_{ij} = \int \dd \theta\,\rho_{\rm p}(\theta) f(\theta)
    v^{\rm eff}(\theta) h_i^{\rm dr}(\theta)h_j^{\rm dr}(\theta)
\end{equation}
where
\begin{equation}
    f(\theta)
    = -\frac{\mathsf F''(\epsilon(\theta))}{\mathsf F'(\epsilon(\theta))}
\end{equation}
depends on the quasiparticle statistics. It is convenient to express it in terms of the distribution function
\begin{equation}
    n(\theta) = \mathsf F'(\epsilon(\theta))
\end{equation}
which gives, depending on the statistics of the quasiparticles 
\begin{equation}\label{eq:qpStatistics2}
    f(\theta) =
\begin{cases}
1- n(\theta), \ \text{ fermions}\\
 1+ n(\theta), \  \text{  bosons}\\
 1, \ \text{ classical particles}\\
 n(\theta), \ \text{ radiative   modes.}\\
\end{cases}
\end{equation}
In \eqref{eq:BCintegrals}, the quantity $\rho_{\rm p}(\theta)$ is the quasiparticle spectral density introduced in sec. \ref{sec:corrFF}. From its basic definition in terms of stationary averages ${\tt q}_i = \int \dd \theta \,\rho_{\rm p}(\theta)h_i(\theta)$ along with \eqref{eq:wgge}, one finds the standard relations $\rho_{\rm p}(\theta) = (p')^{\rm dr}(\theta) n(\theta) / (2\pi) = \rho_{\rm tot}(\theta) n(\theta)$.

Expressions \eqref{eq:BCintegrals} are most conveniently expressed in terms of integral operator kernels, $\mathsf C_{ij} = h_i \cdot \mathsf C h_j$, $\mathsf B_{ij} = h_i \cdot  \mathsf B h_j$ (with $h_1\cdot h_2 = \int \dd \theta\, h_1(\theta)h_2(\theta)$) giving
\begin{equation}\label{eq:BCkernel}
     \mathsf C = (1-nT)^{-1} \rho_{\rm p}f (1-Tn)^{-1},\quad
     \mathsf B = (1-nT)^{-1} v^{\rm eff} \rho_{\rm p}f (1-Tn)^{-1}.
\end{equation}
Finally, using the relation $\mathsf B = \mathsf A\mathsf C$ interpreted in terms of integral kernels,
\begin{equation}\label{eq:Akernel}
     \mathsf A = (1-nT)^{-1} v^{\rm eff}(1-nT)
\end{equation}
(see e.g.~\cite[sec 3.5]{10.21468/SciPostPhysLectNotes.18}). We note also that the expression \eqref{eq:entropy} for the entropy density agrees with the usual Yang-Yang entropy (generalised to arbitrary statistics) \cite{Yang1969}.

\subsection{Drude weights and dynamical structure factors from hydrodynamics} \label{subsec:drudedynamical}

Two main quantities are here of interest at the Euler scale of hydrodynamics: the Drude weights \eqref{drudeKMB}, which are fundamental in characterising ballistic transport, and the dynamical structure factor \eqref{eq:defS}, or more generally \eqref{eq:defSoo}, which describe correlations. For these objects, the results that lend themselves directly to an implementation in integrable systems are the exponential solution \eqref{eq:Ssolution} to the Euler-scale dynamical structure factor, and the projection results of sec. \ref{subsec:projectionsEuler}: formulae  \eqref{drudeAsquareC} (coming from \eqref{eq:drudeexpansion}) for the Drude weights, and the result \eqref{eq:Seulerfull} of the Boltzmann-Gibbs principle.

The general formula \eqref{drudeAsquareC} for the Drude weights can immediately be implemented using the operator-kernel form of the $\mathsf A$ matrix, \eqref{eq:Akernel}, and of the $\mathsf C$ matrix, \eqref{eq:BCkernel}. This gives
\begin{equation}\label{eq:Drude-Projections}
    \mathsf D_{ij} = h_i\cdot (1-nT)^{-1}\big(v^{\rm eff}\big)^2\rho_{\rm p}f h_j^{\rm dr}
    = \int \dd\theta \,\rho_{\rm p}(\theta) f(\theta) v^{\rm eff}(\theta)^2h_i^{\rm dr}(\theta)h_j^{\rm dr}(\theta).
\end{equation}
This is one of the most important formulae in the study of ballistic transport in integrable systems. The fully general expression \eqref{eq:Drude-Projections} was first obtained, by hydrodynamic projections in GHD, in \cite{SciPostPhys.3.6.039}. In fact, the formula  \eqref{eq:Drude-Projections} was obtained earlier in the special cases of spin, charge and energy diagonal Drude weights in various quantum integrable models, by using different techniques \cite{Fujimoto_1998,PhysRevLett.82.1764,Klumper_2002,Sakai_2003}. The first time the more general Drude weight was considered was using GHD in \cite{IN_Drude,Bulchandani2018}, where it was studied by linear response on the partitioning protocol. The fact that this linear response formulation does indeed lead to \eqref{eq:Drude-Projections} was shown in \cite{SciPostPhys.3.6.039}. Formula \eqref{eq:Drude-Projections} supersedes the bounds from Mazur inequalities combined with appropriate choices of quasi-local charges, emphasising how the TBA encodes the complete space of conserved charges (see the review \cite{ilievski2016quasilocal}).

As the formula \eqref{eq:Drude-Projections} is derived from general hydrodynamic projection principles along with the general forms of the specific free energy and the free energy flux, it is expected to hold in a large family of quantum and classical integrable models, including for instance classical particle systems and solition gases. Note in particular that the derivation of this formula {\em does not require any hydrodynamic approximation}: it is based on mathematically rigorous projection principles. The most important non-rigorous step is the assumption on the {\em space of emergent ballistic degrees of freedom}: the Hilbert space $\mathcal H_{\rm bal}$ is assumed to be an appropriate space of functions of the rapidity, allowing for the integral-kernel formulae $\mathsf B = \mathsf{AC}$ to hold.

The general formula \eqref{eq:Seulerfull} for the dynamical structure factors also has a direct implementation in integrable systems using integral kernels:
\begin{equation}
    \lim_{\rm Eul} S_{ij}(\kappa) = h_i\cdot (1-nT)^{-1}e^{\ii v^{\rm eff} \kappa}\rho_{\rm p}f h_j^{\rm dr}
    = \int \dd\theta \,\rho_{\rm p}(\theta) f(\theta) e^{\ii v^{\rm eff}(\theta) \kappa}h_i^{\rm dr}(\theta)h_j^{\rm dr}(\theta)
\end{equation}
or, in space-time, using \eqref{eq:boltzmanngibbsxt} and performing the rapidity integral using the delta-function,
\begin{equation}
    \lim_{\rm Eul} S_{ij}(x,t) 
    = t^{-1}\sum_{\theta\in\theta_*(\xi)} \frac{\rho_{\rm p}(\theta) f(\theta)}{|(v^{\rm eff})'(\theta)|} h_i^{\rm dr}(\theta)h_j^{\rm dr}(\theta).
\end{equation}
Here we introduced the set $\theta_*(\xi) = \{\theta:v^{\rm eff}(\theta) = \xi\}$ of rapidities for which the effective velocity corresponds to the space-time ray $v$ of the Euler scaling limit.

Finally, for large-scale correlation functions of generic observables we apply the TBA to Eq.~\eqref{eq:Seulerfull}. In order to do this, we need the projections of the observable onto conserved quantities, which can be expressed as derivatives:
\begin{equation}
    \aveI{o,q_i} = -\frac{\partial \ave{o}}{\partial \beta^i}.
\end{equation}
This requires the knowledge of the average $\ave{o}$ in maximal entropy states. This is in general a difficult object to evaluate exactly, and techniques beyond the TBA are required. Of particular importance, the Leclair-Mussardo formula  \cite{Leclair1996, Saleur:1999hq, Pozsgay:2010xd} gives averages in generalised Gibbs ensembles as an infinite, but relatively well convergent, form factor series. Other results for such one-point averages exist, see for instance \cite{Delfino2001,Kitanine2002,LeClair1999,Kormos2010,Pozsgay:2010xd}. Once the average $\ave{o}$ is known, then it is always possible to write the result of the derivative in a form that mimics \eqref{eq:BCintegrals}:
\begin{equation}\label{eq:Vohydro}
    -\frac{\partial \ave{o}}{\partial \beta^i} = \int \dd\theta\, \rho_{\rm p}(\theta) f(\theta) h_i^{\rm dr}(\theta) V^o(\theta).
\end{equation}
The function $V^o(\theta)$ characterises the observable $o$. In particular, performing the $\beta^i$ derivative on the Leclair-Mussardo formula gives exactly formula \eqref{V_o}, as shown in \cite{10.21468/SciPostPhys.5.5.054}. Thus $V^o(\theta)$ is identified with the equal-rapidity one-particle-hole form factor of $o$ introduced in sec. \ref{subsec:bootstrap} (with the same notation). It is important to emphasise, however, the definition \eqref{eq:Vohydro} is independent from any Bethe ansatz structure of the underlying model, and holds in the full generality of the TBA and GHD formalisms\footnote{In fact, it can also be argued that the Leclair-Mussardo formula enjoys the same level of generality.}.

In integral-kernel form we have
\begin{equation}
    \aveI{o,q_i} = V^o \cdot \rho_{\rm p} f (1-Tn)^{-1} h_i
\end{equation}
and this gives
\begin{eqnarray}\label{eq:So1o2integrable}
    \lim_{\rm Eul} S_{o_1,o_2}(\kappa) &=& \int \dd\theta \,\rho_{\rm p}(\theta) f(\theta) e^{\ii v^{\rm eff}(\theta) \kappa}V^{o_1}(\theta)V^{o_2}(\theta)\\
    \lim_{\rm Eul} S_{o_1,o_2}(x,t) 
    &=&  t^{-1}\sum_{\theta\in\theta_*(\xi)} \frac{\rho_{\rm p}(\theta) f(\theta)}{|(v^{\rm eff})'(\theta)|} V^{o_1}(\theta)V^{o_2}(\theta).\nonumber
\end{eqnarray}
In the expression for $\lim_{\rm Eul} S_{o_1,o_2}(x,t)$, it is possible that $\xi$ be in the direction of a supremum of the effective velocity, where $(v^{\rm eff})'(\theta)=0$. This may happen, for instance, in lattice models, where the effective velocity is bounded. In this case, the Euler scaling limit gives infinity. This also happens in conventional hydrodynamics, where \eqref{eq:Ssolutionxt} indicates that it is a distribution supported on the available hydrodynamic velocities. This means that the correlation function itself does not decay as $1/t$ at large times $t$, but rather with a slower power law. In the present case, a form factor analysis suggests $t^{-2/3}$, see next the sec. \ref{subsec:ffEuler}.

The results \eqref{eq:So1o2integrable} were first obtained in \cite{10.21468/SciPostPhys.5.5.054}, by the techniques explained here. They are extremely general asymptotic results for correlation functions in integrable models, perhaps surprising given the technical complexity of the objects being studied. Interestingly, they can also be obtained using the new form-factor theory reviewed in sec. \ref{sec:corrFF}, as we explain next.

\subsection{Form factor expansion at the Euler scale} \label{subsec:ffEuler}

We will now show how one can obtain the GHD predictions for the dynamic correlation functions at the Euler scale from the form-factor expansion, including only \textit{one} particle-hole excitations. This provides a verification for two non-trivial sets of ideas: the form-factor theory reviewed in sec. \ref{sec:corrFF}, and the hydrodynamic ideas, at least at the level of linearised hydrodynamics, reviewed in sec. \ref{sec:hydro}. In particular, the latter are extremely difficult to independently verify in generic systems. To address the Euler scale we need to perform fluid-cell averaging introduced in sec.~\ref{subsec:einstein}, and here for definiteness we consider Eq.~\eqref{eq:Seulerscale}.

We consider the connected two-point function in its spectral representation
\begin{equation}
    S^{o_1, o_2} (\xi, t) = \langle \rho_{\rm p}|o_1(\xi t, t) o_2(0,0)|\rho_{\rm p} \rangle - \langle \rho_{\rm p}|o_1|\rho_{\rm p} \rangle \langle \rho_{\rm p}|o_2|\rho_{\rm p} \rangle,
\end{equation}
and are interested in its large-time limit. In this limit the leading contributions to the correlation function comes from intermediate states involving small number of particle and holes excitations. The leading terms involve single particle or hole excitations. Such terms, involving unequal number of particles and holes, as we will show later, vanish under the fluid-cell averaging. The relevant contributions come then from particle-hole excited states, with single particle-hole excitations determining the leading contribution in $1/t$. Such contribution is
\begin{align}\label{eq:ffsaddle}
    S_{\rm 1ph}^{o_1, o_2} (\xi, t) &= \int \frac{\dd\theta^+}{2\pi}\frac{\dd\theta^-}{2\pi} n(\theta^+)(1-n(\theta^-)) f_n^{o_1}(\theta^+, \theta^- + \ii \pi) f_n^{o_2}(\theta^+, \theta^- + \ii \pi)^* \nonumber \\
    &\times \exp \left[ \ii t \left(\xi(k(\theta^+) - k(\theta^-)) - (\varepsilon(\theta^+) - \varepsilon(\theta^-)) \right)\right],
\end{align}
and in the large $t$ limit can be evaluated using a two-dimensional stationary phase approximation. The stationary points are $\vec{\theta}_0 = (\theta_*(\xi), \theta_*(\xi))$ where $\theta_*(\xi)$ obey $v^{\rm eff}(\theta_*(\xi)) = \xi$. This results in
\begin{equation}
    S_{\rm 1ph}^{o_1, o_2} (\xi, t) = t^{-1} \sum_{\theta\in\{\theta_*(\xi)\}} \frac{\rho_{\rm p}(\theta)(1 - n(\theta))}{|(v^{\rm eff})'(\theta)|} V^{o_1}(\theta) V^{o_2}(\theta).
\end{equation}
 Crucially, these expressions are smooth in $t$ and $x$ and therefore they are invariant under the fluid-cell averaging. In the following we show that this is not the case for contributions coming from excited states with different number of particles and holes.

The contribution from one particle excited state instead average to zero at large times. These read
\begin{equation}
    S_{1p}^{o_1, o_2}(\xi, t) = \int \frac{{\rm d}\theta^+}{2\pi} (1 - n(\theta)) f_n^{o_1}(\theta) f_n^{o_2}(\theta)^* e^{\ii t \left[ \xi k(\theta) - \varepsilon(\theta)\right]}.
\end{equation}
At late times we can again use the saddle point approximation which this time yields contribution oscillatory in time,
\begin{equation}
    S_{\rm 1p}^{o_1, o_2}(\xi, t) = \sum_{\theta\in\{\theta_*(\xi)\}} (1 - n(\theta)) \sqrt{\frac{1}{\rho_s(\theta) (v^{\rm eff})'(\theta)}} f_n^{o_1}(\theta) f_n^{o_2}(\theta)^* e^{\ii\left[\xi k(\theta) - \varepsilon(\theta) \right] \pm \ii \pi/4}.
\end{equation}
The oscillations are centered around zero and make the whole contribution vanish under the fluid-cell averaging.

The saddle point in \eqref{eq:ffsaddle} also allows to inspect the correlator exactly at the light-cone, namely whenever $(v^{\rm eff})'(\theta_*)=0$. This is what happens in lattice system where there is a finite maximal velocity $v^{\rm eff}(\theta_*)=v_{\rm max}$. In this case we have, at large times 
\begin{align} 
    & S_{\rm 1ph}^{o_1, o_2} (x = v_{\rm max} t + \zeta, t) =  {\rho_{\rm p}(\theta_*)(1 - n(\theta_*)}{} V^{o_1}(\theta_*) V^{o_2}(\theta_*) \int \frac{\dd\theta^+}{2\pi}\frac{\dd\theta^-}{2\pi}  \nonumber \\
    &\times e^{ -\ii t (k'(\theta_*)(v^{\rm eff})''(\theta_*))/6(\theta^+)^3 - (\theta^-)^3) )} e^{\ii\zeta k'(\theta_*) ((\theta^+) -  (\theta^-) )}.
\end{align}
Using the integral form of the Airy function we obtain the following scaling form for the correlation functions at the light-cone
\begin{align} \label{eq:decayairy}
     S_{\rm 1ph}^{o_1, o_2} & (x = v_{\rm max} t + \zeta t^{1/3}, t ) =  t^{-2/3} \frac{2^{2/3}V^{o_1}(\theta_*) V^{o_2}(\theta_*) n(\theta_*)(1 - n(\theta_*))}{[ (k'(\theta_*)(v^{\rm eff})''(\theta_*))]^{2/3}}  \nonumber \\& \times \left( {\rm Ai}\Big( \frac{2^{1/3}\zeta}{ [ (k'(\theta_*)(v^{\rm eff})''(\theta_*))]^{1/3}} \Big) \right)^2 .
\end{align}
This form of the correlator at the light-cone is expected to be the same in free theories, where only one particle-hole are presents. However in interacting system diffusive spreading coming from two particle-hole contribution to the correlations, gives a slower decay term, as $1/\sqrt{t}$, which is leading compared to the contribution \eqref{eq:decayairy}.

Let us now look at the Drude weights \eqref{drudeKMB}
\begin{eqnarray} 
    \mathsf D_{ij}&=&\lim_{t\to\infty}\frac{1}{2t}\int_{-t}^t \dd s \int \dd\lambda \int_{-\infty}^\infty\dd x\, \langle j_{i}(x,\boldsymbol{e} s - i \boldsymbol{\beta} \lambda ),j_j(0,0)\rangle   
\end{eqnarray}
Inserting the particle-hole resolution of identity and including only one particle-hole excitations we obtain 
\begin{align} 
    \mathsf D_{ij} & = (2 \pi)\int_0^1 \dd\lambda \lim_{t \to \infty} \frac{1}{2 t}\int_{-t}^{t} ds  \int \frac{\dd\theta^+}{2\pi}\frac{\dd\theta^-}{2\pi} \delta(k(\theta^+) - k(\theta^-))  n(\theta^+)(1-n(\theta^-)) \nonumber \\
    &\times f_n^{j_{i}}(\theta^+, \theta^- + \ii \pi) f_n^{j_{j}}(\theta^+, \theta^- + \ii \pi)^*  \exp \left[ - \lambda \beta^i (h^{\rm dr}_i(\theta^+) - h^{\rm dr}_i(\theta^-)) - \ii s (\varepsilon(\theta^+) - \varepsilon(\theta^-)) \right],  
\end{align}
where the $\delta(k(\theta^+) - k(\theta^-)) $ is generated by the integration over full space $x$. Integrating the latter gives $\theta^+ = \theta^-$, and a trivial integral over time and $\lambda$. We then obtain, using \eqref{LL_1ph_limit-currents},  
\begin{align} 
    \mathsf D_{ij} & =  \int \dd\theta\, \rho_{\rm p}(\theta) (1-n(\theta)) [v^{\rm eff}(\theta)]^2 h^{\rm eff}_i(\theta) h^{\rm eff}_j(\theta),
\end{align}
recovering this way the result obtained by hydrodynamic projection \eqref{eq:Drude-Projections}.

\subsection{The space of ballistic waves}\label{subsec:spaceballistic}

We are now in a position to discuss slightly more explicitly the structure of the Hilbert space of conserved quantities, or ballistic waves, $\mathcal H_{\rm bal}$ introduced in sec. \ref{subsec:projectionsEuler}, and its physical interpretation.

First, it is clear that every conserved quantity $Q_i\in\mathcal H_{\rm bal}$ is, in the TBA, identified with a function $h_i(\theta)$ as per \eqref{eq:wgge}, and the inner product is exactly $\mathsf C_{ij}$, see sec. \ref{subsec:TBA_hydro}. In fact, in the context of describing the Hilbert space $\mathcal H_{\rm bal}$, it is not necessary to discuss the functions $h_i(\theta)$ themselves, as only the dressed versions are involved, $h_i^{\rm dr}(\theta)$. Therefore, the space $\mathcal H_{\rm bal}$, for a maximal entropy state described by $w(\theta)$ as in sec. \ref{subsec:TBA_hydro} (equivalently $\epsilon(\theta)$, $\rho_{\rm p}(\theta)$ or $n(\theta)$), is the space of all functions $\tilde h$ with finite norm under the induced inner product\footnote{In order to have the Hilbert space structure, we take complex functions, but we expect a basis to lie in the real subspace. As usual, we also ask for Lebesgue-integrability of the integrand on $\RaR$, and the space is in fact a space of equivalence classes under modifications on sets of measure zero.}:
\begin{equation}\label{eq:QL2}
    \mathcal H_{\rm bal} \cong L^2(\RaR,\dd\theta\,\rho_{\rm p} f) = \Big\{\tilde h:\RaR\to\CC \;\Big|\;
    \int \dd\theta\, \rho_{\rm p}(\theta) f(\theta) |\tilde h(\theta)|^2 <\infty \Big\}.
\end{equation}
Note that it is expected that $\rho_p(\theta) f(\theta)\geq 0$ for maximal entropy states, hence this makes sense. The dressing is only involved in relating the one-particle charge $h(\theta)$  (for instance, the energy $h(\theta) = m\cosh\theta$ in a relativistic model) to the particular element $\tilde h = h^{\rm dr}$ of $\mathcal H_{\rm bal}$ as parametrised by functions in $L^2(\RaR,\dd\theta\,\rho_{\rm p} f)$. That is, $L^2(\RaR,\dd\theta\,\rho_{\rm p} f)\ni h^{\rm dr} \equiv Q\in \mathcal H_{\rm bal}$ if $h$ is the one-particle eigenvalue of the conserved charge $Q$.

Clearly, from the above, if $h_i\equiv Q_i$ then the projection of the total current $J_i = \int \dd x\, j_i(x)$ onto $\mathcal H_{\rm bal}$ is
\begin{equation}
    L^2(\RaR,\dd\theta\,\rho_{\rm p} f) \ni v^{\rm eff} h_i^{\rm dr} \equiv \mathbb PJ_i.
\end{equation}
Note that even if $h_i^{\rm dr}\in L^2(\RaR,\dd\theta\,\rho_{\rm p} f)$, there is not guarantee that $v^{\rm eff} h_i^{\rm dr}\in L^2(\RaR,\dd\theta\,\rho_{\rm p} f)$: there may be conserved charges whose currents do not lie in the Hilbert space $\mathcal H'$. However, it is expected that local conserved charges have well-behaved currents, see the theorem about the existence of currents in \cite{hydroprojectionsEuler}.

Further, as can be deduced from the results of sec. \ref{subsec:drudedynamical}, the equal-momenta one-particle-hole form factors $V^o(\theta)$ introduced in \eqref{V_o} represent the projection of the extensive observable $O$ onto $\mathcal H_{\rm bal}$,
\begin{equation}
    L^2(\RaR,\dd\theta\,\rho_{\rm p} f) \ni V^o \equiv \mathbb PO.
\end{equation}
From this, the form factor analysis of sec. \ref{subsec:ffEuler} suggests a clear physical interpretation of the space $\mathcal H_{\rm bal}$ of ballistic modes in integrable models: the modes should be identified with excitations by particle-hole pairs on top of generalised Gibbs ensembles. Particle-hole pairs naturally represent small disturbances of the state, with a small cost in terms of all conserved quantities admitted by the model. In this interpretation, instead of characterising linear waves as small propagating disturbances of conserved densities, linear waves are seen as co-propagating pairs of a particle and a hole, which can carry any rapidity $\theta\in\RaR$. Connecting with the $L^2$ structure, one may see these travelling particle-hole pairs as forming a scattering-wave ``basis", in the sense of spectral theory, for the space of ballistic waves (or conserved quantities) $\mathcal H_{\rm bal}$, much like asymptotic states of fixed momenta $p$ represent a scattering basis for square-integrable functions in quantum mechanics. Connecting with the geometric picture of maximal entropy states reviewed in sec. \ref{sec:hydro}, this implies that co-propagating particle-hole pairs in integrable systems span the tangent space to the manifold of maximal entropy states.

\subsection{Dynamical correlation functions in non-stationary backgrounds}\label{subsec:nonstatintegrable}

The discussion of linearised hydrodynamics in sec. \ref{subsec:einstein} and \ref{subsec:projectionsEuler}, and of correlation functions in sec. \ref{sec:corrFF}, is mostly based on the assumption that the background state is stationary and homogeneous. Evaluating correlation functions on top of non-stationary backgrounds -- with large-scale fluid motion -- is an extremely difficult problem. There is no general method known in the universal formulation of hydrodynamics, except for the linear-response dynamical equation \eqref{eq:twopointfunctionnonstat} which is in general hard to solve. However, surprisingly, it is possible to go much further in integrable systems, and to obtain a set of integral equations that describes dynamical two-point functions at the Euler scale, on top of an arbitrary background with Euler-scale fluid motion \cite{10.21468/SciPostPhys.5.5.054}.

\begin{figure}
    \centering
    \includegraphics[width=5cm]{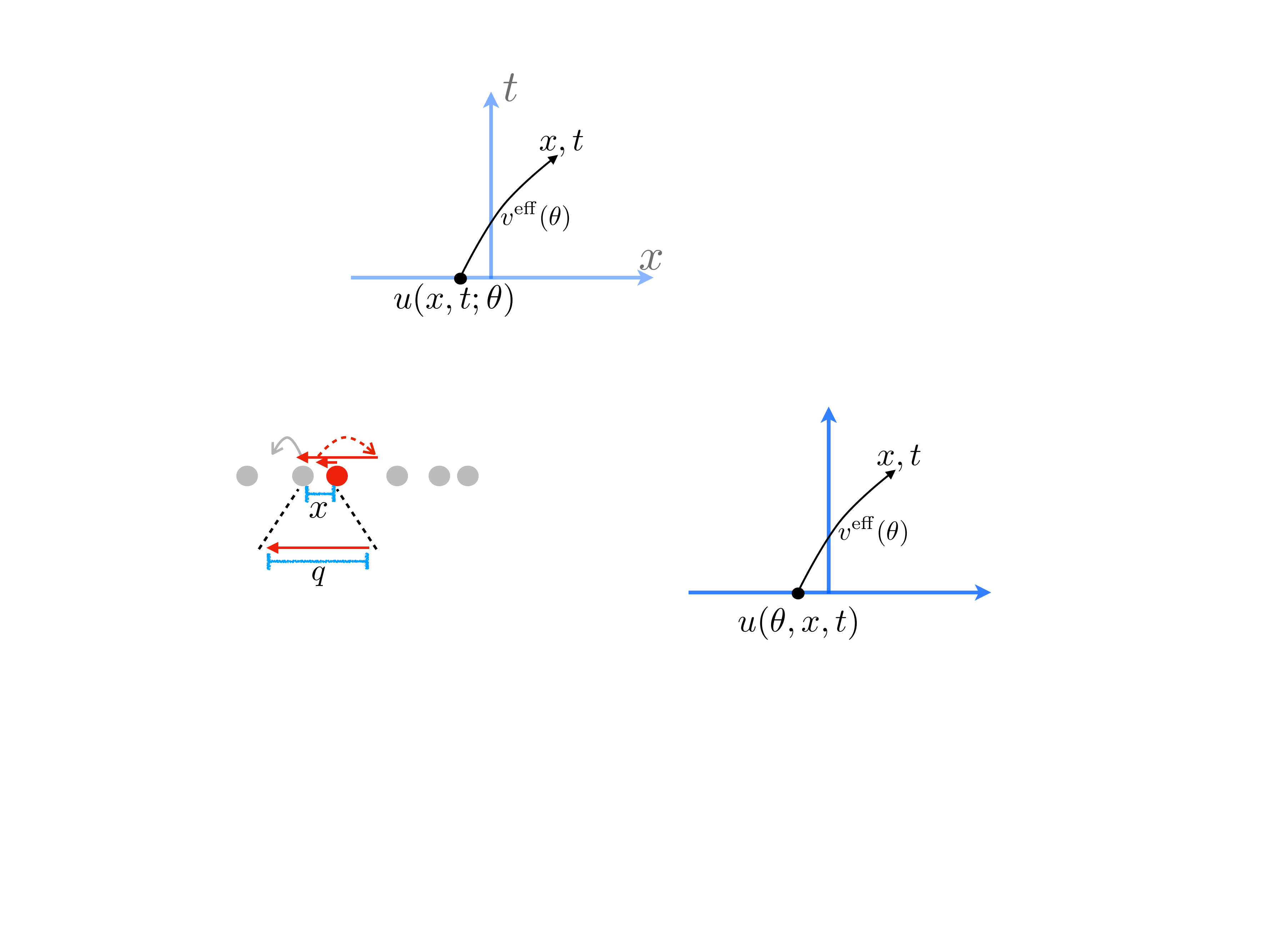}\ 
    \includegraphics[width=5cm]{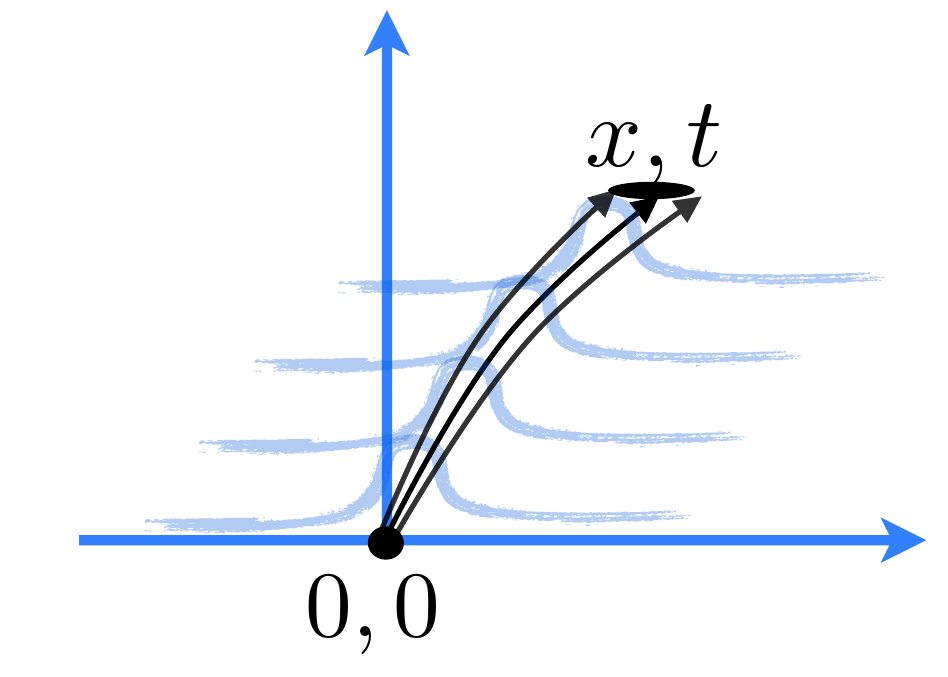}
    \caption{A characteristics is a curve in space-time that goes along the changing effective velocity in the non-stationary fluid. Euler-scale correlations in a non-stationary fluid are due to linear waves following the curved characteristics, and are affected by the change of density of characteristics due to the nonlinearity of the fluid's equations of motion (that is, the nontrivial interactions in the underlying model).}
    \label{fig:characteristics}
\end{figure}

The idea is that of linear response, as in sec. \ref{subsec:einstein}. However, this is applied not to the differential equations of Euler hydrodynamics, but to the {\em solution by characteristics} to the GHD equations. The solution by characteristics to GHD is presented as a set of integral equations with integrals on space and rapidities, where time appears as an explicit, fixed parameter; it was obtained in \cite{Doyon2018} (see also \cite[subsec 4.3]{10.21468/SciPostPhysLectNotes.18}). The solution by characteristics fixes the state of the fluid cell at $(x,t)$ by fixing its occupation function,
\begin{equation}
    n(\theta,x,t) = n(\theta,u,0),
\end{equation}
in a self-consistent fashion, where $u=u(\theta,x,t)$ describes the characteristics (see fig.~\ref{fig:characteristics} and solves
\begin{equation}
    \int_{z}^x \dd y\,\rho_{\rm tot}(\theta,y,t)
    -
    \int_z^u \dd y\,\rho_{\rm tot}(\theta,y,0)
    = \rho_{\rm tot}(\theta,z,0)\,
    v^{\rm eff}(\theta,z,0)\,t\quad (z\to-\infty).
\end{equation}
The fluid is assumed to approach sufficiently fast a homogeneous and stationary state in the far left region $z\to-\infty$.

We apply the functional derivative $\delta /\delta\beta^j(0)$ to the average $\ave{q_i(\ell x,\ell t)}_\ell$ for the inhomogeneous initial state \eqref{eq:nonstatdensity}. In the limit $\ell\to\infty$, this gives the non-stationary Euler-scale two point function for the initial state whose fluid description has space-dependent Lagrange parameters $\beta^i(x)$, as per Eq.~\eqref{eq:Snonstateulerscale},
\begin{equation}
    \lim_{\ell\to\infty} \frac{\delta \ave{q_i(\ell x,\ell t)}_\ell}{\delta \beta^j(0)} = \lim_{\rm Eul} S_{ij}(x,t).
\end{equation}
We then use the above solution by characteristics in order to evaluate the functional derivative. The result is
\begin{equation}\label{eq:Snonstatintegral}
    \lim_{\rm Eul} S_{ij}(x,t) =
    \int \dd\theta\dd\alpha\,\mathsf \Gamma_{(0,0)\to(x,t)}(\theta,\alpha)\rho_{\rm p}(\theta,x,t)f(\theta,x,t)
    h_i^{\rm dr}(\theta,x,t)h_j^{\rm dr}(\theta,x,t)
\end{equation}
where the propagator is a contribution of a direct term and an ``indirect propagator":
\begin{equation}
    \mathsf \Gamma_{(0,0)\to(x,t)}(\theta,\alpha) = \delta(u(\theta,x,t))\delta(\theta-\alpha)
    + \Delta_{(0,0)\to(x,t)}(\theta,\alpha).
\end{equation}
The direct propagator $\delta(u(\theta,x,t))\delta(\theta-\alpha)$ represents the contribution to the large-scale correlation by linear waves travelling along the curbed characteristics, with the fluid motion. The indirect propagator $\Delta_{(0,0)\to(x,t)}(\theta,\alpha)$ itself satisfies a self-contained linear integral equation; this becomes rather involved, and we refer to \cite[Eqs 3.21, 3.22]{10.21468/SciPostPhys.5.5.054} for details. It may be interpreted as the Euler-scale modifications of the linear waves due to the varying density of characteristics, and is nonzero only if there is nontrivial interactions (see fig.~\ref{fig:characteristics}). The observation of correlations in moving fluid therefore offers a way of distinguishing between interacting and non-interacting models at the Euler scale. It is shown in \cite{10.21468/SciPostPhys.5.5.054} that \eqref{eq:Snonstatintegral} indeed solves the evolution equation \eqref{eq:twopointfunctionnonstat}.

Two comments are in order.

First, Eq.~\eqref{eq:Snonstatintegral}, although involving rather complicated objects, can nevertheless be evaluated numerically \cite{10.21468/SciPostPhysCore.3.2.016} using the {\tt iFluid} package \cite{10.21468/SciPostPhys.8.3.041}, and has been checked against numerical simulations in the hard-rod gas \cite{10.21468/SciPostPhysCore.3.2.016}. It is observed there that the indirect propagator gives a relatively small correction to the effects of direct wave propagation, but significant enough to be numerically observable, and stronger than diffusion.  Eq.~\eqref{eq:Snonstatintegral} is a nontrivial prediction that is, up to now, not obtainable by any other techniques, and that does not have an equivalent beyond integrability. 

Second, we note that a similar combination of direct wave propagation and a small, indirect modification when non-trivial interactions are present, is also observed for correlation functions due to quantum fluctuations on top of moving fluids at zero temperature \cite{PhysRevLett.124.140603}, see the review \cite{AlbaReview} in this special issue.

\subsection{Higher-point correlation functions}

\begin{figure}
    \centering
    \includegraphics[width=4cm]{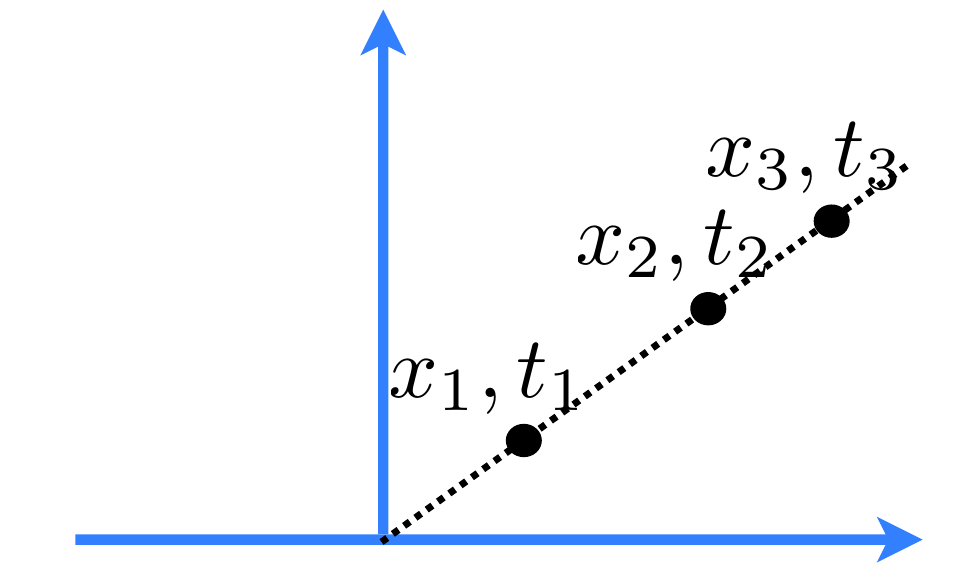}
    \caption{In ``free" models, the correlation functions of many conserved densities at the Euler scale $\lim_{\rm Eul}\ave{q_{i_1}(x_1,t_1)\cdots q_{i_n}(x_n,t_n)}^{\rm c}$ are nonzero only if the observables are located co-linearly. With nontrivial interactions this is not expected to be the case anymore.}
    \label{fig:colinear}
\end{figure}

The techniques just reviewed are immediately generalisable to the insertion of more local fields, and in particular can be used to obtain higher-point correlation functions in stationary states. For free theories, this is simple to do, and one finds that higher-point Euler-scale correlation functions of local conserved densities $\lim_{\rm Eul}\ave{q_{i_1}(x_1,t_1)\cdots q_{i_n}(x_n,t_n)}^{\rm c}$ are nonzero only if positioned colinearly (see fig.~\ref{fig:colinear}), $(x_k-x_1)/(t_k-t_1) = (x_2-x_1)/(t_2-t_1)$; the exact formula is presented in \cite{10.21468/SciPostPhys.5.5.054}. When interactions are present, because of the nontrivial indirect propagator, one expects the presence of {\em non-colinear} Euler-scale correlations, although this has not been fully analysed yet.

The general three-point formulae \eqref{hydro_3point}, \eqref{hydro_3point_2} of sec. \ref{subsec:higher} can be written in the TBA formalism using the integral kernel formalism, at least for the currents $o = j_i$ where all necessary quantities are known in all generality. In particular the three-point coupling \eqref{eq:3pointcoupling} is directly calculable from the average current $\ave{j_i}$. The result is rather involved however, and we skip the details.

A correlation functions with multiple field insertions can also be analysed in integrable models directly using form factors. The approach towards the hydrodynamic regime, from the point of view that the spectral sum, acts as a filter, filtering out higher particle-hole contributions. We have seen example of this, for the two-point function, in sec. \ref{subsec:ffEuler}. There the Euler scale correlator had contributions only from 1 particle-hole excited states. We can expect similar phenomena occurring also for higher point functions. Whereas this approach is yet to be fully developed, we sketch here the results for $3$-point functions.

We consider the connected, correlation function $\ave{o_1(\bfx) o_2(\bfy) o_3(\bfz)}^{\rm c}$ in a homogeneous, maximal entropy state described by $\rho_{\rm p}(\theta)$. Here, we introduced a notation $\bfx = (x,t)$ for the space-time coordinates and we consider local operators, that commute with the total particle number operator. The K\"all\'en–Lehmann spectral sums are then over the particle-hole excited states,
\begin{equation}
	\ave{o_1(\bfx) o_2(\bfy) o_3(\bfz)}^{\rm c} = \sum_{\alpha \neq \beta} \langle \rho_{\rm p} |o_1|\alpha\rangle \langle \alpha | o_2 | \beta \rangle \langle \beta | o_3 | \rho_{\rm p} \rangle e^{i \left( (\bfx-\bfy) \bfk(\alpha) + (\bfy - \bfz) \bfk(\beta) \right)}, \label{3point_ff}
\end{equation}
where $\bfx \bfk(\alpha) = t \varepsilon(\alpha) - x k(\beta)$ and $|\alpha\rangle = |\rho_{\rm p}; \bft_{\alpha}^+, \bft_{\alpha}^- \rangle$, $|\beta\rangle = |\rho_{\rm p}; \bft_{\beta}^+, \bft_{\beta}^-\rangle$ denote an excited state on top of $|\rho_{\rm p}\rangle$ characterized by particle-hole excitations $\{\bft_{\alpha}^+, \bft_{\alpha}^-\}$ and $\{\bft_{\beta}^+, \bft_{\beta}^-\}$ respectively.  We observe that $|\alpha\rangle, |\beta\rangle$ cannot be equal to the averaging state $|\rho_{\rm p}\rangle$ nor to each other - such cases correspond to disconnected contributions. This also implies that at least one of the form-factors must be $2{\rm ph}$, or higher, form-factor. We will now see the implications of these general structure of the correlator on the hydrodynamic regime. 
Following sec.~\ref{subsec:higher}, we set $\bfx = (x,t)$, $\bfy = (y,0)$ and $\bfz = ( z,0)$ and consider limit of large space-time separations between the three points. In such case we can expect that the sums localize on the stationary points of the phase factors,
\begin{equation}
   (\bfx-\bfy) \bfk(\alpha) + (\bfy - \bfz) \bfk(\beta) = t \varepsilon(\alpha) - (x-y) k(\alpha) - (y-z) k(\beta).
\end{equation}
The spectral sum in~\eqref{3point_ff} has contributions from intermediate states composed of various number of particle-hole excitations. We expect that the smaller the number of particle-hole excitations involved, the larger the contribution in the large space-time separation. Given the above discussion the leading contribution comes from such intermediate states that in~\eqref{3point_ff} there are two $1{\rm ph}$ form-factors and one $2{\rm ph}$ form-factor. One class of intermediate states that provide this is with $|\alpha \rangle = |\rho_{\rm p}; \theta_1^+, \theta_2^+, \theta_1^-, \theta_2^-\rangle$ and $|\beta\rangle = |\rho_{\rm p}; \theta_1^+,\theta_1^-\rangle$. The contribution to the phase factor from the first pair is then $t \varepsilon(\theta_1^+, \theta_1^-) - (x-y) k(\theta_1^+, \theta_1^-)$ while from the second pair is  $t \varepsilon(\theta_2^+, \theta_2^-) - (x-z) k(\theta_2^+, \theta_2^-)$. In the limit of large space-time separations, in parallel to computations presented in sec.~\ref{subsec:ffEuler}, the spectral sum can be evaluated with the saddle point approximation. Then the saddle point for $\theta_i^{\pm}$ is $\theta_i^*$ where $v^{\rm eff}(\theta_1^*) = (x-y)/t$ and $v^{\rm eff}(\theta_2^*) = (x-z)/t$. The contribution to the Euler scale correlation from this class of intermediate states is of the following structure\footnote{In writing the expression we assumed that stationary point equations have a single solution. If there are more solutions the result is a sum over contributions from each of them.}
\begin{equation}
    \lim_{\rm Eul}\ave{o_1(\bfx) o_2(\bfy) o_3(\bfz)}^{\rm c} \sim t^{-2} f_{\rho_{\rm p}}^{o_1} (\theta_1^*, \theta_1^* + i\pi, \theta_2^*, \theta_2^* + i\pi) V^{o_2} (\theta_1^*) V^{o_3}(\theta_2^*).
\end{equation}
This expression can be compared with the hydrodynamics predictions of sec.~\ref{subsec:higher}, where a picture of similar structure arises. This shows that the form-factor approach also has the potential to address the Euler scale properties of higher point functions. However, further studies in this direction are required in order to turn the heuristic analysis presented here into quantitative predictions that could be compared with results of hydrodynamic projections.


Finally, the theory of ballistic fluctuation reviewed in sec. \ref{subsec:higher} has been implemented in integrable models, see \cite{10.21468/SciPostPhys.8.1.007}. With the flow parameter $\lambda$ and the direction parameter $\alpha$ as in sec. \ref{subsec:higher}, the pseudoenergy satifies the flow equation
\begin{equation}
    \partial_\lambda \epsilon_{\lambda,\alpha}(\theta)
    ={\rm sgn}\big(\sin\alpha - \cos\alpha\,v^{\rm eff}_{\lambda,\alpha}(\theta)\big)\,
    h_{i;\,\lambda,\alpha}^{\rm dr}(\theta)
\end{equation}
and the full counting statistics (scaled cumulant generating function) is\footnote{This generalises  \cite[Eq 25]{10.21468/SciPostPhys.8.1.007} to arbitrary $\alpha$ and corrects the factor of 2 missing there.}
\begin{eqnarray}
    \mathsf f_{\lambda,\alpha} &=& 
    \int \frac{\dd\theta}{2\pi}\,
    \big(
    \cos\alpha\,e'(\theta)
    -
    \sin \alpha \, p'(\theta)\big)
    \times \nonumber\\ && \qquad
    \times\;\Big[\;
    {\rm sgn}\big(\sin\alpha - \cos\alpha\,v^{\rm eff}_{\lambda,\alpha}(\theta)\big)\,\big(\mathsf F(\epsilon_{\lambda,\alpha}) - 
    \mathsf 
    F(\epsilon_{0,\alpha})\big)
    \quad  \\ && \qquad +\; 2\sum_{\tilde\lambda\in[0,\lambda]\,:\atop v_{\tilde\lambda,\alpha}^{\rm eff}(\theta)=\cot\alpha} {\rm sgn}\big(\partial_\lambda
    v_{\lambda,\alpha}^{\rm eff}(\theta)|_{\tilde\lambda}\big)\,
    \big(\mathsf F(\epsilon_{\tilde\lambda,
    \alpha}) - 
    \mathsf 
    F(\epsilon_{0,\alpha})\big)
    \;\Big]\nonumber
\end{eqnarray}
(recall that $\alpha\in[0,2\pi]$). A generalisation of this theory on non-stationary, Euler-scale backgrounds has also been developed in integrable models, based on the results for correlation functions on non-stationary background reviewed in sec. \ref{subsec:nonstatintegrable}, see \cite{perfetto2020euler}.

\section{Diffusive scale}\label{sec:diffusion}

The scope of this section is to derive an analytical formula for the Onsager matrix $\mathfrak{L}_{ab}$ on a generic stationary thermodynamic state and generic integrable (quantum or classical) model, characterised solely in terms of its scattering shift $T(\theta,\alpha)$. We will first show how this can be obtained by inserting a resolution of identity in the current-current correlator and summing over the particle-hole excitations weighted by the form factors. It turns out that the sum over excitations truncate to a small number of particle and holes, due to the integration over time and space, and the final result has a simple analytical expression. We then show how the diagonal part of the diffusion matrix can be obtained by a simple kinetic argument, and finally how the complete Onsager matrix is reproduced by projecting the current operators on the complete basis of products of local and quasi-local conserved quantities.

The Onsager matrix is seen to be determined by the contributions of two particle-hole pairs in thermodynamic form factors. Conceptually, this ties in with the various ideas that we have discussed in this review. In particular, this corresponds to the contribution to diffusion due to the scattering of ballistic waves, as the latter, as explained above, are identified with single particle-hole pairs.

\subsection{Diffusion from form factor expansion of dynamical response function}
We have seen in previous section how one-particle hole excitations give the full ballistic response of the system. However, the local operators, and current densities, in particular, in fully interacting systems couple to more than the one particle-hole spectrum, as opposed to the free systems. In particular, the presence of two particle-hole excitations proves essential for diffusion and entropy increase.

Diffusive scale hydrodynamics is governed by the Onsager matrix \eqref{onsager}. It is shown in sec. \ref{subsec:ffEuler}, how the single particle-hole excitations of the current-current correlator give the full Drude weights contributions. Let us now consider higher order contributions to the space-integrated current-current correlator 
\begin{equation}
    \int \dd x\,
    (j_i(x,s),j_k(0,0)) = \sum_{n \geq 1}  ({\rm jj})_{ik}^{n-\rm ph} (s).
\end{equation}
Specialising to the time integrated current-current correlation function at the second order $n=2$, results in
\begin{align}\label{eq:blabla2}
 \lim_{T \to \infty} \int_{-T}^T \dd t\,  & ({\rm jj})_{ik}^{2-\rm ph} (t)  = \lim_{T \to \infty} \frac{ (2\pi)^2}{2!^2}  \int \dd \theta_1^- \dd  \theta_2^-  \rho_{\text{p}}(\theta_1^-)  \rho_{\text{p}}(\theta_2^-)    \fint \dd \theta_1^+ \dd  \theta_2^+ \rho_{\text{h}}(\theta_1^+)   \rho_{\text{h}}(\theta_2^+)    \no \\& \times \delta(k[\bfp,\bfh])  \delta_T(\varepsilon[\bfp,\bfh])   \langle \rho_{\rm p} | j_i |  \theta^+_1, \theta^+_2, \theta^-_1, \theta^-_2 \rangle \langle \theta^+_1, \theta^+_2, \theta^-_1, \theta^-_2  | j_k | \rho_{\text{p}}   \rangle,
\end{align}
where the finite-time delta function is defined as 
\begin{equation}
    \delta_T(x)=   \sin(T x \pi)/\pi x .
\end{equation}
The integration over particle momenta in \eqref{eq:blabla2} can then be performed using the two kinematic constraints. Moreover, it is easy to show that higher particle-hole contribution are zero
\begin{equation}
    \lim_{T \to \infty} \int_{-T}^T \dd t\,   ({\rm jj})_{ik}^{n-\rm ph} (t) =0,\quad n >2.
\end{equation}
The reason is the following: the matrix elements of the current operators are proportional to the total energy of the excitations $\varepsilon$, which is forced to be zero by kinematic constraint. The contribution therefore vanishes in the infinite time limit, unless the divergence of the matrix elements compensates for it. The only points in the phase space of particle-hole positions where matrix elements are divergent are those where at least one of the particles shares the same rapidity of one of the holes, which is a zero-measure set of points in $\mathbb{R}^{2n}$. However, this set has a non-zero measure when the number of particle and holes is equal to two, as conservation of energy and momentum $\delta(k) \delta(\varepsilon)$ collapse the integral exactly on the two points $\theta_1^+ = \theta_1^-$, $\theta_2^+ = \theta_2^-$ and $\theta_1^+ = \theta_2^-$, $\theta_2^+ = \theta_1^-$. At these points the divergence from the form factors \eqref{eq:FFcurrents} cancels the zero from the factor $(\varepsilon[\bfp,\bfh])^2$ and the result is finite. Therefore we obtain, using the expressions \eqref{eq:FFcurrents}, 
\begin{align}\label{Lmatrix-final}
   &  \mathfrak{L}_{ik}= \lim_{t \to \infty} \int_{-T}^T \dd t\,   ({\rm jj})_{ik}^{2-\rm ph} (t)\no \\&  = (2 \pi)^2 \int \frac{\dd \theta_1 \dd \theta_2}{2} \rho_{\rm p }(\theta_1) (1-n(\theta_1)) \rho_{\rm p}(\theta_2) (1-n(\theta_2)) |v^{\text{eff} } (\theta_1) - v^{\text{eff}} (\theta_2)|   \no \\&  \times  
  \Big( \frac{ T^{\text{dr}}(\theta_2,\theta_1) h_{i}^{\text{dr}}{} (\theta_2) }{k'(\theta_2)}    - \frac{  T^{\text{dr}}(\theta_1,\theta_2) h_{i}^{\text{dr}}(\theta_1)  }{  k'(\theta_1)}   \Big)   \no \\&  \times   \Big( \frac{ T^{\text{dr}}(\theta_2,\theta_1) h_{k}^{\text{dr}}{} (\theta_2) }{k'(\theta_2)}    - \frac{ T^{\text{dr}}(\theta_1,\theta_2)h_{k}^{\text{dr}}(\theta_1)  }{  k'(\theta_1)}   \Big),
\end{align}
where integration over the particles has been performed, and the two variables $\theta_1$ and $\theta_2$ denote the remaining integration over the positions of the two holes. 

The same result can be obtained by introducing an intermediate step, that is considering the 2 particle-hole contribution at the small energy and momentum of the current-current correlation. The Onsager matrix can then be computed from 
\begin{equation} \label{Onsager_DSF_relation}
    \mathfrak{L}_{ik} = \lim_{\omega \to 0} \lim_{k \to 0} S_{\rm 2ph}^{\rm j_i j_k} (k, \omega),
\end{equation}
where $S_{\rm 2ph}^{\rm j_i j_k} (k, \omega)$ is the 2 particle-hole contribution to the Fourier transform of $S^{\rm j_i j_k} (x, t) \equiv \langle {\rm j}_i(x,t) {\rm j}_k(0,0) \rangle$. For the Lieb-Liniger model and  using the form-factor expansion of the correlator the two particle-hole contribution reads~\cite{Panfil2021}
\begin{align} \label{S_jj_2ph}
 & S_{\rm 2ph}^{\rm j_i j_k} (k, \omega) = (2 \pi)^2 \fint \frac{\dd \theta_1 \dd \theta_2}{2} \rho_{\rm p }(\theta_1) (1-n(\theta_1)) \rho_{\rm p}(\theta_2) (1-n(\theta_2)) |v^{\text{eff} } (\theta_1) - v^{\text{eff}} (\theta_2)|   \no \\&  \times  
   v^{\rm eff}(\theta^*) \Big( \frac{ T^{\text{dr}}(\theta_2,\theta_1) h_{i}^{\text{dr}}{} (\theta_2) }{k'(\theta_2) (v^{\rm eff}(\theta_2) - v^{\rm eff}(\theta^*))}    - \frac{  T^{\text{dr}}(\theta_1,\theta_2) h_{i}^{\text{dr}}(\theta_1)  }{  k'(\theta_1) (v^{\rm eff}(\theta_1) - v^{\rm eff}(\theta^*))}   \Big)   \no \\&  
   \times  v^{\rm eff}(\theta^*)  \Big( \frac{ T^{\text{dr}}(\theta_2,\theta_1) h_{k}^{\text{dr}}{} (\theta_2) }{k'(\theta_2)(v^{\rm eff}(\theta_2) - v^{\rm eff}(\theta^*))} - \frac{ T^{\text{dr}}(\theta_1,\theta_2)h_{k}^{\text{dr}}(\theta_1)  }{  k'(\theta_1) (v^{\rm eff}(\theta_1) - v^{\rm eff}(\theta^*))}   \Big), 
\end{align}
where $\theta^*$ is defined through $v^{\rm eff}(\theta^*) = \omega/k$. This expression constitutes the leading order in $k$ with fixed $\omega/k$. The double Hadamard integral (denoted by $\fint$) regularizes the poles at $\theta_i = \theta^*$. The limits in~\eqref{Onsager_DSF_relation} correspond then to $\theta^* \to \infty$, under which~\eqref{S_jj_2ph} becomes~\eqref{Lmatrix-final}. The integrand of the limiting expression is a regular function, hence no regularization appears in~\eqref{Lmatrix-final}. 

The result~\eqref{Lmatrix-final} can be promoted to generic integrable models (and it does recover the previously known expression for hard-rod gases, see \cite{Spohn1991,Boldrighini1997}), provided the integration over rapidities is also seen as a sum over different particle species in case of string solutions and nested structure. In this general case, labelling with the two labels $s_1,s_2$ the different string types, the expression reads 
\begin{align}\label{eq:DCgeneralmain}
&\mathfrak{L}_{ik}= \sum_{s_1,s_2} \int \frac{\dd \theta_1 \dd \theta_2}{2} \rho_{{\rm p} ;s_1}(\theta_1) f_{s_1}(\theta_1)\rho_{{\rm p}; s_2}(\theta_2) f_{s_2}(\theta_2) |v_{s_1}^{\text{eff} } (\theta_1) - v_{s_2}^{\text{eff}} (\theta_2)|   \no \\&  \times
  \Big(\ \frac{T_{s_1,s_2}^{\text{dr}}(\theta_2,\theta_1) h_{i;s_2}^{\text{dr}}{} (\theta_2) }{ k'_{s_2}(\theta_2)}    - \frac{T_{s_2,s_1}^{\text{dr}}(\theta_1,\theta_2)  h_{i;s_1}^{\text{dr}}(\theta_1)  }{   k'_{s_1}(\theta_1)}   \Big) \no \\&  \times   \Big(\ \frac{T_{s_1,s_2}^{\text{dr}}(\theta_2,\theta_1) h_{k;s_2}^{\text{dr}}{} (\theta_2) }{ k'_{s_2}(\theta_2)}    - \frac{T_{s_2,s_1}^{\text{dr}}(\theta_1,\theta_2)  h_{k;s_1}^{\text{dr}}(\theta_1)  }{   k'_{s_1}(\theta_1)}   \Big),
\end{align} 
where we have used the generic symbol $f(\theta)$ incorporating the statistics of the excitations, see Eq. \eqref{eq:qpStatistics2}.
The scattering shift $T_{s_1,s_2}(\theta,\alpha)$ is now also a matrix in the string indices.  
Equation \eqref{Lmatrix-final} can be written as quadratic form as following
  \begin{equation}
	(\mathfrak{D} \mathsf C)_{ik} = h_i \cdot \frac{1}{2 \pi} (1-\sigma nT)^{-1} k'\widetilde{\mathfrak{D}} n f \ (1-T^{\rm T}n\sigma)^{-1} h_k. \label{DC_kernel}
  \end{equation}
  with the following Markov-type matrix 
  \begin{equation}\label{eq:dtildemain}\frac{1}{2}\widetilde{{\mathfrak{D}}}_{s_1,s_2}(\theta,\alpha) = \delta_{s_1,s_2}\delta(\theta-\alpha) \widetilde{w}_{s_1}(\theta) - \widetilde{W}_{s_1,s_2}(\theta,\alpha),
\end{equation}
with rates given by
 \begin{equation}\label{eq:off-diago-diff}
\widetilde{W}_{s_1,s_2}(\theta,\alpha) = \frac{(2 \pi)^2}{2} \chi_{s_1}(\theta) \frac{T^{\rm dr}_{s_1,s_2}(\theta,\alpha)T^{\rm dr}_{s_2,s_1}(\alpha,\theta)}{k'_{s_1 }(\theta)^2   } |v^{\rm eff}_{s_1}(\theta)-v^{\rm eff}_{s_2}(\alpha)|,
\end{equation}
and the diagonal part obtained by the sum over all jump rates
 \begin{align}\label{eq:wtilde}
 \widetilde{w}_{s_1}(\theta) =   & \frac{(2 \pi)^2}{2}   \sum_{s_2}\int \dd \alpha\, \chi_{s_2}(\theta) \left( \frac{T^{\rm dr}_{s_1,s_2}(\theta,\alpha)}{k'_{s_1}(\theta)   } \right)^2 | v^{\rm eff}_{s_1}(\theta) - v^{\rm eff}_{s_2}(\alpha)|,
\end{align}
where we used the definition of quasiparticle susceptibilities
\begin{equation}
    \chi_{s}(\theta)= k_s'(\theta) n_s(\theta)f_s(\theta) /(2 \pi).
\end{equation}
The quantity \eqref{eq:wtilde} can be seen as the effective variance for the quasiparticle fluctuations with rapidity $\theta$ inside the local stationary state, caused by the random scattering processes (see next section), where the ratio $  {T^{\rm dr}_{s_1,s_2}(\theta,\alpha)}/k'_{s_1}(\theta)  $ can be seen as the average displacement of the quasiparticle of type $s_1$ and rapidity $\theta$ scattering many times with all the other quasiparticles present in the local stationary state.

Finally, we point out that there is a clear relation between the particle-hole hierarchy in transport coefficients and the hydrodynamic projection introduced in sec. \ref{sec:hydroCorr}, see fig. \ref{fig:phProj}. 
\begin{figure}[h]
\begin{center}
\includegraphics[width=0.7\textwidth]{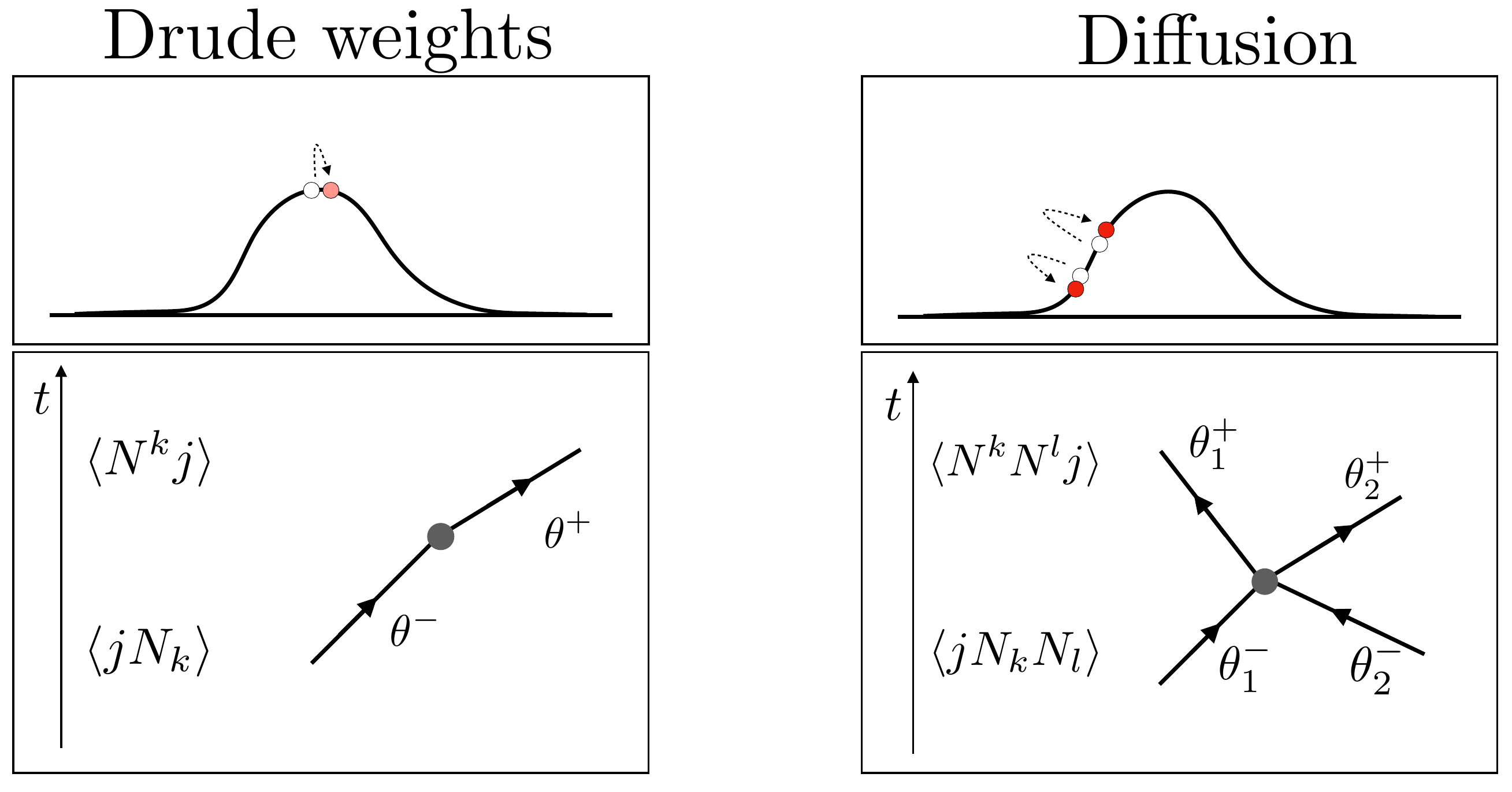}
\end{center}
\caption{Left: ballistic Drude weights are given by one particle-hole excitations in the limit of small momentum that can be seen as a single quasiparticle propagating through the system and undergoing a zero-momentum scattering with the background, or alternatively as the projection of the current operator on each normal mode, propagated and projected back. Right: Onsager matrix is instead given by the two-particle hole contribution at low momentum, that can be seen as a scattering of two quasiparticles, or as the projection of the current operator on two normal modes, propagated and projected back, summed over all different modes.      } 
\end{figure}\label{fig:phProj}

\subsection{Kinetic picture of quasiparticle diffusion}
The diagonal part (in quasiparticles modes) of the diffusion matrix, Eq. \eqref{eq:wtilde}, can be derived in a quite simple fashion by employing some arguments from standard kinetic theory of quasiparticles, introduced in \cite{ghkv}. One indeed should follow one of the quasiparticles, for example the one with rapidity $\theta$ (and of type $k$ in general, which we omit in the following) moving through a thermodynamic state with density $\rho_{\rm p }(\alpha)$ with its effective velocity $v^{\rm eff}(\theta)$.  Its velocity is a functional of $\rho_{\rm p }(\theta)$ due to scatterings with other quasiparticles. However, their density is not constant but it fluctuates on any finite sub-system, leading to fluctuations in the velocity and therefore to to diffusive spreading of the quasiparticle trajectory. Specifically, a quasiparticle with rapidity $\theta$ will collide with another quasiparticle with rapidity $\alpha$ if they started out in a spatial window of size $|v^{\rm eff}(\theta) - v^{\rm eff}(\alpha)|t$. We therefore compute the variance of the fluctuation of the tagged quasipartcle $\langle [\delta x(\theta)]^2 \rangle$ (where $\langle \cdot \rangle$ here denotes an average over ensemble realisations) as given by the fluctuations in density of all scattered quasiparticles. Namely 
\begin{equation}
    \langle [\delta x(\theta)]^2 \rangle = t^2 \int \dd\alpha \left( \frac{ \partial v^{\rm eff}(\theta)}{\partial n(\alpha)}\right)^2 \langle  [\delta n(\alpha)]^2 \rangle
\end{equation}
with $n(\alpha)$ the filling functions. One then uses the expression for the fluctuations of filling functions in a sub-system of size $\ell$ \cite{PhysRevB.54.10845}
\begin{equation}
    \langle [ \delta n(\alpha)]^2   \rangle =  \chi(\alpha)/\ell
\end{equation}
by definition of quasiparticle susceptibility. 
Using that the fluctuations should be computed on the spatial interval $\ell$ of size 
\begin{equation}
    \ell = |v^{\rm eff}(\alpha) - v^{\rm eff}(\theta)| t,
\end{equation}
where particles could actually scatter, gives
\begin{equation}\label{eq:kinetic-displa}
 \lim_{t \to \infty}   \langle [\delta x(\theta)]^2 \rangle/t =  \int \dd\alpha \left( \frac{ \partial v^{\rm eff}(\theta)}{\partial n(\alpha)}\right)^2  \frac{\chi(\alpha)}{|v^{\rm eff}(\alpha) - v^{\rm eff}(\theta)|}
\end{equation}
Finally, using the expression for the effective velocity, \eqref{eq:effective-velo} one can show that 
\begin{equation}
 \frac{ \partial v^{\rm eff}(\theta)}{\partial n(\alpha)} = \frac{2\pi T^{\rm dr}(\theta,\alpha)}{k'(\theta)} (v^{\rm eff}(\theta)-v^{\rm eff}(\alpha))
\end{equation}
such that equation \eqref{eq:kinetic-displa} is equivalent to 
\begin{equation}
\langle[\delta x(\theta)]^2 \rangle  = 2 t      \tilde{w}(\theta),
\end{equation}
namely the fluctuations of trajectories of each quasiparticle grow as $\sqrt{t}$ and their effective diffusion constant is provided by the function $\tilde{w}(\theta)$ in \eqref{eq:wtilde}. Despite the simplicity of the argument, the inclusion of the off-diagonal components of diffusion, namely the term \eqref{eq:off-diago-diff}, requires a slightly more elaborate treatment. Indeed quasiparticles do not only witness a spreading in their trajectories but they also randomly exchange their rapidities or momenta. A more complete kinetic picture is therefore provided by using the formalism of hydrodynamic projection on charges and therefore on normal modes, namely the quasiparticles. 

\subsection{Diffusion in integrable systems from hydrodynamic projection}\label{difhydro}
In this section we will show that expression obtained from the form factor expansion can be exactly reproduced by the projection onto the products of bilocal charges considered in sec. \ref{subsec:projectionsdiffusive}.

In order to explicitly evaluate the lower bound we need to discern the forms of the $G$-matrix \eqref{Gmat}, and transformation ${\mathsf R}$ that takes us to the normal mode basis, which will allow us to evaluate the expression \eqref{ons_bound}.

We first focus on $G$-matrix, which was evaluated for integrable systems in \cite{PhysRevB.101.041411} and \cite{10.21468/SciPostPhys.9.5.075}.
The connection between the projection result \eqref{ons_bound} can be established by evaluating the $G$-matrix in integrable systems. We will use a compactified notation, with the indices $i$ denoting both rapidity and particle type $i=(\theta,s)$, with the Einsteins summation convention $a^i b_i=\sum_s\int \dd\theta\, a_s(\theta) b_s(\theta)$. $G$-matrix is related to the Hessian
\begin{equation}
{\mathsf G}_i^{\,\,jk}=\frac{1}{2}{\mathsf R}_i^{\,\,l}\big({\mathsf R}^{-\mathrm{T}}{\mathsf H}_l {\mathsf R}^{-1}\big)^{jk},
\end{equation}
by definition. Hessian describes the curvature of the currents on the manifold of steady states wrt expectation values of charges and can be obtained by taking the derivative of the flux Jacobian wrt quasiparticle densities
\begin{equation}
    {\mathsf H}_i^{jk}=\frac{\partial {\mathsf A}_i^j}{\partial \rho_k}.
\end{equation}
The transformation which diagonalizes the flux Jacobian, taking us to the normal mode basis is (see \eqref{eq:Akernel})
\begin{equation}
    {\mathsf R}_i^j=\frac{1}{\sqrt{\chi_i}}(1-nT)_i^j.
\end{equation}

Evaluating the derivative of the flux Jacobian yields the following expression for the $G$-matrix
\begin{align}
\label{gmatrix}
{\mathsf G}_i^{\,\,jk}&=\frac{1}{2\rho^\mathrm{tot}_i}\sqrt{\frac{\chi_j\chi_k}{\chi_i}}\Big[(v^\mathrm{eff}_j-v^\mathrm{eff}_i)(T^\dr)_i^{\,\,j}\delta_i^{\,\,k}+(v^\mathrm{eff}_k-v^\mathrm{eff}_i)(T^\dr)_i^{\,\,k}\delta_i^{\,\,j}\Big].\end{align}
Equiped with this expression, the expression for the Onsager coefficients \eqref{dif_nm_od} can be readily evaluated, reproducing the exact result obtained from the form factor expansion \eqref{eq:DCgeneralmain}.

Two comments are in order. First of all, notice that the diagonal elements of $G$-matrix, ${\mathsf G}_i^{jj}$, vanish, since they are proportional to the difference of effective velocities, implying that all contributions to the Onsager coefficients from the products of normal modes in expression \eqref{dif_nm_bound} are finite. This observation is consistent with the prediction of phenomenological theory of nonlinear fluctuating hydrodynamics which states that superdiffusive behaviour should emerge, provided that any diagonal elements of $G$ are non-vanishing. Nevertheless, superdiffusion does occur also in integrable systems, but it can be viewed as a consequence of diverging summation over the finite mode contributions \cite{2103.01976}. 

Secondly, the observation that the products of normal modes yield exact result for Onsager coefficients indicates that they might represent a complete set of quadratic charges and a complete basis of diffusive subspace in integrable systems. Special care needs to be taken in the case of integrable models with additional non-abelian symmetries, where the Onsager coefficients can diverge, and additional quadratic charges arising from the Yangian symmetry can be present \cite{Prosen_2014}.
\section{Applications}\label{sec:applications}
\subsection{$T \bar{T}$ deformed CFTs}
Conformal field theories are ubiquitous in nature, as they describe critical phenomena, low temperature and large scale physics.
They can be characterized by the left and right moving  quasiparticles that do not interact among each-other, meaning that the scattering between the left and the right movers is trivial. In many CFTs, such as minimal models \cite{Bazhanov1996}, or Liouville CFTs \cite{zamolodchikov2012quantum}, each of the sectors admits an integrable description. Despite their interacting nature, CFTs exhibit some nongeneric behaviour, for instance, vanishing bulk viscosity (momentum-momentum Onsager coefficient). In order to study these properties as we depart from a strict zero temperature CFT description, $T\bar{T}$ deformation \cite{Zamolodchikov:2004ce,Dubovsky2012,Caselle2013,Cavagli2016,Conti2019} is especially useful as it preserves integrability. The $T\bar{T}$ deformation introduces integrable scattering between the left and the right movers, modifying the scattering matrix through the CDD factor $\Sigma(\theta)$, $S^{(\sigma)}(\theta)=e^{\ii \Sigma(\theta)}S^{(0)}(\theta)$. In case of massless theories the CDD factor reads $\Sigma(\theta)=-\sigma \frac{M^2}{4}\exp(\theta)$, where $\sigma$ corresponds to the deformation strength. In what follows we will discuss how GHD can be used to obtain transport coefficients \cite{PhysRevLett.126.121601,PhysRevD.103.066012}, note, however, that $T\bar{T}$ deformation also admits a holographic dual \cite{McGough:2016lol,Guica:2017lia}, allowing for the comparison of two methods \cite{PhysRevLett.126.121601,PhysRevD.103.066012}.
On the level of Lagrangian the $T\bar{T}$-deformation manifests itself as a continuous deformation
\begin{equation}
\partial_{\sigma} \mathcal{L}(\sigma)=\frac{1}{2\pi^2}T^{0\mu}T^{1\nu}\epsilon_{\mu\nu},
\end{equation}
where $T_{\mu\nu}$ are the components of the stress-energy tensor and  $\mathfrak{L}(0)$ corresponds to the undeformed Lagrangian. Importantly, deformation preserves Lorentz invariance. In classical systems the effects of CDD factor are manifested in the increased width of fundamental particles \cite{cardy2021toverline}.

In the following we will focus on the energy and momentum diffusion constants and Drude weights in $T\bar{T}$ deformed CFTs.

\subsubsection{Drude weights in $T\bar{T}$-deformed CFTs}

Let us specialize to the boosted thermal states $\rho=\frac{1}{Z}\exp(-\beta H+\nu P)$, where $P$ corresponds to the momentum operator. 

Relativistic invariance has important consequences for thermal transport, as the energy current is preserved. Firstly, this implies that the energy Drude weight is a static quantity related to the momentum susceptibility
\begin{equation}\label{energy_dv}
D_{EE}=-\left.\p^2 f/\p \nu^2\right|_{\nu\to0},
\end{equation}
where $f$ is the free energy, which can be evaluated explicitly. Secondly the components of Onsager matrix associated with the energy current vanish $\mathfrak{L}_{E\bullet}=\mathfrak{L}_{\bullet E}=0$.
The second quantity we will compute is the bulk viscosity, which is related to the energy Drude weight in a simple manner. In $T\bar{T}$ deformed CFTs pseudoenergies for the left and right movers $\epsilon_{\pm}(\theta)$ satisfy two integral equations 
\begin{equation}
    \epsilon_{\pm}(\theta)=\beta_{\pm}E_{\pm}(\theta)-T L_{\pm}(\theta)-\tilde{T}_{\pm\mp} L_{\mp}(\theta).
\end{equation}
$E_{\pm}(\theta)=\frac{M}{2}\exp(\pm\theta )$ and $p_{\pm}(\theta)=\pm \frac{M}{2}\exp(\pm\theta )$ are the bare energy and momentum, $L_{\pm}(\theta)=\log(1+\exp(-\epsilon_\pm(\theta)))$ and $\beta_{\pm}=\beta\mp \nu$. $T_\pm(\theta,\theta')$ codes for the interaction between the same particles in CFT, while the phase shift $\tilde{T}_{\pm\mp}(\theta,\theta')=-\frac{\sigma}{2\pi} p_+(\theta)p_{-}(\theta')$ arises due to the $T\bar{T}$ deformation. After taking into account that the phase shift is factorizable, and  the relation between the bare energies and momentum, one can observe that $T\bar{T}$ deformation results in the renormalization of temperatures
\begin{eqnarray}\label{pseudo_def}
    \epsilon_{\pm}(\theta)=\tilde{\beta}_{\pm}E_{\pm}(\theta)-T L_{\pm}(\theta),\\
    \tilde{\beta}_{\pm}=\frac{\beta_{\pm}}{2}\left(1+\sqrt{1-\frac{\pi\sigma c}{3\beta_+\beta_-}}\right),
\end{eqnarray}
where $c$ is the central charge.
In order to obtain the bulk viscosity we need to evaluate the effective velocities 
 \begin{equation}\label{effective-v}
     v^\eff_\pm(\theta)\equiv v^\eff_\pm\equiv\pm v_\eff=\pm\sqrt{1-\frac{\pi\sigma c}{3\beta^2}},
 \end{equation}
which can be deduced by noticing that equations 
\begin{eqnarray}
    (p'_\pm)^\dr(\theta)&=&\p_\beta\tilde{\beta}_\pm E_\pm(\theta)+[T n(p'_\pm)^\dr](\theta),\\
   (E'_\pm)^\dr(\theta)&=&-\p_\nu\tilde{\beta}_\pm E_\pm(\theta)+[T n(E'_\pm)^\dr](\theta)
\end{eqnarray}
are related by transformation $(E'_\pm)^\dr(\theta)\to v^\eff_\pm(\theta)(p'_\pm)^\dr(\theta) $.
Considering a general expression for Drude weights in integrable systems,
and the relation $p^\dr(\theta)=v^\eff E^\dr(\theta)$, we see that the bulk viscosity is related to the energy Drude weight by
    $D_{PP}=(v^\eff)^2 D_{EE}$. This allows us to obtain the closed form expressions of the energy and momentum Drude weights 
\begin{equation}
    {\mathsf D}_{EE}=\frac{\pi c}{3\beta^3}\frac{1}{v^\eff},\quad {\mathsf D}_{PP}=\frac{\pi c}{3\beta^3}v^\eff.
\end{equation}

\subsubsection{Diffusion constants in $T\bar{T}$ deformed CFTs}
Let us proceed by evaluating the momentum Onsager coefficients $\mathfrak{L}_{PP}$.  
A general expression for the Onsager matrix again simplifies significantly. This can be viewed as a consequence of degenerate dressed velocities of left and right movers, which do not depend on the rapidity variable $\theta$, but only on the type of particles. This means that the only non-vanishing contributions will arise from the scatterings between the left and the right movers
\begin{equation}
    \mathfrak{L}_{PP}=8\pi^2(v^\eff)^3\int\dd\theta\dd\lambda\chi(\theta)\chi(\lambda)(T^\dr_{+-}(\theta,\lambda)+T^\dr_{-+}(\lambda,\theta))^2,
\end{equation}
with the quasiparticle susceptibility $\chi(\theta)$.
As  the dressed kernel factorizes $T^\dr_{\pm\mp}(\theta,\lambda)=-\sigma p^\dr_\pm(\theta)p^\dr_\mp(\lambda)/(2\pi)$ \cite{PhysRevD.103.066012}, and the dressed energies and momenta are related in a simple manner, the diffusion constant can be expressed in terms of the Drude weights
\begin{equation} \label{LPP}
  \mathfrak{L}_{PP}=\frac{\sigma^2}{2}v_c {\mathsf D}_{EE}^2=\frac{\sigma^2}{2}\frac{{\mathsf D}_{PP}^2}{v^3_c}.
\end{equation}

Above results on Drude matrix can also be obtained by considering a bipartition protocol in holographic CFTs. Due to the large $c$ limit in the semi-classical holographic calculation, however, the corresponding bulk viscosity vanishes.
\subsection{Heisenberg XXZ spin chain}
Heisenberg XXZ spin chain is one of the most well studied integrable models, describing the dynamics of one dimensional magnets, and spinless fermions 
\begin{equation}\label{eq:XXZchain}
    H = \sum_j \Big[\vec{S}_j \cdot \vec{S}_{j+1} + (\Delta-1) S^z_j S^z_{j+1} + h \sum_j S^z_j \Big].
\end{equation}
Despite its long history and the richness of its dynamics, its out-of-equilibrium properties were largely unexplored until very  recently. Here, we will focus on two particularly important aspects, the celebrated fractal spin Drude weight, and the comparison of the low temperature diffusion constant obtained through GHD and semi-classical theory.
\subsubsection{Fractal Drude weight of the gapless XXZ chain}
The question of quantifying the spin transport in the gapless XXZ spin chain
at finite temperatures
has a long history. It has long been understood that Drude weights can, in general, be bounded by quasi-local conservation laws, by means of Mazur bound \cite{Mazur1969, PhysRevB.55.11029}. However, in all states with no magnetisation imbalance, i.e. finite temperature ensemble with $h=0$, the standard local integrals of motion do not provide a non-zero bounds.  Therefore, it came as a surprise that a thermodynamic Bethe ansatz calculation \cite{PhysRevLett.82.1764} predicted a finite value of the Drude weight for $\Delta<1$. Appropriate quasi-local conservation laws, providing the bound on the Drude weight were constructed only much later \cite{PhysRevLett.106.217206,Ilievski2012,PhysRevLett.111.057203,ilievski2016quasilocal}. Curiously, these result indicated that the lower bound on the Drude weight at infinite temperature could be a nowhere continuous function of the anisotropy $\Delta$. Naively, one might expect that as the value of the parameter $\Delta$ is slightly altered, the correlation functions can change only slightly, and therefore cannot be discontinuous. While such an argument is valid when we are dealing with the finite size systems at finite times, it can fail when the thermodynamic and long-time limits are considered. 

\begin{figure}[h]
\begin{center}
\includegraphics[width=0.45\textwidth]{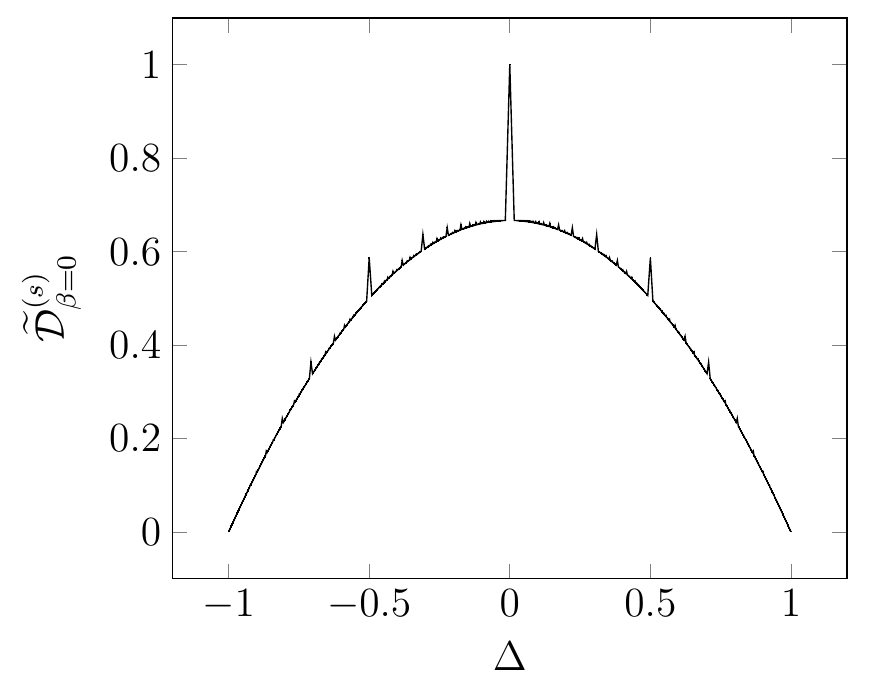}\includegraphics[width=0.45\textwidth]{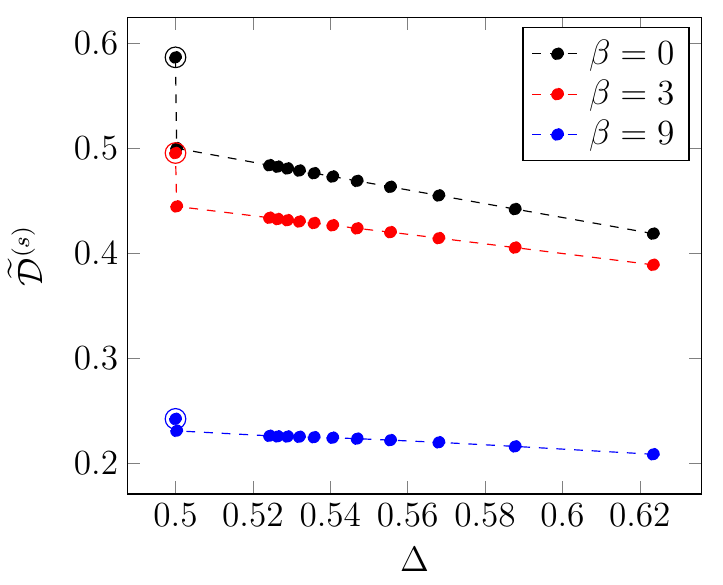}
\end{center}
\caption{In the figure on the left the rescaled infinite temperature Drude weight ($\tilde{\mathcal{D}}^{(s)} = {\mathsf D}_{ss} 16/\beta$) is shown. In the figure on the right the dependence of the Drude weight on anisotropy $\Delta$ for different temperatures is shown. \label{fig:Drude} \cite{IN_Drude}}
\end{figure}

GHD can be used to show that the infinite temperature bound on the Drude weight is indeed saturated, and, furthermore, that the fractal structure persists even at finite temperatures \cite{IN_Drude}. Such behaviour is a consequence of fine tuned dependence of the particle content of the Heisenberg spin chain on anisotropy. While for $\Delta\geq1$ the number of particle species is infinite, their number $N_p=\sum_1^l \nu_l$ in $\Delta<1$ regime can be deduced by considering the continued fraction representation of the anisotropy parameter $\Delta=\cos(\frac{m}{\ell}\pi)$, $\frac{m}{\ell}=\frac{1}{\nu_1+\frac{1}{\nu_2+\cdots}}$, and only the last two particle species contribute to the spin Drude weight. The infinite temperature $\beta \to 0$ expression for the Drude weight
\begin{equation}\label{eq:drudexxz}
{\mathsf D}_{ss} = \frac{\beta}{16} \frac{\sin^{2}{(\pi m/\ell)}}{\sin^{2}{(\pi/\ell)}}\left(1-\frac{\ell}{2\pi}\sin{(2\pi/\ell)}\right),
\end{equation}
was then shown to be in perfect agreement with the GHD prediction, \eqref{eq:Drude-Projections}, in \cite{IN_Drude}.
Furthermore, in the scope of GHD finite-temperature Drude weight can be considered on the same footing, showing the same kind of the discontinuous behaviour, see fig. \ref{fig:Drude}, which becomes less pronounced as temperature is lowered. This is indeed in agreement with the Drude weight computed within the emergent bosonic field theory, which gives an exact prediction at zero temperature \cite{Benz2005}. Finally, we shall stress that while the value of the Drude weight \eqref{eq:drudexxz} is valid for all values of $\Delta = \cos(m \pi /\ell)$, we can also take the limit of infinite $m$ and $\ell$ to predict its value at irrational values of ${\rm arcos} (\Delta)/\pi$. This limit gives 
\begin{equation}\label{eq:drudexxz2}
{\mathsf D}_{ss} \Big|_{\Delta = \cos (\pi x)} = \frac{\beta}{24} \sqrt{1 - \Delta^2} ,
\end{equation}
with $x$ being an irrational number within the interval $(0,1)$ (notice that this excludes the point $\Delta=0$, where instead equation \eqref{eq:drudexxz} should be used). 
Therefore the spin Drude weight takes the above value for almost all values of $\Delta$, expect for a zero-measure set of points. 
\subsubsection{Spin diffusion in the XXZ chain and comparison with Sachdev-Damle semi-classical theory}
The Drude weights of the majority of conserved quantities in integrable models are finite, namely charge transport is in a first approximation usually ballistic and their diffusive spreading is only a finite-time correction in most cases. However, there are settings where the ballistic channel can be switched off, leaving the diffusive spreading as the leading transport behaviour at late times. This is the case of the charges associated to the generators of the global symmetry in integrable systems whenever the reference states are Gibbs states with particle-hole symmetry. In particular, in the XXZ spin-1/2 chain \eqref{eq:XXZchain},
whenever $\Delta>1$ and magnetic field is zero $h=0$, the spin charge $S^z_j$ is transported diffusely with a diffusion constant that depends non-trivially on $\Delta$ and temperature $\beta^{-1}$, as confirmed by numerical simulations \cite{PhysRevB.90.155104,PhysRevB.91.115130,Ljubotina2017}. This can be obtained by computing the Onsager matrix element for the spin charge, by employing the equation \eqref{eq:DCgeneralmain}. Note that in thermal states at half filling $h=0$ the spin-flip symmetry is present. Since magnetisation is the only conserved charge which changes the sign if the spin-flip transformation is performed, the associated matrix elements of diffusion and susceptibility matrix factorize, resulting in
\begin{equation}
    \mathfrak{L}_{\rm spin,\rm spin} = \mathfrak{D}_{\rm spin, \rm spin} \chi_{\rm spin}
\end{equation}
where $ \chi_{\rm spin}= C_{\rm spin, \rm spin}$. If we specialize to the spin in equation \eqref{eq:DCgeneralmain}, we find therefore the dressed magnetisation $h_{\rm spin}^{\rm dr} = m^{\rm dr} = (1- T n)^{-1} m$, where the bare magnetisation is a constant function in the rapidities that only depend on the length of the string, namely $m_k(\theta) = k$. Its dressed version is instead a function that vanishes at zero magnetic field. This can be understood as a form of effective dressing: the quasiparticles of the model at finite temperature are effectively para-magnetic, they do not retain any magnetization in the absence of an externally applied magnetic field due to interactions among them. However the matrix element 
$\mathfrak{L}_{\rm spin,\rm spin}$ is not zero at zero magnetic limit, but a careful limit of large string number $k \to \infty$ has to be taken in order to find the correct result in the limit of zero external magnetic field. In this limit numerous simplifications occur, one of them dubbed as "magic formula" \cite{PhysRevLett.123.186601}, leading to the spin diffusion constant that reads 
\begin{equation}
\mathfrak{D}_{\rm spin, \rm spin}  =  \sum_{s \geq 1} \int {\dd\theta}{}   \left| v^{\rm eff}_s(\theta)\right|  \frac{\chi_s(\theta)}{16 \chi_{\rm spin}^2}\left. \partial^2_h (m^{\rm dr}_s)^2\right|_{h=0},
\end{equation}
where $s$ denotes the sum over all the string types and $\chi_s(\theta) = \rho_{\rm p}{}_{s}(\theta) (1- n_s(\theta))$ the quasiparticle susceptibilities. The magic formula is a non-trivial relation between thermodynamic functions, but it comes quite natural when looking at diffusion from the perspective of hydrodynamic projection, as one of the normal modes in the product should be magnetization due to the spin flip asymmetry of the current. The dressed magnetisation is computed at finite $h$, and in the limit $h \to 0$ we obtain
\begin{equation}
\left. \partial^2_h (m^{\rm dr}_s)^2\right|_{h=0}  = 2 \lim_{h \to 0} ( \frac{m^{\rm dr}_s}{h})^2
\end{equation}

In the limiting case of Gibbs ensemble with infinite temperature $\beta \to 0$ this function, as well as all the other thermodynamic functions, can be expressed analytically, leading to the following expression for the diffusion constant \cite{gv_superdiffusion} 
\begin{equation}
\lim_{\beta \to 0}D_{\rm spin}  = \frac{2 \sinh \eta}{9 \pi} \sum_{s \geq 1} (1+s) \left( \frac{s+2}{\sinh \eta s} - \frac{s}{\sinh \eta (s+2)} \right)
\end{equation}
where $\Delta = \cosh \eta$. 
In the low-temperature limit $\beta \to \infty$, instead, one finds a diverging diffusion constant due to the exponential vanishing of the spin susceptibility $\chi_{\rm spin}$ \cite{PhysRevLett.123.186601}
\begin{equation}
\lim_{\beta \to \infty} D_{\rm spin}  = \frac{1}{2} \frac{\partial v_1^{\rm eff}}{\partial k} \Big|_{k=0}
e^{\beta \mathfrak{m } } \left( 1 + \mathcal{O}(e^{- \beta \mathfrak{m }}) \right) \end{equation}
and where the spectral gap of the chain $\mathfrak{m}$ is given in terms of $\Delta$ by the equation \cite{Takahashi1999}
\begin{equation}
\mathfrak{m} =   \sinh{(\eta)} \sum_{k \geq 0} \frac{(-1)^k}{\cosh{(k \eta)}},
\end{equation}
and where $v_1^{\rm eff}(k)$ is the velocity of string-1 quasiparticles, which coincide with the spinon velocity in the limit of low temperatures \cite{Takahashi1999}. This expression for the diffusion constant coincide with the one obtained two decades ago by Sachdev and Damle \cite{PhysRevB.57.8307} by a different, semi-classical approach, which is here therefore validated in the small temperature regime. The semi-classical approach, however,  gives incorrect prediction in the massless limit $\mathfrak{m}  \to 0$, namely for the isotropic Heisenberg chain $\Delta \to 1$ \cite{PhysRevLett.123.186601}. While the semi-classical theory predicts normal diffusion, it is now well established that spin transport is in fact superdiffusive, as discussed in more details in a complementary review \cite{2103.01976}.

\subsection{Diffusion in quasiparticle space and scrambling of momenta}\label{subsec:diffusionscrambling}

As another example we consider the dynamics of an initially inhomogeneous state (here in the Lieb-Liniger gas model) and show the different features brought by the ballistic and diffusive dynamics in the hydrodynamic evolution. Specifically we will see that whereas the ballistic evolution causes the fast spatial homogenization of the initial state, the diffusive dynamics is responsible for the redistribution and the smoothening of the particles rapidities.

The Lieb-Liniger model~\cite{1963_Lieb_PR_130_1} describes a gas of non-relativistic bosons by the following Hamiltonian (with $\hbar = 2m = 1$)
\begin{equation}
  H_{\rm LL} = - \sum_{j=1}^N \partial_x^2 + 2c \sum_{i<j}^N \delta(x_i - x_j),
\end{equation}
here written for a finite number of particles $N$. The quasi-particle picture of the Lieb-Liniger model, in the repulsive case $c>0$, consists of a single particle type. Its bare momentum and energy are $p(\theta) = \theta$ and $e(\theta) = \theta^2$ and the scattering kernel is
\begin{equation}
  T(\theta, \theta') = \frac{1}{\pi} \frac{c}{c^2 + (\theta - \theta')^2}.
\end{equation}
The thermodynamics of the model follow then the standard TBA structure presented in the previous sections. 

The generalized hydrodynamics describes the evolution of the particles density $\rho_{\rm p}(\theta; x,t)$, including diffusive spreading, through the following equation~\cite{PhysRevLett.121.160603,De_Nardis_2019},
\begin{equation}
    \partial_t \rho_{\rm p} + \partial_x (v^{\rm eff} \rho_{\rm p}) = \frac{1}{2} \partial_x \left( \mathfrak{D} \partial_x \rho_{\rm p} \right), \label{GHD_eq}
\end{equation}
with the diffusion operator in rapidity space given as 
\begin{equation}
    \mathfrak{D} = (1 - n T)^{-1} \rho_s \widetilde{{\mathfrak{D}}} \rho_s^{-1} (1 - n T),
\end{equation}
and where $\widetilde{{\mathfrak{D}}}$ is defined in~\eqref{eq:dtildemain}. The GHD equation~\eqref{GHD_eq} as stated is an equation for the particle density. Equivalently, it can stated as an equation for the filling function $n(\theta; x,t)$. One of the main features of diffusive terms is to smoothen out the microscopic features of the fluid, namely to increase the entropy of each fluid cell $s(x,t)$. Clearly the total entropy of the system is not increasing, so the hydrodynamic entropy increase signals that the hydrodynamics description is neglecting fine microscopic structure.
The GHD equation predicts an entropy production according to \cite{De_Nardis_2019}
\begin{equation}
    \frac{\dd S(t)}{\dd t} = \frac{\dd }{\dd t} \int \dd x \ s(x,t) = \frac{1}{2} \int {\rm d}x \,[ \Sigma \cdot \mathfrak{D}  \mathsf C \Sigma    ]
\end{equation}
where the function $\Sigma = (1- T n)[\partial_x n/n(1-n)]$. Here, the local entropy of each fluid cell is defined according to the Yang-Yang equation
\begin{equation}
	s(x,t) = - \int {\rm d}\theta \rho_s(\theta; x,t) g(n(\theta; x,t)), \label{entropy_local}
\end{equation}
with $g(n) = n \log n + (1 - n) \log(1 - n)$.

\begin{figure}
  \center
  \includegraphics[width=7cm]{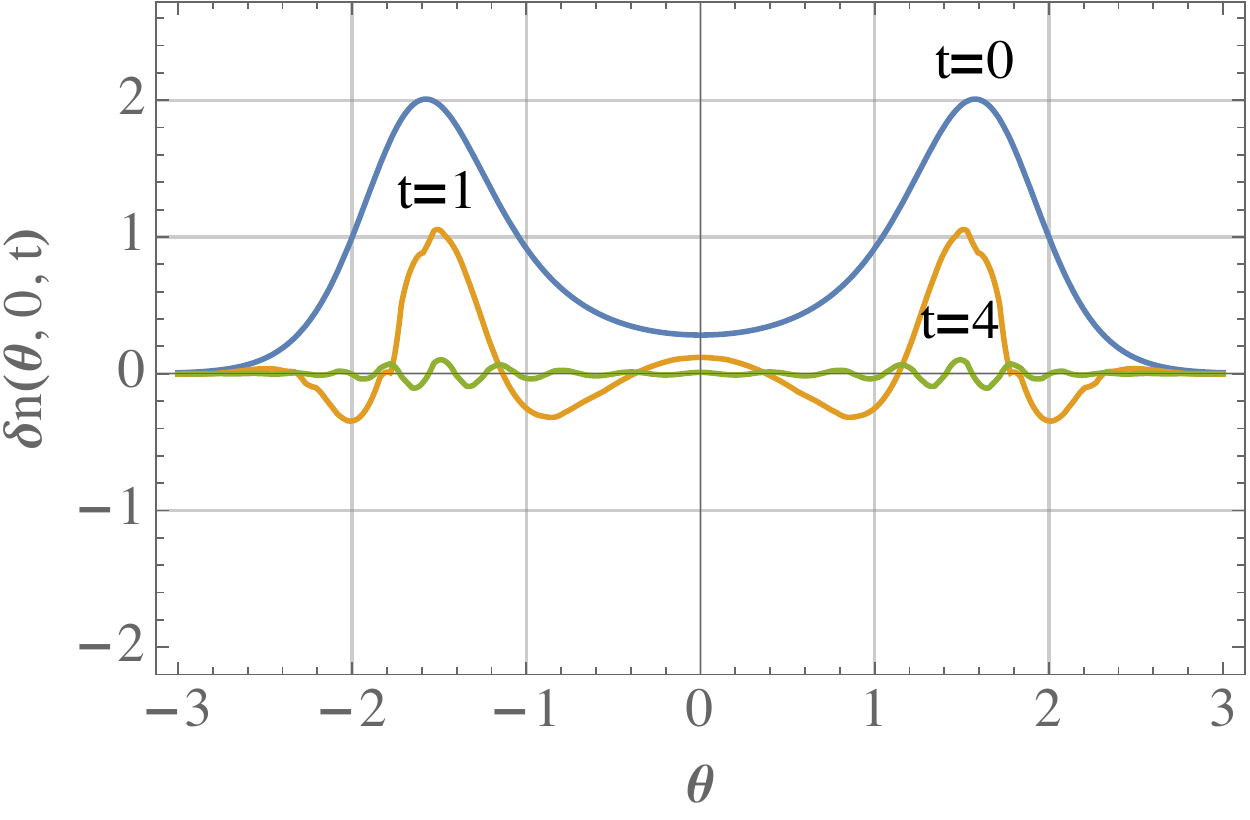}~
  \includegraphics[width=7cm]{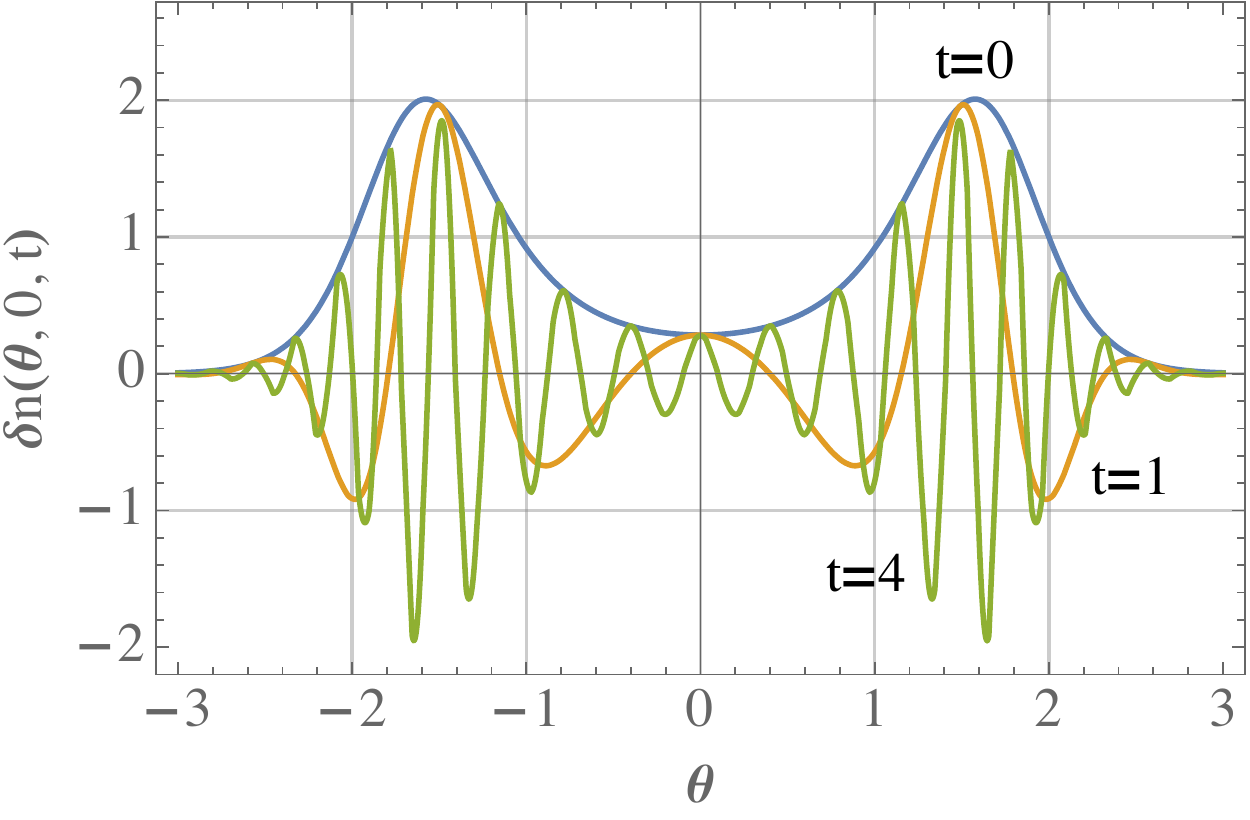}
  \caption{Time evolution of the local filling function $\delta n(\theta, 0, t)$ with (left panel) and without (right panel) the diffusion term in the linearized GHD, as in Eq. \eqref{linear_GHD}. Figure reproduced from~\cite{10.21468/SciPostPhysCore.1.1.002}.}
  \label{fig:rapidities}
\end{figure}

We shall now consider a linear approximation to~\eqref{GHD_eq}, in which we assume that the filling function is close to a space-time independent equilibrium $\underline{n}(\theta)$,
\begin{equation}
    n(\theta; x,t) = \underline{n}(\theta) + \delta n(\theta; x,t). 
\end{equation}
Then, at the linear order in $\delta n$, the hydrodynamic equation reads~\cite{De_Nardis_2019}
\begin{equation}
    \partial_t \delta n(\theta; x,t) + v_{\underline{n}}^{\rm eff} \partial_x \delta n(\theta; x,t) = \frac{1}{2} \int \widetilde{{\mathfrak{D}}}(\theta, \alpha) \partial_x^2 \delta n(\alpha, t, x). \label{linear_GHD}
\end{equation}
This is now a linear integro-differential equation that can be efficiently solved by a mixture of analytical and numerical methods. First, the equation can be transformed to a Fourier space dual to the spatial variable $x$. Second, the resulting equation can be solved by numerically diagonalizing the remaining integral operator. This way of solving guarantees a numerically stable solution valid even at large times. For more details we refer to~\cite{10.21468/SciPostPhysCore.1.1.002}. The entropy production in the linearized regime, and due to the diffusion we have
\begin{equation}
	\partial_t S_{\rm NS}(t) = \frac{1}{2} \int {\rm d}x \int{\rm d\theta}\,{\rm d\alpha} \frac{\partial_x \delta n(\theta; x,t)}{\underline{n}(\theta) (1 - \underline{n}(\theta))} \underline{\rho}_{\rm t}(\theta) \widetilde{{\mathfrak{D}}}(\theta, \alpha) \partial_x \delta n(\alpha, t, x). \label{entropy_production}
\end{equation}
We will analyze the dynamics resulting from~\eqref{linear_GHD} for a simple case of an initial perturbation
\begin{equation}
    \delta n(\theta, x, 0) = \epsilon \cos(k x) \cdot \underline{n}(\theta) ( 1 - \underline{n}(\theta)),
\end{equation}
over the background state $\underline{n}(\theta)$ being a thermal state with $T = 1$, and we are following the dynamics of the filling function $\delta n(\theta, 0, t)$. The results are shown in fig.~\ref{fig:rapidities}.
We observe that in the absence of diffusion $\delta n(\theta, 0, t)$ approaches $0$ in the averaged sense. As a function of $\theta$ it starts to quickly oscillate and averaging over a small window yields zero. On the other hand, the presence of diffusion leads to scattering processes and redistribution of the rapidities that causes an actual, and quick, decay of their density, see fig. \ref{fig:rapidities}. 

\begin{figure}
	\center
	\includegraphics[width=7cm]{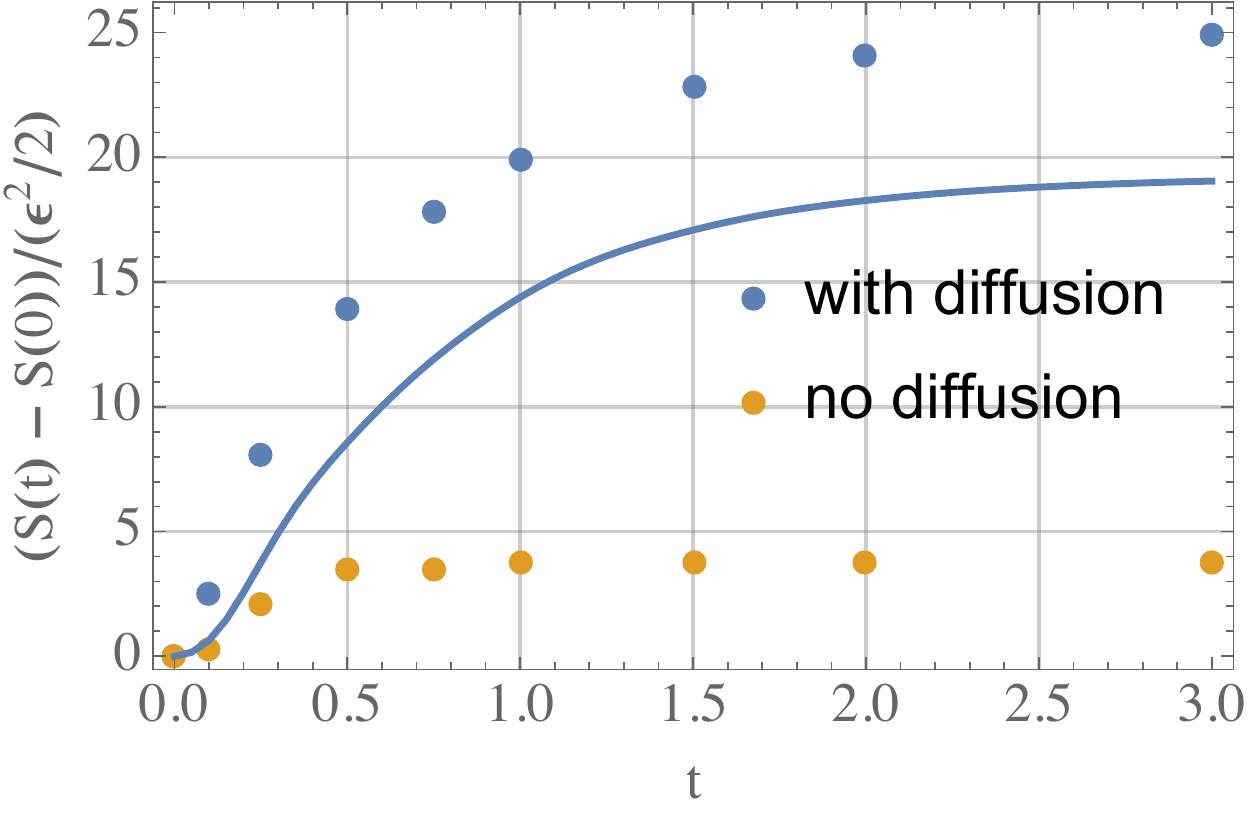}
	\caption{We plot the difference in the entropy between the state at time $t$ and the initial state. The dots corresponds to evaluating the entropy from the definition~\eqref{entropy_local} with the help of the numerical solution considered here. The solid line is computed with the GHD prediction for the entropy production~\eqref{entropy_production}. The discrepancy between the GHD predictions and observed entropy originates from linearization of the full dynamics. Figure reproduced from~\cite{10.21468/SciPostPhysCore.1.1.002}.}
	\label{fig:entropy}
\end{figure}

The difference in these two mechanism can be traced back to the phenomena of entropy production. As explained before, ballistic hydrodynamics conserves entropy and diffusive spreading is necessary to observe entropy production. This phenomenology slightly changes in the linearized hydrodynamics regime. An artefact of the linearization is the presence of entropy production already at the ballistic level, due to higher order terms that are neglected within the linearisation. In order to see this let us recall that at the Euler level the entropy is conserved because the local entropy $s(x,t)$ obeys a continuity equation
\begin{equation}
    \partial_t s(x,t) + \partial_x j_s^{(1)}(x,t) = 0, \label{entropy_continuity}
\end{equation}
where the Euler entropy current is 
\begin{equation}
    j_s^{(1)}(x,t) = \int {\rm d}\theta v^{\rm eff} \rho_s g(n).
\end{equation}
The continuity equation holds by the virtue of the GHD equations~\eqref{GHD_eq}. Indeed, introducing $G(n) = g(n)/n$, we have
\begin{align}
    \partial_t s(x,t) + \partial_x j_s^{(1)}(x,t) &= \int {\rm d}\theta \left(\partial_t(\rho_p G(n)) + \partial_x (v^{\rm eff} \rho_p G(n) \right) \nonumber \\
    &= \int {\rm d}\theta \left[ G(n) \left(\partial_t \rho_p + \partial_x (v^{\rm eff} \rho_p) \right) + \frac{dG}{dn} \left(\partial_t n + v^{\rm eff}\partial_x n \right)\right].
\end{align}
The two terms vanish when the full GHD equations are fulfilled. However, in the case of linearized GHD, the equations for $\rho_p$ and $n$ hold only at the linear level and
\begin{align} 
    \partial_t s(x,t) + \partial_x j_s^{(1)}(x,t)= \int {\rm d}\theta \left[ G(\underline{n}) \partial_x \left( \delta v^{\rm eff} \delta \rho_p \right) + \frac{dG}{d\underline{n}} \delta v^{\rm eff}\partial_x \delta n \right] + \mathcal{O}((\delta n)^3)
\end{align}
The first term is a total spatial derivative so it can be understood as a correction to a current $j_s^{(1)}$ and as such does not change the entropy globally.  Instead, the second term leads to entropy production according to
\begin{equation}
    \partial_t S_{\rm E}(t) = \int {\rm d}x \int {\rm d}\theta \frac{dG}{d\underline{n}} \delta v^{\rm eff} \partial_x \delta n.
\end{equation}
The total entropy production in the linearized, diffusive hydrodynamics is then a sum of $S_{\rm E}(t)$ coming from the Euler scale and $S_{\rm NS}(t)$ originating from the diffusive Navier-Stokes scale.
Still, the diffusive terms are the ones responsible for most of the change in the total entropy as shown in fig.~\ref{fig:entropy}. Moreover, if we compensate for the entropy production at the ballistic level, the resulting entropy following the linearized dynamics agrees with the general GHD prediction~\eqref{entropy_production}.
This shows that scrambling in the rapidity space is related to the total entropy production in the real space and that both processes are driven by the presence of diffusive terms in the hydrodynamics.

\subsection{Classical Sinh-Gordon field theory}

Although most of the discussion until now has centered around quantum integrable models, as mentioned, GHD and the general hydrodynamic principles apply equally well to classical models. Viewing classical field theories as classical limits of quantum field theories, one must understand the meaning of the quasiparticles at the basis of the TBA formulation, and GHD. and ask what is the classical counterpart of the GHD. In integrable QFT’s, quasiparticles are either quantised field modes above a specific vacuum, or quantum versions of solitonic-like excitations in which field configurations connect two vacua. Both types of quasiparticles have similar qualitative features - they are stable excitations of the field. However, in the classical limit the two acquire different qualities. The solitonic excitations stay as particle-like field configurations where energy is concentrated on a small region, whereas quantised field modes turn into classical radiative modes, with extended energy support \cite{Babelon2003}.

The GHD of solitonic gases is that of gases of particles with specific two-body scatterings, see the review \cite{GennadyEl} in this special issue. However, this is not sufficient, in general, to describe maximal entropy states, such as the usual thermal Gibbs states, in classical field theory. Such states are {\em mixed gases}, containing both solitons and radiative modes. It turns out that, although radiative modes do not have localised energy supports, the resulting GHD formulation has the same form; the TBA structure is very rigid, as reviewed in sec.~\ref{subsec:TBA_hydro}. Physically, classical radiative modes can be seen as un-normalisable scattering waves and can be obtained as limits of wave packets, much like the construction of quantum particles in QFT. The TBA, and GHD, applies equally well to the bosonic and fermionic excitations in quantum theories as to solitonic and radiative modes in classical field theories.

Consider a classical field theory with Hamiltonian $H(\Phi, \Pi)$ with $\Phi(x)$ and $\Pi(x)$ are canonically conjugated fields. If the theory is integrable there exists an infinite set of extensive conserved charges $Q_i[\Phi, \Pi]$. In the following we will assume that they can be expressed through local densities, however these assumptions can probably be relaxed by allowing quasi-local charges. For the local ones we have
\begin{equation}
	Q_i[\Phi, \Pi] = \int {\rm d}x\, q_i(x, t),
\end{equation}
with $q_i(x, t)$ a local functional of fields $\Phi$ and $\Pi$ (a functional supported at $x,t$, involving fields and their derivatives), obeying a continuity equation
\begin{equation}
    \partial_t  q_i(x,t) + \partial_t j_i(x,t) = 0,
\end{equation}
where $j_i(x,t)$ is an associated current.

Statistical averages in GGEs are defined through  Euclidean functional integrals,
\begin{equation}
    \langle o_1(x_1) \cdots o_N(x_N) \rangle = \frac{\int \mathcal{D}\Phi \mathcal{D}\Pi\, o_1(x_1) \cdots o_N(x_N) \,e^{-\sum_i \beta^i Q_i[\Phi, \Pi]}}{\int \mathcal{D}\Phi \mathcal{D}\Pi\, e^{-\sum_i \beta^i Q_i[\Phi, \Pi]}} \label{GGE_avg_ShG}
\end{equation}
where $o_i(x_i)$ are local functionals. In order to evaluate this functional integral, one may in principle use the inverse scattering method. This method provides a map from the elementary fields to the scattering data \cite{Babelon2003}. The scattering data may be seen as parametrising the quasiparticle content of a given configuration of the elementary fields. Here a quasiparticle is characterised by its type $a$ and rapidity $\theta$, and falls into the two aforementioned categories: $a$ is either solitonic or radiative. A generic solution to the equations of motion is specified in terms of a finite sets of solitons $\{\theta_l,a_l\}$ ($a_l$ solitonic for all $l$) and continuum of radiative modes with distributions $\rho_{\rm p}(\theta,a)$ (for all $a$ radiative). The conserved charges consists then of contributions from both of them (see e.g.~the discussion in \cite{Bastianello_2018})
\begin{equation}
    	Q_i[\Phi, \Pi] = \sum_l h_i( \theta_l,a_l) + \sum_{a:\ {\rm radiative}} \int \dd\theta\, \rho_{\rm p}(\theta, a) h_i(\theta, a).
\end{equation}
The index $a$ for solitonic modes is not necessarily discrete; there may be more continuous parameters than the rapidity $\theta$ characterising the soliton. From this characterisation of the charges, the evaluation of GGE averages \eqref{GGE_avg_ShG} may in principle proceed. See for instance \cite{DeLuca2016}.

As outlined in sec.~\ref{subsec:TBA_hydro}, although there is no model-independent proof, the result appears to invariably take the general TBA form, and the construction of the GHD follows through. The function $\mathsf F(\epsilon)$ takes one of the two forms
\begin{equation}
    \mathsf{F}_a(\epsilon) = 
        \begin{cases}
            -e^{-\epsilon} \quad &(\textrm{$a$ is a solitonic mode}), \\
            \log \epsilon \quad &(\textrm{$a$ is a radiative mode}).
        \end{cases}
\end{equation}
and consequently, the distribution function $n(\theta, a)$ is
\begin{equation}
    n(\theta, a) = 
        \begin{cases}
            e^{-\epsilon(\theta, a )} \quad &(\textrm{$a$ is a solitonic mode}), \\
            \frac{1}{\epsilon(\theta, a)} \quad &(\textrm{$a$ is a radiative mode}).
        \end{cases}
\end{equation}

In particular, one of the most nontrivial prediction of GHD, the Euler-scale correlation functions Eq.~\eqref{eq:So1o2integrable}, gives
\begin{equation} \label{2pt_Euler_ShG}
    \lim_{\rm Eul}S_{o_1,o_2}(x,t)
    = t^{-1}\sum_{(\theta,a)\in\Theta_*(\xi)} \frac{\rho_{\rm p}(\theta,a) f(\theta,a)}{|(v^{\rm eff})'(\theta,a)|} V^{o_1}(\theta,a)V^{o_2}(\theta,a).
\end{equation}
Here $\Theta_*(\xi) = \{(\theta,a): v^{\rm eff}(\theta,a) = \xi\}$ and is $f(\theta,a)$ the statistical factor~\eqref{eq:qpStatistics2},
\begin{equation}
    f(\theta, a) = 
        \begin{cases}
            1 \quad &(\textrm{$a$ is a solitonic mode}), \\
            n(\theta, a) \quad &(\textrm{$a$ is a radiative mode}),
        \end{cases}
\end{equation}
and $V^o$ is obtained from the knowledge of the GGE average of the observable $o(x,t)$ by \eqref{eq:Vohydro}. To our knowledge, there is no independent derivation of \eqref{2pt_Euler_ShG} by classical integrability techniques.

In order to illustrate these concepts, the classical sinh-Gordon field theory was considered in \cite{Bastianello_2018}. The Sinh-Gordon theory provides a simple setup to explore the hydrodynamics of radiative modes, as solitonic modes are absent in it: it admits a single quasiparticle, of radiative type. Importantly, this allows for an explicit verification of the GHD predictions against a numerical experiment: Monte Carlo simulations of the GGE averages~\eqref{GGE_avg_ShG} combined with a numerical solution of the classical field equations.

The model is described by a single scalar field $\Phi$ with the following Lagrangian,
\begin{equation}
    \mathcal{L}_{\rm ShG} = \frac{1}{2}\partial_{\mu} \Phi \partial^{\mu} \Phi - \frac{m^2}{g^2}\left(\cosh(g\Phi) - 1\right),
\end{equation}
The scattering kernel for the quasiparticle is \cite{DeLuca2016}
\begin{equation}
    \varphi(\theta) = \frac{g^2}{4} \frac{\cosh(\theta)}{\sinh^2(\theta)}.
\end{equation}
and in \cite{Bastianello_2018}, two correlators were considered: the two-point function of stress-energy tensor traces
\begin{equation}
    \langle T^{\mu}_{\,\,\mu}(x, t) T^{\nu}_{\,\,\nu}(0,0) \rangle^{c},
\end{equation}
and the symmetrized two-point function of vertex operators
\begin{equation}
    \langle \cosh(k g\Phi(x, t)) \cosh(k g \Phi(0,0)) \rangle^{c}.
\end{equation}
It is worth mentioning that classical field theories of radiative modes suffer from UV divergences familiar from the classical treatment of the black-body radiation. For example, averages of local densities and currents of energy are diverging for thermal states in the sinh-Gordon model. However the two-point functions of the trace of the stress-energy tensor and of the vertex operators are UV finite. For more details on this issue we refer to \cite{Bastianello_2018}.

\begin{figure}
    \centering
    \includegraphics[width=14cm]{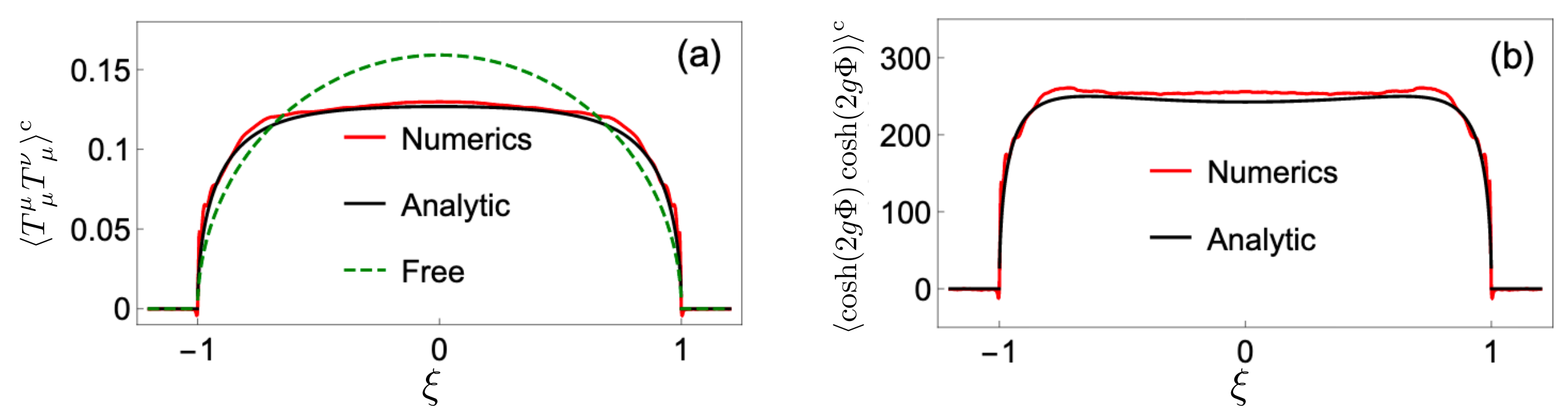}
    \caption{Curves taken from \cite{Bastianello_2018}. Euler scaling limit of two-point functions of (a) stress-energy tensor traces, and (b) vertex operators, as function of the ray $\xi=x/t$. The red curve is the result of the Monte Carlo numerical evaluation, and the black curve is the GHD prediction \eqref{2pt_Euler_ShG}. In (a), comparison is made against the free boson case, in order to show that interactions play an important role and are correctly accounted for by hydrodynamics.}
    \label{fig:ShG_2pt}
\end{figure}

As discussed in sec.~\ref{subsec:einstein} and~\ref{subsec:ffEuler}, in order to relate the microscopic correlation functions with hydrodynamic predictions, fluid-cell averaging may be necessary. It turns out that, for the correlation functions considered here, it is sufficient to average the quantity $t S_{o_1,o_2}(x,t)$, expected to stay bounded and nonzero at large $t$, along the rays with $\xi=x/t$ fixed:
\begin{equation}
    \lim_{\rm Eul} S_{o_1,o_2}(x,t) = t^{-1} \lim_{u \rightarrow \infty} \left( \frac{1}{u} \int_0^{u} \dd s\,s \langle o_1(s \xi,s) o_2(0,0) \rangle^{\rm c}  \right) \quad (\xi = x/t).
\end{equation}
The prediction is \eqref{2pt_Euler_ShG}, with
\begin{align}
    V^{T^{\mu}_{\,\,\mu}}(\theta) &= m \left[\cosh^{\rm dr}(\theta)\left(1 - (v^{\rm eff}(\theta))^2\right) \right], \label{V_T_ShG}\\
    V^{\cosh(kg\Phi)}(\theta) &= \frac{g^2}{4\pi} \mathcal{V}^{k} \sum_{l=0}^{k-1} \frac{\mathcal{V}^l}{\mathcal{V}^{l+1}}(2l+1) n(\theta) \frac{p^l(\theta) d^l(\theta)}{\rho_{\rm p}(\theta)}, \label{V_vertex_ShG}
\end{align}
where $\mathcal{V}^k = \ave{ e^{kg\Phi(x,t)}}$ and 
\begin{equation}
    d^l(\theta) = e^{\theta} + \frac{g^2}{4} \mathcal{P} \int \frac{\dd \gamma}{2\pi}\frac{1}{\sinh(\theta - \gamma)} \left(-2l - \partial_\gamma \right) (n(\gamma) d^l(\gamma)).
\end{equation}
The function $V^{T^{\mu}_{\,\,\mu}}(\theta)$ in~\eqref{V_T_ShG} follows simply from the fact that $T^{\mu}_{\,\,\nu}$ is a tensor of conserved densities and currents, with $V^{ q_i} = 2\pi h_i^{\rm dr}(\theta)$ and $V^{j_i} = 2\pi v^{\rm eff}(\theta) h_i^{\rm dr}(\theta)$. The computation of $V^{e^{kg\Phi}}(\theta)$ is more involved and relies on Eq.~\eqref{eq:Vohydro}, along with the knowledge of the expectation values $\langle e^{kg\Phi(x,t)}\rangle$ in a GGE state, computed in the quantum sinh-Gordon model in \cite{NEGRO2013166,doi:10.1142/S0217751X14501115} and in the classical limit in \cite{Bastianello_2018}. fig.~\ref{fig:ShG_2pt} shows the comparison between the results of the Monte Carlo simulations and the GHD predictions for the large-time behavior of these two-point functions, with an excellent agreement.

\section{Summary and outlook}

There is no doubt that generalized hydrodynamics represents a great leap in our understanding of dynamical phenomena in low-dimensional many-body systems.
We focused on large scale behaviour of correlation functions in the scope of GHD. On one hand this allowed us to discern transport and hydrodynamic properties of integrable systems, and on the other hand, it provided a platform where new methods, which do not rely on the integrability structure, were developed.

We here discussed how the relevant degrees of freedom that affect the hydrodynamics of integrable systems can be identified through complementary approaches, such as hydrodynamic projections, and thermodynamic form factors, the latter obtained either by considering finite size properties or in the scope of the thermodynamic bootstrap program. This effort resulted in explicit expressions for the linear response transport coefficients, which allow us to obtain numerical solutions without any finite size or finite time effects, and, in some cases even admit closed form solutions.

These advances played a major role in addressing a number of open questions. Notably, they presented a firm evidence that the spin Drude weight in the Heisenberg spin chain is a nowhere continuous function of the anisotropy parameter in the Hamiltonian even at finite temperatures. New results also provided a platform where the older approaches, such as the  semi-classical theory of diffusion at low temperatures and holography, could be put to the test. It also allowed us to better understand one of the major questions of statistical physics and thermodynamics, which is how can the entropy production emerge from the microscopically reversible dynamics on large enough scales.  Importantly, the framework can be used to treat quantum and classical systems on the same footing, as exemplified with the classical Sinh-Gordon field theory.

Several open questions constitute exciting new directions for further developments:
From a general perspective the question remains as to what quantities, beyond the quadratic charges, constitute the diffusive Hilbert space, in order to provide bounds on diffusion constants. While in integrable systems we argued that a complete set of such quantities corresponds to the products of quasilocal charges this does not seem to be the case in generic systems with few conservation laws. Although, as we reviewed, these charges usually provide a lower bound on momentum diffusion, in typical quantum chains with space reversal symmetry, without momentum conservation, one can argue that contributions to the Onsager matrix from the products of conservation laws will vanish. This question becomes even more pressing beyond one dimension, where quadratic charges do not provide appropriate bounds. In this case some indications might be obtained from the bounds on diffusion in holography \cite{PhysRevLett.87.081601}.

Moreover it remains to be understood how to go beyond diffusive effects, namely how to compute dynamical correlators at small but finite frequency and momentum. It seems that the approach layed here can constitute a solid basis from where to extend the result, as suggested in \cite{hydroprojectionsEuler} in the context of hydrodynamic projections at the Euler scale. Moreover, recently the effects of large-scale correlations on transport were studied in \cite{jin2021interplay}, arguing that correlations in the initial state should suppress transport when considering the bi-partition protocol. In view of this result a natural question is what effects, if any, do the long-range correlations, i.e. relaxing the ergodicity assumption \eqref{clustering}, have on the linear response transport coefficients.

Finally, in this review we considered setups where relaxation to a stationary state, which is homogeneous in space and time occurs within the hydrodynamic cells, or, if this is not the case appropriately space and time averaging is assumed. Nevertheless, in certain systems maximum entropy states that break the space and time translation symmetry can exist as a consequence of dynamical symmetries \cite{PhysRevB.102.041117,10.21468/SciPostPhys.9.1.003,buca2020quantum}. From a generic standpoint the Euler scale hydrodynamics of such systems in the linear response regime can be understood on a similar footing \cite{10.21468/SciPostPhys.9.1.003,hydroprojectionsEuler}, however at the moment it is not clear what role dynamical symmetries play in concrete examples of integrable systems, such as the Heisenberg chain at the roots of unity.

The concept of the thermodynamic form-factors provided important inputs to development of generalized hydrodynamics and both approaches greatly benefited from fruitful exchange of ideas. Still there are many open questions regarding thermodynamic form-factor approach to dynamic correlation functions. Some of them, like generalization of the bootstrap approach to theories with bound states or predictions for the Euler scale of higher point functions   concern directly hydrodynamic regimes. We also expect future parallel developments in areas where both approaches have not yet been fully addressed. For example, this is the case for correlation functions of semi-local operators (or twist fields). Other questions, of a more fundamental nature, concern the mathematically rigorous definition of the thermodynamic form-factors and execution of the soft-modes summations.

\section*{Acknowledgements} 
We are grateful to many colleagues for collaborations on topics directly connected to this review and for inspiring discussions over the years. Special thanks go
to Denis Bernard, Bruno Bertini, Vir Bulchandani, Olalla Castro Alvaredo, Jean-S\'ebastien Caux, Mario Collura,  Axel Cort{\'e}s Cubero, J\'er\^{o}me Dubail,  Joseph Durnin, Maurizio Fagotti, Sarang Gopalakrishnan, Enej Ilievski, Giuseppe Policastro,
Toma\v{z} Prosen, Romain Vasseur, and Takato Yoshimura. 
\section*{References}
\bibliographystyle{ieeetr.bst}
\bibliography{TC}

\begin{thebibliography}{100}

\bibitem{PhysRevLett.122.090601}
M.~Schemmer, I.~Bouchoule, B.~Doyon, and J.~Dubail, ``Generalized hydrodynamics
  on an atom chip,'' {\em Phys. Rev. Lett.}, vol.~122, p.~090601, Mar 2019.

\bibitem{malvania2020generalized}
N.~Malvania, Y.~Zhang, Y.~Le, J.~Dubail, M.~Rigol, and D.~S. Weiss,
  ``{Generalized hydrodynamics in strongly interacting 1D Bose gases},'' 2020.

\bibitem{scheie2021}
A.~Scheie, N.~E. Sherman, M.~Dupont, S.~E. Nagler, M.~B. Stone, G.~E. Granroth,
  J.~E. Moore, and D.~A. Tennant, ``Detection of
  {Kardar{\textendash}Parisi{\textendash}Zhang} hydrodynamics in a quantum
  {Heisenberg} spin-1/2 chain,'' {\em Nature Physics}, Mar. 2021.

\bibitem{Calabrese2005}
P.~Calabrese and J.~Cardy, ``Evolution of entanglement entropy in
  one-dimensional systems,'' {\em Journal of Statistical Mechanics: Theory and
  Experiment}, vol.~2005, p.~P04010, Apr. 2005.

\bibitem{Pasquale-ed}
P.~Calabrese, ``Quantum integrability in out-of-equilibrium systems,'' {\em J.
  Stat. Mech. Theory Exp}, p.~064001, 2016.

\bibitem{Eisert2015}
J.~Eisert, M.~Friesdorf, and C.~Gogolin, ``Quantum many-body systems out of
  equilibrium,'' {\em Nature Physics}, vol.~11, pp.~124--130, Feb. 2015.

\bibitem{Heisenberg1928}
W.~Heisenberg, ``Zur theorie des ferromagnetismus,'' {\em Zeitschrift f{\"u}r
  Physik}, vol.~49, pp.~619--636, Sept. 1928.

\bibitem{Essler2005_Hubbard}
F.~H.~L. Essler, H.~Frahm, F.~G\"{o}hmann, A.~Kl\"{u}mper, and V.~E. Korepin,
  {\em The One-Dimensional Hubbard Model}.
\newblock Cambridge University Press, Feb. 2005.

\bibitem{PhysRev.130.1605}
E.~H. Lieb and W.~Liniger, ``{Exact Analysis of an Interacting Bose Gas. I. The
  General Solution and the Ground State},'' {\em Phys. Rev.}, vol.~130,
  pp.~1605--1616, May 1963.

\bibitem{Bethe1931}
H.~Bethe, ``Zur theorie der metalle,'' {\em Zeitschrift f{\"u}r Physik},
  vol.~71, pp.~205--226, Mar. 1931.

\bibitem{hep-th/9605187}
L.~D. Faddeev, ``How algebraic {Bethe} ansatz works for integrable model,''
  1996.

\bibitem{Yang1969}
C.~N. Yang and C.~P. Yang, ``Thermodynamics of a one-dimensional system of
  bosons with repulsive delta-function interaction,'' {\em Journal of
  Mathematical Physics}, vol.~10, pp.~1115--1122, July 1969.

\bibitem{Takahashi1999}
M.~Takahashi, {\em Thermodynamics of One-Dimensional Solvable Models}.
\newblock Cambridge University Press, Mar. 1999.

\bibitem{Delfino2001}
G.~Delfino, ``One-point functions in integrable quantum field theory at finite
  temperature,'' {\em Journal of Physics A: Mathematical and General}, vol.~34,
  pp.~L161--L168, Mar. 2001.

\bibitem{Kitanine2002}
N.~Kitanine, J.~M. Maillet, N.~A. Slavnov, and V.~Terras, ``Emptiness formation
  probability of {the XXZ spin-$1/2$ Heisenberg} chain at {$\Delta=-1/2$},''
  {\em Journal of Physics A: Mathematical and General}, vol.~35,
  pp.~L385--L388, June 2002.

\bibitem{LeClair1999}
A.~LeClair and G.~Mussardo, ``Finite temperature correlation functions in
  integrable {QFT},'' {\em Nuclear Physics B}, vol.~552, pp.~624--642, July
  1999.

\bibitem{Kormos2010}
M.~Kormos and B.~Pozsgay, ``One-point functions in massive integrable {QFT}
  with boundaries,'' {\em Journal of High Energy Physics}, vol.~2010, Apr.
  2010.

\bibitem{Pozsgay:2010xd}
B.~Pozsgay, ``{Mean values of local operators in highly excited Bethe
  states},'' {\em J. Stat. Mech.}, vol.~1101, p.~P01011, 2011.

\bibitem{Schlappa2012}
J.~Schlappa, K.~Wohlfeld, K.~J. Zhou, M.~Mourigal, M.~W. Haverkort, V.~N.
  Strocov, L.~Hozoi, C.~Monney, S.~Nishimoto, S.~Singh, A.~Revcolevschi, J.-S.
  Caux, L.~Patthey, H.~M. R{\o}nnow, J.~van~den Brink, and T.~Schmitt,
  ``Spin{\textendash}orbital separation in the quasi-one-dimensional {Mott}
  insulator {Sr$_2$CuO$_3$},'' {\em Nature}, vol.~485, pp.~82--85, Apr. 2012.

\bibitem{PhysRevLett.115.085301}
F.~Meinert, M.~Panfil, M.~J. Mark, K.~Lauber, J.-S. Caux, and H.-C. N\"agerl,
  ``{Probing the Excitations of a Lieb-Liniger Gas from Weak to Strong
  Coupling},'' {\em Phys. Rev. Lett.}, vol.~115, p.~085301, Aug 2015.

\bibitem{Schweigler2017}
T.~Schweigler, V.~Kasper, S.~Erne, I.~Mazets, B.~Rauer, F.~Cataldini,
  T.~Langen, T.~Gasenzer, J.~Berges, and J.~Schmiedmayer, ``Experimental
  characterization of a quantum many-body system via higher-order
  correlations,'' {\em Nature}, vol.~545, pp.~323--326, May 2017.

\bibitem{Langen2013}
T.~Langen, R.~Geiger, M.~Kuhnert, B.~Rauer, and J.~Schmiedmayer, ``Local
  emergence of thermal correlations in an isolated quantum many-body system,''
  {\em Nature Physics}, vol.~9, pp.~640--643, Sept. 2013.

\bibitem{PhysRevA.91.043617}
N.~Fabbri, M.~Panfil, D.~Cl\'ement, L.~Fallani, M.~Inguscio, C.~Fort, and J.-S.
  Caux, ``{Dynamical structure factor of one-dimensional Bose gases:
  Experimental signatures of beyond-Luttinger-liquid physics},'' {\em Phys.
  Rev. A}, vol.~91, p.~043617, 2015.

\bibitem{2102.08376}
A.~Scheie, P.~Laurell, A.~M. Samarakoon, B.~Lake, S.~E. Nagler, G.~E. Granroth,
  S.~Okamoto, G.~Alvarez, and D.~A. Tennant, ``Witnessing entanglement in
  quantum magnets using neutron scattering,'' 2021.

\bibitem{2009_Caux_JMP_50}
J.-S. Caux, ``Correlation functions of integrable models: a description of the
  {ABACUS} algorithm,'' {\em J. Math. Phys.}, vol.~50, no.~9, p.~095214, 2009.

\bibitem{PhysRevA.89.033605}
M.~Panfil and J.-S. Caux, ``{Finite-temperature correlations in the
  Lieb-Liniger one-dimensional Bose gas},'' {\em Phys. Rev. A}, vol.~89,
  p.~033605, Mar 2014.

\bibitem{POZSGAY2008209}
B.~Pozsgay and G.~Tak{\'a}cs, ``Form factors in finite volume ii: Disconnected
  terms and finite temperature correlators,'' {\em Nuclear Physics B},
  vol.~788, no.~3, pp.~209 -- 251, 2008.

\bibitem{Pozsgay:2010cr}
B.~Pozsgay and G.~Takacs, ``{Form factor expansion for thermal correlators},''
  {\em J. Stat. Mech.}, vol.~1011, p.~P11012, 2010.

\bibitem{Pozsgay2018}
B.~Pozsgay and I.~Sz{\'e}cs{\'e}nyi, ``{LeClair-Mussardo series for two-point
  functions in Integrable QFT},'' {\em Journal of High Energy Physics},
  vol.~2018, no.~5, p.~170, 2018.

\bibitem{Essler:2009zz}
F.~H.~L. Essler and R.~M. Konik, ``{Finite-temperature dynamical correlations
  in massive integrable quantum field theories},'' {\em J. Stat. Mech.},
  vol.~0909, p.~P09018, 2009.

\bibitem{1742-5468-2012-09-P09001}
N.~Kitanine, K.~K. Kozlowski, J.~M. Maillet, N.~A. Slavnov, and V.~Terras,
  ``Form factor approach to dynamical correlation functions in critical
  models,'' {\em Journal of Statistical Mechanics: Theory and Experiment},
  vol.~2012, no.~09, p.~P09001, 2012.

\bibitem{doi:10.1063/1.5094332}
K.~K. Kozlowski, ``{Long-distance and large-time asymptotic behaviour of
  dynamic correlation functions in the massless regime of the XXZ spin-1/2
  chain},'' {\em Journal of Mathematical Physics}, vol.~60, no.~7, p.~073303,
  2019.

\bibitem{10.21468/SciPostPhys.9.6.082}
E.~Granet and F.~H.~L. Essler, ``{A systematic $1/c$-expansion of form factor
  sums for dynamical correlations in the Lieb-Liniger model},'' {\em SciPost
  Phys.}, vol.~9, p.~82, 2020.

\bibitem{granet2020lowdensity}
E.~Granet, ``{Low-density limit of dynamical correlations in the Lieb-Liniger
  model},'' 2020.

\bibitem{10.21468/SciPostPhys.9.3.033}
E.~Granet, M.~Fagotti, and F.~H.~L. Essler, ``{Finite temperature and quench
  dynamics in the Transverse Field Ising Model from form factor expansions},''
  {\em SciPost Phys.}, vol.~9, p.~33, 2020.

\bibitem{Dugave_2013}
M.~Dugave, F.~Göhmann, and K.~K. Kozlowski, ``{Thermal form factors of the XXZ
  chain and the large-distance asymptotics of its temperature dependent
  correlation functions},'' {\em Journal of Statistical Mechanics: Theory and
  Experiment}, vol.~2013, p.~P07010, Jul 2013.

\bibitem{G_hmann_2017}
F.~Göhmann, M.~Karbach, A.~Klümper, K.~K. Kozlowski, and J.~Suzuki, ``Thermal
  form-factor approach to dynamical correlation functions of integrable lattice
  models,'' {\em Journal of Statistical Mechanics: Theory and Experiment},
  vol.~2017, p.~113106, Nov 2017.

\bibitem{2011.12752}
C.~Babenko, F.~Göhmann, K.~K. Kozlowski, and J.~Suzuki, ``{A thermal form
  factor series for the longitudinal two-point function of the Heisenberg-Ising
  chain in the antiferromagnetic massive regime},'' 2020.

\bibitem{PhysRevX.6.041065}
O.~A. Castro-Alvaredo, B.~Doyon, and T.~Yoshimura, ``Emergent hydrodynamics in
  integrable quantum systems out of equilibrium,'' {\em Phys. Rev. X}, vol.~6,
  p.~041065, Dec 2016.

\bibitem{PhysRevLett.117.207201}
B.~Bertini, M.~Collura, J.~{De Nardis}, and M.~Fagotti, ``{Transport in
  Out-of-Equilibrium $XXZ$ Chains: Exact Profiles of Charges and Currents},''
  {\em Phys. Rev. Lett.}, vol.~117, p.~207201, Nov 2016.

\bibitem{SD97}
S.~Sachdev and K.~Damle, ``{Low Temperature Spin Diffusion in the
  One-Dimensional Quantum O(3) Nonlinear $\sigma$ Model},'' {\em Physical
  Review Letters}, vol.~78, pp.~943--946, feb 1997.

\bibitem{PhysRevB.57.8307}
K.~Damle and S.~Sachdev, ``{Spin dynamics and transport in gapped
  one-dimensional Heisenberg antiferromagnets at nonzero temperatures},'' {\em
  Phys. Rev. B}, vol.~57, pp.~8307--8339, Apr 1998.

\bibitem{PhysRevB.55.11029}
X.~Zotos, F.~Naef, and P.~Prelov{\v{s}}ek, ``{Transport and conservation
  laws},'' {\em Phys. Rev. B}, vol.~55, pp.~11029--11032, May 1997.

\bibitem{PhysRevLett.82.1764}
X.~Zotos, ``{Finite Temperature Drude Weight of the One-Dimensional Spin- $1/2$
  Heisenberg Model},'' {\em Phys. Rev. Lett.}, vol.~82, pp.~1764--1767, Feb
  1999.

\bibitem{PhysRevB.84.155125}
J.~Herbrych, P.~Prelov\ifmmode~\check{s}\else \v{s}\fi{}ek, and X.~Zotos,
  ``{Finite-temperature Drude weight within the anisotropic Heisenberg
  chain},'' {\em Phys. Rev. B}, vol.~84, p.~155125, Oct 2011.

\bibitem{Ilievski2012}
E.~Ilievski and T.~Prosen, ``{Thermodyamic Bounds on Drude Weights in Terms of
  Almost-conserved Quantities},'' {\em Communications in Mathematical Physics},
  vol.~318, pp.~809--830, oct 2012.

\bibitem{1742-5468-2014-9-P09037}
R.~G. Pereira, V.~Pasquier, J.~Sirker, and I.~Affleck, ``{Exactly conserved
  quasilocal operators for the XXZ spin chain},'' {\em Journal of Statistical
  Mechanics: Theory and Experiment}, vol.~2014, no.~9, p.~P09037, 2014.

\bibitem{Doyon_2007}
B.~Doyon, ``Finite-temperature form factors: a review,'' {\em Symmetry,
  Integrability and Geometry: Methods and Applications}, Jan 2007.

\bibitem{DeNardisP2018}
J.~D. Nardis and M.~Panfil, ``{Particle-hole pairs and density-density
  correlations in the Lieb-Liniger model},'' {\em Journal of Statistical
  Mechanics: Theory and Experiment}, vol.~2018, p.~033102, Mar. 2018.

\bibitem{Bootstrap_JHEP}
A.~{Cort{\'e}s Cubero} and M.~{Panfil}, ``{Thermodynamic bootstrap program for
  integrable QFT's: form factors and correlation functions at finite energy
  density},'' {\em Journal of High Energy Physics}, vol.~2019, no.~1, p.~104,
  2019.

\bibitem{Spohn1982}
H.~Spohn, ``Hydrodynamical theory for equilibrium time correlation functions of
  hard rods,'' {\em Ann. Phys.}, vol.~141, pp.~353--364, jul 1982.

\bibitem{Spohn1991}
H.~Spohn, {\em Large Scale Dynamics of Interacting Particles}.
\newblock Springer Berlin Heidelberg, 1991.

\bibitem{PhysRevLett.120.144101}
M.~D. Maiden, D.~V. Anderson, N.~A. Franco, G.~A. El, and M.~A. Hoefer,
  ``Solitonic dispersive hydrodynamics: Theory and observation,'' {\em Phys.
  Rev. Lett.}, vol.~120, p.~144101, Apr 2018.

\bibitem{Medenjak17}
M.~Medenjak, K.~Klobas, and T.~Prosen, ``Diffusion in deterministic interacting
  lattice systems,'' {\em Phys. Rev. Lett.}, vol.~119, p.~110603, Sep 2017.

\bibitem{medenjak2019two}
M.~Medenjak, V.~Popkov, T.~Prosen, E.~Ragoucy, and M.~Vanicat, ``Two-species
  hardcore reversible cellular automaton: matrix ansatz for dynamics and
  nonequilibrium stationary state,'' {\em SciPost physics}, vol.~6, no.~art
  074, 2019.

\bibitem{klobas2018exactly}
K.~Klobas, M.~Medenjak, and T.~Prosen, ``Exactly solvable deterministic lattice
  model of crossover between ballistic and diffusive transport,'' {\em Journal
  of Statistical Mechanics: Theory and Experiment}, vol.~2018, no.~12,
  p.~123202, 2018.

\bibitem{Klobas2019}
K.~Klobas, M.~Medenjak, T.~Prosen, and M.~Vanicat, ``Time-dependent matrix
  product ansatz for interacting reversible dynamics,'' {\em Communications in
  Mathematical Physics}, vol.~371, pp.~651--688, July 2019.

\bibitem{10.21468/SciPostPhysCore.2.2.010}
K.~Klobas and T.~Prosen, ``{Space-like dynamics in a reversible cellular
  automaton},'' {\em SciPost Phys. Core}, vol.~2, p.~10, 2020.

\bibitem{PhysRevLett.123.170603}
A.~J. Friedman, S.~Gopalakrishnan, and R.~Vasseur, ``{Integrable Many-Body
  Quantum Floquet-Thouless Pumps},'' {\em Phys. Rev. Lett.}, vol.~123,
  p.~170603, Oct 2019.

\bibitem{ilievski2016quasilocal}
E.~Ilievski, M.~Medenjak, T.~Prosen, and L.~Zadnik, ``{Quasilocal charges in
  integrable lattice systems},'' {\em Journal of Statistical Mechanics: Theory
  and Experiment}, vol.~2016, p.~064008, jun 2016.

\bibitem{Essler_2015_GGE}
F.~H.~L. Essler, G.~Mussardo, and M.~Panfil, ``{Generalized Gibbs ensembles for
  quantum field theories},'' {\em Physical Review A}, vol.~91, May 2015.

\bibitem{De_Nardis_2019}
J.~D. Nardis, D.~Bernard, and B.~Doyon, ``{Diffusion in generalized
  hydrodynamics and quasiparticle scattering},'' {\em SciPost Phys.}, vol.~6,
  p.~49, 2019.

\bibitem{convexity}
R.~B. Israel, {\em Convexity in the Theory of Lattice Gases}.
\newblock Princeton University Press, 1979.

\bibitem{hydroprojectionsEuler}
B.~Doyon, ``Hydrodynamic projections and the emergence of linearised euler
  equations in one-dimensional isolated systems,'' 2020.

\bibitem{Lieb1972}
E.~H. Lieb and D.~W. Robinson, ``The finite group velocity of quantum spin
  systems,'' {\em Communications in Mathematical Physics}, vol.~28,
  pp.~251--257, Sep 1972.

\bibitem{Vidmar2016}
L.~Vidmar and M.~Rigol, ``{Generalized Gibbs ensemble in integrable lattice
  models},'' {\em Journal of Statistical Mechanics: Theory and Experiment},
  vol.~2016, p.~064007, June 2016.

\bibitem{pseudolocality}
B.~Doyon, ``Thermalization and pseudolocality in extended quantum systems,''
  {\em Communications in Mathematical Physics}, vol.~351, no.~1, pp.~155--200,
  2017.

\bibitem{freeenergyfluxes}
B.~Doyon and J.~Durnin, ``Free energy fluxes and the kubo-martin-schwinger
  relation,'' 2020.

\bibitem{Ben}
B.~Doyon, ``Diffusion and superdiffusion from hydrodynamic projection,'' 2019.

\bibitem{2103.01976}
V.~B. Bulchandani, S.~Gopalakrishnan, and E.~Ilievski, ``Superdiffusion in spin
  chains,'' 2021.

\bibitem{robinson1997operator}
D.~Robinson and O.~Bratteli, ``Operator algebras and quantum statistical
  mechanics 2,'' 1997.

\bibitem{SciPostPhys.2.2.014}
B.~Doyon and T.~Yoshimura, ``{A note on generalized hydrodynamics:
  inhomogeneous fields and other concepts},'' {\em SciPost Phys.}, vol.~2,
  p.~014, 2017.

\bibitem{Spohn_JMP}
H.~Spohn, ``Interacting and noninteracting integrable systems,'' {\em Journal
  of Mathematical Physics}, vol.~59, p.~091402, sep 2018.

\bibitem{Spohn2014}
H.~Spohn, ``Nonlinear fluctuating hydrodynamics for anharmonic chains,'' {\em
  Journal of Statistical Physics}, vol.~154, pp.~1191--1227, feb 2014.

\bibitem{Ilievski2018}
E.~Ilievski, J.~De~Nardis, M.~Medenjak, and T.~Prosen, ``Superdiffusion in
  one-dimensional quantum lattice models,'' {\em Phys. Rev. Lett.}, vol.~121,
  p.~230602, Dec 2018.

\bibitem{1812.02701}
S.~Gopalakrishnan and R.~Vasseur, ``{Kinetic Theory of Spin Diffusion and
  Superdiffusion in $XXZ$ Spin Chains},'' {\em Phys. Rev. Lett.}, vol.~122,
  p.~127202, Mar 2019.

\bibitem{PhysRevLett.123.186601}
J.~De~Nardis, M.~Medenjak, C.~Karrasch, and E.~Ilievski, ``Anomalous spin
  diffusion in one-dimensional antiferromagnets,'' {\em Phys. Rev. Lett.},
  vol.~123, p.~186601, Oct 2019.

\bibitem{PhysRevB.101.041411}
V.~B. Bulchandani, ``{Kardar-Parisi-Zhang universality from soft gauge
  modes},'' {\em Phys. Rev. B}, vol.~101, p.~041411, Jan 2020.

\bibitem{PhysRevLett.124.210605}
J.~De~Nardis, M.~Medenjak, C.~Karrasch, and E.~Ilievski, ``Universality classes
  of spin transport in one-dimensional isotropic magnets: The onset of
  logarithmic anomalies,'' {\em Phys. Rev. Lett.}, vol.~124, p.~210605, May
  2020.

\bibitem{SciPostPhys.3.6.039}
B.~Doyon and H.~Spohn, ``{Drude Weight for the Lieb-Liniger Bose Gas},'' {\em
  SciPost Phys.}, vol.~3, p.~039, 2017.

\bibitem{toth2003onsager}
B.~T{\'o}th and B.~Valk{\'o}, ``Onsager relations and eulerian hydrodynamic
  limit for systems with several conservation laws,'' {\em Journal of
  Statistical Physics}, vol.~112, no.~3, pp.~497--521, 2003.

\bibitem{grisi2011current}
R.~M. Grisi and G.~M. Sch{\"u}tz, ``Current symmetries for particle systems
  with several conservation laws,'' {\em Journal of Statistical Physics},
  vol.~145, no.~6, pp.~1499--1512, 2011.

\bibitem{10.21468/SciPostPhys.6.6.068}
D.~Karevski and G.~M. Schütz, ``{Charge-current correlation equalities for
  quantum systems far from equilibrium},'' {\em SciPost Phys.}, vol.~6, p.~68,
  2019.

\bibitem{10.21468/SciPostPhysLectNotes.18}
B.~Doyon, ``{Lecture Notes On Generalised Hydrodynamics},'' {\em SciPost Phys.
  Lect. Notes}, p.~18, 2020.

\bibitem{Fagotti_2016}
M.~Fagotti, ``Charges and currents in quantum spin chains: late-time dynamics
  and spontaneous currents,'' {\em Journal of Physics A: Mathematical and
  Theoretical}, vol.~50, p.~034005, dec 2016.

\bibitem{doyonFluctuations2020}
B.~Doyon and J.~Myers, ``Fluctuations in ballistic transport from euler
  hydrodynamics,'' {\em Annales Henri Poincar{\'e}}, vol.~21, no.~1,
  pp.~255--302, 2020.

\bibitem{Einstein1905}
A.~Einstein, ``\"{U}ber die von der molekularkinetischen theorie der w\"{a}rme
  geforderte bewegung von in ruhenden fl\"{u}ssigkeiten suspendierten
  teilchen,'' {\em Annalen der Physik}, vol.~322, no.~8, pp.~549--560, 1905.

\bibitem{Rudin1991}
W.~Rudin, {\em Functional Analysis}.
\newblock McGraw-Hill International Editions, 1991.

\bibitem{Mazur1969}
P.~Mazur, ``Non-ergodicity of phase functions in certain systems,'' {\em
  Physica}, vol.~43, pp.~533--545, Sept. 1969.

\bibitem{bertini2020finite}
B.~Bertini, F.~Heidrich-Meisner, C.~Karrasch, T.~Prosen, R.~Steinigeweg, and
  M.~Znidaric, ``Finite-temperature transport in one-dimensional quantum
  lattice models,'' 2020.

\bibitem{KipnisLandimBook}
C.~Kipnis and C.~Landim, {\em Scaling Limits of Interacting Particle Systems}.
\newblock Springer-Verlag Berlin Heidelberg, 1999.

\bibitem{inprepa}
``in preparation.''

\bibitem{Shiraishi_2019}
N.~Shiraishi, ``Proof of the absence of local conserved quantities in the {XYZ}
  chain with a magnetic field,'' {\em {EPL} (Europhysics Letters)}, vol.~128,
  p.~17002, nov 2019.

\bibitem{DurninBhaseenDoyon2020}
J.~Durnin, M.~J. Bhaseen, and B.~Doyon, ``Non-equilibrium dynamics and weakly
  broken integrability,'' 2019.

\bibitem{Prosen_2014}
T.~Prosen, ``Lower bounds on high-temperature diffusion constants from
  quadratically extensive almost-conserved operators,'' {\em Phys. Rev. E},
  vol.~89, p.~012142, Jan 2014.

\bibitem{MKP17}
M.~Medenjak, C.~Karrasch, and T.~Prosen, ``{Lower Bounding Diffusion Constant
  by the Curvature of Drude Weight},'' {\em Phys. Rev. Lett.}, vol.~119,
  p.~080602, Aug 2017.

\bibitem{10.21468/SciPostPhys.9.5.075}
M.~Medenjak, J.~D. Nardis, and T.~Yoshimura, ``{Diffusion from Convection},''
  {\em SciPost Phys.}, vol.~9, p.~75, 2020.

\bibitem{tanikawaExact2021}
Y.~Tanikawa, K.~Takasan, and H.~Katsura, ``{Exact results for nonlinear Drude
  weights in the spin-1/2 XXZ chain},'' 2021.

\bibitem{favaHydrodynamic2021}
M.~Fava, S.~Biswas, S.~Gopalakrishnan, R.~Vasseur, and S.~A. Parameswaran,
  ``Hydrodynamic non-linear response of interacting integrable systems,'' 2021.

\bibitem{Bernard_2016}
D.~Bernard and B.~Doyon, ``Conformal field theory out of equilibrium: a
  review,'' {\em Journal of Statistical Mechanics: Theory and Experiment},
  vol.~2016, p.~064005, jun 2016.

\bibitem{doyonFinite2007}
B.~Doyon, ``Finite-temperature form factors: a review,'' {\em SIGMA}, vol.~3,
  p.~011, 2007.

\bibitem{Chen_2014}
Y.~Chen and B.~Doyon, ``Form factors in equilibrium and non-equilibrium mixed
  states of the ising model,'' {\em Journal of Statistical Mechanics: Theory
  and Experiment}, vol.~2014, p.~P09021, sep 2014.

\bibitem{chetriteNonequilibrium2015}
R.~Chetrite and H.~Touchette, ``Nonequilibrium markov processes conditioned on
  large deviations,'' {\em Annales Henri Poincar{\'e}}, vol.~16, no.~9,
  pp.~2005--2057, 2015.

\bibitem{levitovlesovik}
S.~Levitov and G.~Lesovik, ``Charge distribution in quantum shot noise,'' {\em
  JETP Letters}, vol.~58, p.~230, 1993.

\bibitem{avronFredholm2008}
J.~E. Avron, S.~Bachmann, G.~M. Graf, and I.~Klich, ``Fredholm determinants and
  the statistics of charge transport,'' {\em Communications in Mathematical
  Physics}, vol.~280, no.~3, pp.~807--829, 2008.

\bibitem{gawedzkikozlowski1}
K.~Gawedzki and K.~K. Kozlowski, ``Full counting statistics of energy transfers
  in inhomogeneous nonequilibrium states of (1+1)d cft,'' 2019.

\bibitem{gawedzkikozlowski2}
K.~Gawedzki and K.~K. Kozlowski, ``Large deviations of energy transfers in
  nonequilibrium cft and asymptotics of non-local riemann-hilbert problems,''
  2020.

\bibitem{RevModPhys.81.1665}
M.~Esposito, U.~Harbola, and S.~Mukamel, ``Nonequilibrium fluctuations,
  fluctuation theorems, and counting statistics in quantum systems,'' {\em Rev.
  Mod. Phys.}, vol.~81, pp.~1665--1702, Dec 2009.

\bibitem{Bernard_2013}
D.~Bernard and B.~Doyon, ``Time-reversal symmetry and fluctuation relations in
  non-equilibrium quantum steady states,'' {\em Journal of Physics A:
  Mathematical and Theoretical}, vol.~46, p.~372001, aug 2013.

\bibitem{McCoyWuBOOK}
B.~McCoy and T.~T. Wu, {\em {The two-dimensional Ising model}}.
\newblock Harvard Univ. Press, Cambridge, Mass., 1973.

\bibitem{PhysRevA.4.2331}
B.~M. McCoy, E.~Barouch, and D.~B. Abraham, ``Statistical mechanics of the
  $\mathrm{XY}$ model. iv. time-dependent spin-correlation functions,'' {\em
  Phys. Rev. A}, vol.~4, pp.~2331--2341, Dec 1971.

\bibitem{MussardoBOOK}
G.~Mussardo, {\em Statistical Field Theory}.
\newblock Oxford University Press, 2010.

\bibitem{KorepinBOOK}
V.~E. Korepin, N.~M. Bogoliubov, and A.~G. Izergin, {\em Quantum Inverse
  Scattering Method and Correlation Functions}.
\newblock Cambridge: Cambridge Univ. Press, 1993.

\bibitem{10.21468/SciPostPhysLectNotes.6}
S.~Pakuliak, E.~Ragoucy, and N.~Slavnov, ``{Nested Algebraic Bethe Ansatz in
  integrable models: recent results},'' {\em SciPost Phys. Lect. Notes}, p.~6,
  2018.

\bibitem{GERRARD2020115021}
A.~Gerrard and V.~Regelskis, ``{Nested algebraic Bethe ansatz for deformed
  orthogonal and symplectic spin chains},'' {\em Nuclear Physics B}, vol.~956,
  p.~115021, 2020.

\bibitem{Kozlowski_2017}
K.~Kozlowski, E.~K. Sklyanin, and A.~Torrielli, ``Quantization of the
  kadomtsev–petviashvili equation,'' {\em Theoretical and Mathematical
  Physics}, vol.~192, p.~1162–1183, Aug 2017.

\bibitem{Maillet_2018}
J.~M. Maillet and G.~Niccoli, ``On quantum separation of variables,'' {\em
  Journal of Mathematical Physics}, vol.~59, p.~091417, Sep 2018.

\bibitem{1984_Izergin_CMP_94}
A.~G. Izergin and V.~E. Korepin, ``The quantum inverse scattering method
  approach to correlation functions,'' {\em Commun. Math. Phys.}, vol.~94,
  pp.~67--92, 1984.

\bibitem{1989_Slavnov_TMP_79}
N.~A. Slavnov, ``Calculation of scalar products of wave functions and form
  factors in the framework of the algebraic {B}ethe {A}nsatz,'' {\em Theor.
  Math. Phys.}, vol.~79, p.~502, 1989.

\bibitem{Kojima_1997}
T.~Kojima, V.~E. Korepin, and N.~A. Slavnov, ``Determinant representation for
  dynamical correlation functions of the quantum nonlinear schrödinger
  equation,'' {\em Communications in Mathematical Physics}, vol.~188,
  p.~657–689, Oct 1997.

\bibitem{1990_Slavnov_TMP_82}
N.~A. Slavnov, ``{Nonequal-time current correlation function in a
  one-dimensional Bose gas},'' {\em Theor. Math. Phys.}, vol.~82, p.~273, 1990.

\bibitem{1999_Kitanine_NPB_554}
N.~Kitanine, J.~M. Maillet, and V.~Terras, ``Form factors of the {XXZ}
  {H}eisenberg finite chain,'' {\em Nucl. Phys. B}, vol.~554, no.~3, pp.~647 --
  678, 1999.

\bibitem{Belliard2013}
S.~Belliard, S.~Pakuliak, E.~Ragoucy, and N.~A. Slavnov, ``Form factors {in
  SU}(3)-invariant integrable models,'' {\em Journal of Statistical Mechanics:
  Theory and Experiment}, vol.~2013, p.~P04033, Apr. 2013.

\bibitem{Fuksa2017}
J.~Fuksa and N.~A. Slavnov, ``Form factors of local operators in supersymmetric
  quantum integrable models,'' {\em Journal of Statistical Mechanics: Theory
  and Experiment}, vol.~2017, p.~043106, Apr. 2017.

\bibitem{Pakuliak2014}
S.~Z. Pakuliak, E.~Ragoucy, and N.~A. Slavnov, ``Determinant representations
  for form factors in quantum integrable models with the {GL}(3)-invariant
  r-matrix,'' {\em Theoretical and Mathematical Physics}, vol.~181,
  pp.~1566--1584, Dec. 2014.

\bibitem{Hutsalyuk2016}
A.~Hutsalyuk, A.~Liashyk, S.~Pakuliak, E.~Ragoucy, and N.~Slavnov, ``Form
  factors of the monodromy matrix entries ingl(2$\vert$1)-invariant integrable
  models,'' {\em Nuclear Physics B}, vol.~911, pp.~902--927, Oct. 2016.

\bibitem{KAROWSKI1978455}
M.~Karowski and P.~Weisz, ``Exact form factors in (1 + 1)-dimensional field
  theoretic models with soliton behaviour,'' {\em Nuclear Physics B}, vol.~139,
  no.~4, pp.~455 -- 476, 1978.

\bibitem{SmirnovBOOK}
F.~A. Smirnov, {\em Form Factors in Completely Integrable Models of Quantum
  Field Theory}.
\newblock World Scientific, 1992.

\bibitem{Leclair1996}
A.~Leclair, F.~Lesage, S.~Sachdev, and H.~Saleur, ``Finite temperature
  correlations in the one-dimensional quantum ising model,'' {\em Nuclear
  Physics B}, vol.~482, pp.~579--612, dec 1996.

\bibitem{Saleur:1999hq}
H.~Saleur, ``{A Comment on finite temperature correlations in integrable
  QFT},'' {\em Nucl. Phys.}, vol.~B567, pp.~602--610, 2000.

\bibitem{Leclair_1996}
A.~Leclair, F.~Lesage, S.~Sachdev, and H.~Saleur, ``{Finite temperature
  correlations in the one-dimensional quantum Ising model},'' {\em Nuclear
  Physics B}, vol.~482, p.~579–612, Dec 1996.

\bibitem{Castro_Alvaredo_2002}
O.~Castro-Alvaredo and A.~Fring, ``Finite temperature correlation functions
  from form factors,'' {\em Nuclear Physics B}, vol.~636, p.~611–631, Aug
  2002.

\bibitem{Doyon_2005}
B.~Doyon, ``Finite-temperature form factors in the free majorana theory,'' {\em
  Journal of Statistical Mechanics: Theory and Experiment}, vol.~2005,
  p.~P11006–P11006, Nov 2005.

\bibitem{Pozsgay:2007kn}
B.~Pozsgay and G.~Takacs, ``{Form-factors in finite volume I: Form-factor
  bootstrap and truncated conformal space},'' {\em Nucl. Phys.}, vol.~B788,
  pp.~167--208, 2008.

\bibitem{Caux_2016}
J.-S. Caux, ``The quench action,'' {\em Journal of Statistical Mechanics:
  Theory and Experiment}, vol.~2016, p.~064006, Jun 2016.

\bibitem{Gamayun_2021}
O.~Gamayun, N.~Iorgov, and Y.~Zhuravlev, ``Effective free-fermionic form
  factors and the xy spin chain,'' {\em SciPost Physics}, vol.~10, Mar 2021.

\bibitem{babenko2020exact}
C.~Babenko, F.~Göhmann, K.~K. Kozlowski, J.~Sirker, and J.~Suzuki, ``{Exact
  real-time longitudinal correlation functions of the massive XXZ chain},''
  2020.

\bibitem{1742-5468-2015-2-P02019}
J.~D. Nardis and M.~Panfil, ``{Density form factors of the 1D Bose gas for
  finite entropy states},'' {\em Journal of Statistical Mechanics: Theory and
  Experiment}, vol.~2015, no.~2, p.~P02019, 2015.

\bibitem{SciPostPhys.1.2.015}
J.~D. Nardis and M.~Panfil, ``{Exact correlations in the Lieb-Liniger model and
  detailed balance out-of-equilibrium},'' {\em SciPost Phys.}, vol.~1, p.~015,
  2016.

\bibitem{Bootstrap_SciPost}
A.~C. Cubero and M.~Panfil, ``{Generalized hydrodynamics regime from the
  thermodynamic bootstrap program},'' {\em SciPost Phys.}, vol.~8, p.~4, 2020.

\bibitem{cubero2020generalized}
A.~C. Cubero, ``How generalized hydrodynamics time evolution arises from a form
  factor expansion,'' 2020.

\bibitem{TakatoReview}
T.~Yoshimura, A.~C. Cubero, and H.~Spohn, ``to appear,'' 2021.

\bibitem{Panfil_2021}
M.~Panfil, ``The two particle–hole pairs contribution to the dynamic
  correlation functions of quantum integrable models,'' {\em Journal of
  Statistical Mechanics: Theory and Experiment}, vol.~2021, p.~013108, Jan
  2021.

\bibitem{10.21468/SciPostPhys.5.5.054}
B.~Doyon, ``{Exact large-scale correlations in integrable systems out of
  equilibrium},'' {\em SciPost Phys.}, vol.~5, p.~54, 2018.

\bibitem{Mossel_2012}
J.~Mossel and J.-S. Caux, ``{Generalized {TBA} and generalized Gibbs},'' {\em
  Journal of Physics A: Mathematical and Theoretical}, vol.~45, p.~255001, may
  2012.

\bibitem{PhysRevLett.95.204101}
G.~A. El and A.~M. Kamchatnov, ``Kinetic equation for a dense soliton gas,''
  {\em Phys. Rev. Lett.}, vol.~95, p.~204101, Nov 2005.

\bibitem{PhysRevLett.120.045301}
B.~Doyon, T.~Yoshimura, and J.-S. Caux, ``Soliton gases and generalized
  hydrodynamics,'' {\em Phys. Rev. Lett.}, vol.~120, p.~045301, Jan 2018.

\bibitem{GennadyEl}
G.~El, ``to appear,'' 2021.

\bibitem{Doyon2018}
B.~Doyon, H.~Spohn, and T.~Yoshimura, ``A geometric viewpoint on generalized
  hydrodynamics,'' {\em Nucl. Phys. B}, vol.~926, pp.~570--583, jan 2018.

\bibitem{Fujimoto_1998}
S.~Fujimoto and N.~Kawakami, ``Exact drude weight for the one-dimensional
  hubbard model at finite temperatures,'' {\em Journal of Physics A:
  Mathematical and General}, vol.~31, pp.~465--474, jan 1998.

\bibitem{Klumper_2002}
A.~Klümper and K.~Sakai, ``The thermal conductivity of the spin-1/2 {XXZ}
  chain at arbitrary temperature,'' {\em Journal of Physics A: Mathematical and
  General}, vol.~35, pp.~2173--2182, feb 2002.

\bibitem{Sakai_2003}
K.~Sakai and A.~Klümper, ``Non-dissipative thermal transport in the massive
  regimes of {the XXZ chain},'' {\em Journal of Physics A: Mathematical and
  General}, vol.~36, pp.~11617--11629, nov 2003.

\bibitem{IN_Drude}
E.~Ilievski and J.~De~Nardis, ``Microscopic origin of ideal conductivity in
  integrable quantum models,'' {\em Phys. Rev. Lett.}, vol.~119, p.~020602, Jul
  2017.

\bibitem{Bulchandani2018}
V.~B. Bulchandani, R.~Vasseur, C.~Karrasch, and J.~E. Moore, ``{Bethe-Boltzmann
  hydrodynamics and spin transport in the XXZ chain},'' {\em Phys. Rev. B},
  vol.~97, p.~045407, Jan 2018.

\bibitem{10.21468/SciPostPhysCore.3.2.016}
F.~S. Møller, G.~Perfetto, B.~Doyon, and J.~Schmiedmayer, ``{Euler-scale
  dynamical correlations in integrable systems with fluid motion},'' {\em
  SciPost Phys. Core}, vol.~3, p.~16, 2020.

\bibitem{10.21468/SciPostPhys.8.3.041}
F.~S. Møller and J.~Schmiedmayer, ``{Introducing iFluid: a numerical framework
  for solving hydrodynamical equations in integrable models},'' {\em SciPost
  Phys.}, vol.~8, p.~41, 2020.

\bibitem{PhysRevLett.124.140603}
P.~Ruggiero, P.~Calabrese, B.~Doyon, and J.~Dubail, ``Quantum generalized
  hydrodynamics,'' {\em Phys. Rev. Lett.}, vol.~124, p.~140603, Apr 2020.

\bibitem{AlbaReview}
V.~Alba, B.~Bertini, M.~Fagotti, L.~Piroli, and P.~Ruggiero,
  ``Generalized-hydrodynamic approach to inhomogeneous quenches: Correlations,
  entanglement and quantum effects,'' 2021.

\bibitem{10.21468/SciPostPhys.8.1.007}
J.~Myers, M.~J. Bhaseen, R.~J. Harris, and B.~Doyon, ``{Transport fluctuations
  in integrable models out of equilibrium},'' {\em SciPost Phys.}, vol.~8,
  p.~7, 2020.

\bibitem{perfetto2020euler}
G.~Perfetto and B.~Doyon, ``Euler-scale dynamical fluctuations in
  non-equilibrium interacting integrable systems,'' 2021.

\bibitem{Panfil2021}
M.~Panfil, ``The two particle{\textendash}hole pairs contribution to the
  dynamic correlation functions of quantum integrable models,'' {\em Journal of
  Statistical Mechanics: Theory and Experiment}, vol.~2021, p.~013108, Jan.
  2021.

\bibitem{Boldrighini1997}
C.~Boldrighini and Y.~Suhov, ``One-dimensional hard-rod caricature of
  hydrodynamics: "navier-stokes correction" for local equilibrium initial
  states,'' {\em Communications in Mathematical Physics}, vol.~189,
  pp.~577--590, Nov. 1997.

\bibitem{ghkv}
S.~Gopalakrishnan, D.~A. Huse, V.~Khemani, and R.~Vasseur, ``Hydrodynamics of
  operator spreading and quasiparticle diffusion in interacting integrable
  systems,'' {\em Phys. Rev. B}, vol.~98, p.~220303, Dec 2018.

\bibitem{PhysRevB.54.10845}
P.~Fendley and H.~Saleur, ``Nonequilibrium dc noise in a luttinger liquid with
  an impurity,'' {\em Phys. Rev. B}, vol.~54, pp.~10845--10854, Oct 1996.

\bibitem{Bazhanov1996}
V.~V. Bazhanov, S.~L. Lukyanov, and A.~B. Zamolodchikov, ``Integrable structure
  of conformal field theory, quantum {KdV} theory and thermodynamic bethe
  ansatz,'' {\em Communications in Mathematical Physics}, vol.~177,
  pp.~381--398, Apr. 1996.

\bibitem{zamolodchikov2012quantum}
A.~Zamolodchikov, A.~Belavin, A.~Zamolodchikov, and Y.~Pugai, {\em Quantum
  Field Theories in Two Dimensions: Collected Works of Alexei Zamolodchikov}.
\newblock No.~v. 2 in Collected works of Alexei Zamolodchikov, World
  Scientific, 2012.

\bibitem{Zamolodchikov:2004ce}
A.~B. Zamolodchikov, ``{Expectation value of composite field T anti-T in
  two-dimensional quantum field theory},'' 1 2004.

\bibitem{Dubovsky2012}
S.~Dubovsky, R.~Flauger, and V.~Gorbenko, ``Solving the simplest theory of
  quantum gravity,'' {\em Journal of High Energy Physics}, vol.~2012, Sept.
  2012.

\bibitem{Caselle2013}
M.~Caselle, D.~Fioravanti, F.~Gliozzi, and R.~Tateo, ``Quantisation of the
  effective string with {TBA},'' {\em Journal of High Energy Physics},
  vol.~2013, July 2013.

\bibitem{Cavagli2016}
A.~Cavagli{\`{a}}, S.~Negro, I.~M. Sz{\'{e}}cs{\'{e}}nyi, and R.~Tateo,
  ``{$T\overline{T}$}-deformed 2d quantum field theories,'' {\em Journal of
  High Energy Physics}, vol.~2016, Oct. 2016.

\bibitem{Conti2019}
R.~Conti, S.~Negro, and R.~Tateo, ``The {$T\overline{T}$} perturbation and its
  geometric interpretation,'' {\em Journal of High Energy Physics}, vol.~2019,
  Feb. 2019.

\bibitem{PhysRevLett.126.121601}
M.~Medenjak, G.~Policastro, and T.~Yoshimura, ``{$T\overline{T}$}-deformed
  conformal field theories out of equilibrium,'' {\em Phys. Rev. Lett.},
  vol.~126, p.~121601, Mar 2021.

\bibitem{PhysRevD.103.066012}
M.~Medenjak, G.~Policastro, and T.~Yoshimura, ``Thermal transport in
  {$T\overline{T}$}-deformed conformal field theories: From integrability to
  holography,'' {\em Phys. Rev. D}, vol.~103, p.~066012, Mar 2021.

\bibitem{McGough:2016lol}
L.~McGough, M.~Mezei, and H.~Verlinde, ``{Moving the CFT into the bulk with $
  T\overline{T} $},'' {\em JHEP}, vol.~04, p.~010, 2018.

\bibitem{Guica:2017lia}
M.~Guica, ``{An integrable Lorentz-breaking deformation of two-dimensional
  CFTs},'' {\em SciPost Phys.}, vol.~5, no.~5, p.~048, 2018.

\bibitem{cardy2021toverline}
J.~Cardy and B.~Doyon, ``$t{\overline t}$ deformations and the width of
  fundamental particles,'' 2021.

\bibitem{PhysRevLett.106.217206}
T.~Prosen, ``{Open $XXZ$ Spin Chain: Nonequilibrium Steady State and a Strict
  Bound on Ballistic Transport},'' {\em Phys. Rev. Lett.}, vol.~106, p.~217206,
  May 2011.

\bibitem{PhysRevLett.111.057203}
T.~Prosen and E.~Ilievski, ``Families of quasilocal conservation laws and
  quantum spin transport,'' {\em Phys. Rev. Lett.}, vol.~111, p.~057203, Aug
  2013.

\bibitem{Benz2005}
J.~Benz, T.~Fukui, A.~Kl\"{u}mper, and C.~Scheeren, ``{On the Finite
  Temperature Drude Weight of the Anisotropic Heisenberg Chain},'' {\em Journal
  of the Physical Society of Japan}, vol.~74, pp.~181--190, Jan. 2005.

\bibitem{PhysRevB.90.155104}
C.~Karrasch, D.~M. Kennes, and J.~E. Moore, ``Transport properties of the
  one-dimensional hubbard model at finite temperature,'' {\em Phys. Rev. B},
  vol.~90, p.~155104, Oct 2014.

\bibitem{PhysRevB.91.115130}
C.~Karrasch, D.~M. Kennes, and F.~Heidrich-Meisner, ``Spin and thermal
  conductivity of quantum spin chains and ladders,'' {\em Phys. Rev. B},
  vol.~91, p.~115130, Mar 2015.

\bibitem{Ljubotina2017}
M.~Ljubotina, M.~{\v{Z}}nidari{\v{c}}, and T.~Prosen, ``Spin diffusion from an
  inhomogeneous quench in an integrable system,'' {\em Nature Communications},
  vol.~8, July 2017.

\bibitem{gv_superdiffusion}
S.~Gopalakrishnan and R.~Vasseur, ``{Kinetic Theory of Spin Diffusion and
  Superdiffusion in $XXZ$ Spin Chains},'' {\em Phys. Rev. Lett.}, vol.~122,
  p.~127202, Mar 2019.

\bibitem{1963_Lieb_PR_130_1}
E.~H. Lieb and W.~Liniger, ``{Exact Analysis of an Interacting Bose Gas. I. The
  General Solution and the Ground State},'' {\em Phys. Rev.}, vol.~130, no.~4,
  pp.~1605--1616, 1963.

\bibitem{PhysRevLett.121.160603}
J.~De~Nardis, D.~Bernard, and B.~Doyon, ``Hydrodynamic diffusion in integrable
  systems,'' {\em Phys. Rev. Lett.}, vol.~121, p.~160603, Oct 2018.

\bibitem{10.21468/SciPostPhysCore.1.1.002}
M.~Panfil and J.~Pawełczyk, ``{Linearized regime of the generalized
  hydrodynamics with diffusion},'' {\em SciPost Phys. Core}, vol.~1, p.~2,
  2019.

\bibitem{Babelon2003}
O.~Babelon, D.~Bernard, and M.~Talon, {\em Introduction to Classical Integrable
  Systems}.
\newblock Cambridge University Press, 2003.

\bibitem{Bastianello_2018}
A.~Bastianello, B.~Doyon, G.~Watts, and T.~Yoshimura, ``Generalized
  hydrodynamics of classical integrable field theory: the sinh-gordon model,''
  {\em SciPost Physics}, vol.~4, Jun 2018.

\bibitem{DeLuca2016}
A.~D. Luca and G.~Mussardo, ``Equilibration properties of classical integrable
  field theories,'' {\em Journal of Statistical Mechanics: Theory and
  Experiment}, vol.~2016, p.~064011, June 2016.

\bibitem{NEGRO2013166}
S.~Negro and F.~Smirnov, ``On one-point functions for sinh-gordon model at
  finite temperature,'' {\em Nuclear Physics B}, vol.~875, no.~1, pp.~166--185,
  2013.

\bibitem{doi:10.1142/S0217751X14501115}
S.~Negro, ``On sinh–gordon thermodynamic bethe ansatz and fermionic basis,''
  {\em International Journal of Modern Physics A}, vol.~29, no.~20, p.~1450111,
  2014.

\bibitem{PhysRevLett.87.081601}
G.~Policastro, D.~T. Son, and A.~O. Starinets, ``Shear viscosity of strongly
  coupled $n\phantom{\rule{0ex}{0ex}}=\phantom{\rule{0ex}{0ex}}4$
  supersymmetric yang-mills plasma,'' {\em Phys. Rev. Lett.}, vol.~87,
  p.~081601, Aug 2001.

\bibitem{jin2021interplay}
T.~Jin, T.~Gautié, A.~Krajenbrink, P.~Ruggiero, and T.~Yoshimura, ``Interplay
  between transport and quantum coherences in free fermionic systems,'' 2021.

\bibitem{PhysRevB.102.041117}
M.~Medenjak, B.~Bu\ifmmode~\check{c}\else \v{c}\fi{}a, and D.~Jaksch,
  ``{Isolated Heisenberg magnet as a quantum time crystal},'' {\em Phys. Rev.
  B}, vol.~102, p.~041117, Jul 2020.

\bibitem{10.21468/SciPostPhys.9.1.003}
M.~Medenjak, T.~Prosen, and L.~Zadnik, ``{Rigorous bounds on dynamical response
  functions and time-translation symmetry breaking},'' {\em SciPost Phys.},
  vol.~9, p.~3, 2020.

\bibitem{buca2020quantum}
B.~Buca, A.~Purkayastha, G.~Guarnieri, M.~T. Mitchison, D.~Jaksch, and
  J.~Goold, ``Quantum many-body attractors,'' 2020.

\end{thebibliography}

\end{document}